\documentclass[apj]{emulateapj}
\usepackage[colorlinks,linkcolor={blue},citecolor={blue},urlcolor={red}]{hyperref}
\usepackage{epsfig,graphicx,natbib,amsmath,amsfonts,amssymb}%,xfrac}
\usepackage{color}
\usepackage{multirow}
\usepackage{booktabs}
\usepackage{amssymb}
\usepackage{threeparttablex} % for "ThreePartTable" environment
\usepackage{booktabs} 

\usepackage{longtable}
\usepackage{array} % for extrarowheight
\usepackage[dvipsnames]{xcolor}  

\newcommand{\re}{\Reff}

\newcommand{\dindex}{D$_n$(4000)}

\newcommand{\Reff}{{$R_{\rm e}$}}

\newcommand{\Msun}{${\rm  M}_{\odot}$}

\newcommand{\HI}{H{\sc{I}}}

\newcommand{\sfr}{SFR$_{\rm <Re}$}
\newcommand{\mstar}{$M_{\rm \ast, <Re}$}
\newcommand{\lgmstar}{$\log_{10}M_{\rm \ast, <Re}$}
\newcommand{\lgsfr}{$\log_{10}$SFR$_{\rm <Re}$}

\newcommand{\sigsfr}{$\Sigma_{\rm SFR}$}
\newcommand{\sigmstar}{$\Sigma_{\ast}$}

\newcommand{\dsigsfr}{$\Delta\Sigma_{\rm SFR}$}

\newcommand{\myemail}{\email{enci.wang@phys.ethz.ch}}

\shorttitle{Elevation and suppression of star formation within galaxies}
\shortauthors{Wang et al.}

\graphicspath{{fig/}}
\begin{document}

\title{On the elevation and suppression of star formation within galaxies} 
\author{
Enci Wang\altaffilmark{1},
Simon J. Lilly\altaffilmark{1}, 
Gabriele Pezzulli\altaffilmark{1},
and Jorryt Matthee\altaffilmark{1}
} \myemail

\altaffiltext{1}{Department of Physics, ETH Zurich, Wolfgang-Pauli-strasse 27, CH-8093 Zurich, Switzerland}

\begin{abstract}  
To understand star formation in galaxies, we investigate the star formation rate (SFR) surface density (\sigsfr) profiles for galaxies, based on a well-defined sample of 976 star-forming MaNGA galaxies. 
We find that the typical \sigsfr\ profiles within 1.5\re\ of normal SF galaxies can be well described by an exponential function for different stellar mass intervals, while the sSFR profile shows positive gradients, especially for more massive SF galaxies. This is due to the more pronounced central cores or bulges rather than the onset of a ``quenching'' process. 
While galaxies that lie significantly above (or below) the star formation main sequence (SFMS) show overall an elevation (or suppression) of \sigsfr\ at all radii, this central elevation (or suppression) is more pronounced in more massive galaxies.  The degree of central enhancement and suppression is quite symmetric, suggesting that both the elevation and suppression of star formation are following the same physical processes.  Furthermore, we find that the dispersion in \sigsfr\ within and across the population is found to be tightly correlated with the inferred gas depletion time, whether based on the stellar surface mass density or the orbital dynamical time.  This suggests that we are seeing the response of a simple gas-regulator system to variations in the accretion rate.  This is explored using a heuristic model that can quantitatively explain the dependence of $\sigma$(\sigsfr) on gas depletion timescale. Variations in accretion rate are progressively more damped out in regions of low star-formation efficiency leading to a reduced amplitude of variations in star-formation.  % This idea is also supported by re-examination of some other quite independent relations in the literature, including the comparison of compact and extended star-forming galaxies in \cite{Wang-Kong-Pan-18}. 
  
\end{abstract}

\section{Introduction}
\label{sec:introduction}

%A, the build-up of the SFMS of galaxies  .... definition of SF and quenched galaxies ...
%     while where and how SF boosting or quenching occurs in galaxies still not well understood. 
It has long been established that most star-forming (SF) galaxies lie on a narrow sequence in the stellar mass-star formation rate (SFR) plane, with a scatter of $\sim$0.3 dex in the specific star-formation rate (sSFR).  This has been established both in the local universe and up to at least $z \sim 3$.  This tight relation is widely known as the star formation Main Sequence \citep[SFMS; e.g.][]{Brinchmann-04, Daddi-07, Noeske-07,
Salim-07, Elbaz-11}.   Galaxies that lie significantly above, or below, the SFMS are forming stars more, or less, actively than ``normal'' SF galaxies of the same stellar mass.  There is also a large population of galaxies with SFR that are one or more dex below the SFMS.  They are usually defined as quenched or ``passive'' galaxies.  
%The prevalence of the quenched population has significantly increased since redshift of 1 with respect to the SF population \citep[][]{Muzzin-13, Tomczak-14},  suggesting a strong evolution from SF galaxies and quenched population over the past 8 Gyrs.  
Despite this progress in defining the star-formation properties of galaxies across the population, the questions of how, when and where the SFR of galaxies is elevated,  suppressed or even quenched are still not well understood, either observationally or theoretically. 

%B, the driven mechanisms of SF boosting or quenching .... 
%      SF boosting: the bar, interaction, mergers, disk instabilities ... 
%      SF quenching: internal and external  ... 
Many mechanisms have been proposed to account for the elevation or suppression (including quenching) of star formation in galaxies. These mechanisms can be generally categorized into two types: internal processes and externally driven environmental effects.  Both of these may produce either a change in the cold gas content or a change in the star formation efficiency (SFE), defined as the SFR per unit mass of cold gas.  Internal processes that may enhance the star formation include disk instabilities \citep{Dekel-Burkert-14, Zolotov-15, Tacchella-16a}, the existence of bars \citep{ Wang-12, Lin-17, Chown-18}, and positive feedback from an AGN \citep{Silk-Nusser-10, Mahoro-Povic-Nkundabakura-17, Kalfountzou-17}, usually via an increase in the SFE.  For instance, the radial inflow of cold gas would lead to an increase of gas density in the inner region of galaxies and further an increase of SFE, according to the Kennicutt-Schmidt law \citep{Kennicutt-98}. However, internal processes may also suppress star formation, either through galactic winds driven by stellar feedback \citep[e.g.][]{Ceverino-Klypin-09, Muratov-15, El-Badry-16}, through negative AGN feedback \citep{Nulsen-05, McNamara-Nulsen-07, Fabian-12, Cicone-14} and through so-called morphological quenching \citep{Martig-09}, either by stripping (or heating) the cold gas, or stabilizing the gas disk against star formation.  Environmental processes, such as tidal or ram-pressure stripping \citep[e.g.][]{Gunn-Gott-72, Moore-96, Abadi-Moore-Bower-99, Poggianti-17}, or the interactions and mergers of galaxies \citep[e.g.][]{Conselice-Chapman-Windhorst-03, Cox-06, Smethurst-15}, can both enhance the star formation on a relative short timescale by compressing the cold gas and causing the gas disk instabilities to  enhance the SFE, but subsequently then suppress the star formation on longer timescales by removing the cold gas reservoir and/or exhausting the cold gas. 

%C, the link between quenching and structural properties, and the argument from Lilly et al. 2016. 
%the negative sSFR may have nothing with quenching, and using SFR profile is more direct way to study the star formation and quenching in galaxies.  
Observationally,  the degree of star formation activity or quiescence of galaxies has been found to be closely related to many galaxy properties, such as the stellar mass, the host halo mass, the position in the host halo and structural properties \citep[e.g.][]{Weinmann-06, vandenBosch-08, Peng-10, Peng-12, Knobel-15, Wang-18b, Wang-18c}.  However, it is still unclear which is the driving physical parameter for quenching. On the one hand, the fraction of quenched galaxies in the population is found to be most closely related to the bulge mass or the central stellar mass surface density \citep[e.g.][]{Bell-12, Wake-vanDokkum-Franx-12, Fang-13, Omand-Balogh-Poggianti-14, Barro-17}, which is often taken as evidence that the internal structure of galaxies must play a major role in  quenching star formation.  On the other hand, \cite{Lilly-Carollo-16} have suggested that the apparent close relationship between the inner structural properties and the quenched fraction of galaxies may not be fundamental at all. They showed that this close correlation can be naturally generated from the observed evolution of the size-mass relation of SF galaxies without invoking any physical links at all between inner structural properties and quenching.  Interestingly, their toy model can also broadly reproduce the observed gradients of specific star formation within galaxies caused by the inside-out assembly of star-forming galaxies.  
%The radial gradients of sSFR overall reflect the mode of the build-up of stellar mass (inside-out or outside-in) as well as the star formation in galaxies, which should not be directly treated as the star formation indicator.   

This emphasizes that, although the overall sSFR of a galaxy is clearly the (only) indicator of quenching, study of the sSFR {\it gradients} within galaxies may not be relevant for understanding the quenching process, because they can arise from quite unrelated processes.  In this paper, we therefore focus primarily on the SFR gradients (i.e. the radial dependence of the surface density of SFR), rather than sSFR gradients, as a more reasonable and direct way to investigate the star formation, and possible quenching, in galaxies.

%D, Indeed,  the powerful of IFU surveys, enable us to study galaxies of different galactic radius statistically.  
%the findings of ``inside-out'' assembly mode of galaxies, based on the IFU surveys or long-slit spectra. 
%E,  In spite of the inside-out growth picture, whether the bulge can account for the negative sSFR profile? 
%This leads to a ``inside-out'' quenching scenario, which is different from, but related to ``inside-out'' growth. 
%The previous  results using SFR profile, Ellison et al. 2018, Nelson et al. 2016, Gonzalez Delgado et al. 2016, Spindler et al. 2018, Medling et al. 2018 ...  SF boost and suppression from the center outwards.  
The widespread deployment of integral field spectrographs (IFS) is producing many spatially resolved spectroscopic surveys of relatively nearby galaxies, such as MaNGA \citep{Bundy-15, Blanton-17}, CALIFA \citep{Sanchez-12} and SAMI \citep{Croom-12}.  These all show the well-known fact that galaxies with stellar mass greater than $10^{10}$\Msun\ generally exhibit older stellar population in galactic centers than outward regions \citep[e.g.][]{Perez-13, Li-15, Ibarra-Medel-16,  Goddard-17, Wang-18a, Rowlands-18}, suggesting an inside-out build-up of massive galaxies. This inside-out growth scenario is also required by the observed fact that the size of galaxies has been increasing over time \citep[e.g.][]{Toft-07, Williams-10, Newman-12, vanderWel-14, Shibuya-Ouchi-Harikane-15}.    The question then is whether the bulge or stellar core can fully account for the positive sSFR profile in massive SF galaxies. %, or what is the relation between central bulges and the star formation statuses in galaxies? 
By studying the sSFR profile for 814 SF and partially quenched galaxies, \cite{Belfiore-18} found that the sSFR profiles in partially quenched galaxies are suppressed at all galactocentric distance out to 2 effective radius compared to the SF galaxies of same mass.  They further pointed out that the sSFR suppression is not simply due to the contribution of a massive bulge. 
Consistent with this, \cite{Wang-18a} found that massive SF galaxies, with or without significant bulges, show strong gradients of sSFR, as indicated by the 4000 \AA\ break and H$\alpha$ equivalent width.   This suggested that the presence of a central dense object is not a main driving parameter of quenching, and that an inside-out cessation of star formation  (distinct from the inside-out assembly picture) might be indicated.  A critical examination of this idea is the major motivation for the current study. 

Thanks to the new IFS surveys, there have been many recent studies of the radial profile of the surface density of star-formation, \sigsfr(r), in galaxies \citep[e.g.][]{GonzalezDelgado-16, Nelson-16, Ellison-18, Spindler-18, Guo-19, Medling-18}. By studying the \sigsfr\ of several hundred galaxies selected from MaNGA, \cite{Ellison-18} found that galaxies significantly above the SFMS exhibit SFR enhancement at all galactocentric radii with the highest enhancement in the center. They also find that galaxies below 1 dex of the SFMS exhibit SFR suppression at all radii with the greatest suppression at the center.  
Similarly, \cite{Spindler-18} identified a bimodal distribution of sSFR profiles of MaNGA galaxies, namely centrally suppressed and centrally unsuppressed galaxies.  Similar results are also found in high redshift galaxies.  
By using the H$\alpha$ maps for a sample of 3200 SF galaxies at redshift of $\sim$1, \cite{Nelson-16} found a ``coherent star formation'' across the SFMS: galaxies above the SFMS exhibit enhanced H$\alpha$ emission at all radii, especially in the center, while galaxies below the SFMS exhibit suppressed H$\alpha$ emission at all radius especially in the center with respect to normal SFMS galaxies. 

%F, the basic idea of this work, and the relation with previous findings. What's new from the previous result and interpretation? \\
% - more general explanation of star formation and quenching; 
% - explain both elevation and suppression in galaxies, and the dependence on stellar mass and radius; 
% - providing the explanation of the structural dependence of Star formation main sequence; 
% - naturally producing the inside-out quenching, with no need of introducing black hole feedback and morphological quenching; 

Despite many works on the resolved star formation of galaxies, a satisfactory picture to understand the boosting and suppression, or even quenching, of star formation in galaxies is still lacking. In this paper, we construct the SFR surface density profiles in galaxies across the SFMS using the MaNGA released data \citep{Abolfathi-18} and present a more general perspective to understand both the elevation and the suppression of SFR in galaxies, and which can simultaneously explain the dependence of the fluctuation of \sigsfr\ on global stellar mass and on galactocentric distance.  We construct a toy model to quantitatively explain the variation of star formation rates in galaxies, both from galaxy to galaxy and radially within galaxies. While we do not explicitly consider quenching {\it per se}, our model may naturally explain the apparent inside-out quenching without invoking AGN feedback or other morphology-related processes, in terms of driving processes that are not themselves radially dependent.  Our main observational result in star formation rate suggests that both increases and decreases are primarily governed by the same general process of gas-regulation. 

Much of the interpretation developed in this paper will be based on the ``gas regulator'' model of galaxies from \cite{Lilly-13} and its response to variations in the inflow rates.   This gas regulator model is somewhat different from the so-called ``bath-tub'' model of galaxies \citep[e.g.][]{Bouche-10, Dave-Finlator-Oppenheimer-12}.  Although both are based on the same fundamental continuity equation, reflecting the conservation of mass, in the ``bath-tub'' model, the mass of gas in the ``reservoir'' in the galaxy is not allowed to change, i.e. $\dot{M}_{\rm gas} = 0$.  It is therefore not able to {\it regulate} the star-formation rate.  The gas regulator model allows the gas mass to change.  One consequence of this difference is that the gas regulator model of \cite{Lilly-13} produces a mass-metallicity relation that analytically has the SFR (or equivalently the gas fraction) as a second parameter. Therefore, it produces an analytic form of a Fundamental Metallicity Relation \citep[FMR;][]{Mannucci-10}, in which the form of the FMR reflects only the parameters of the regulator system, specifically the star-formation efficiency and the mass-loading of a wind.  In the current paper, we will see that it is precisely the dynamic response of the gas regulator to changes in the inflow rate that is of interest and which will underly the interpretation of the paper. 
 
%G., the paper organization
The paper is structured as follows. In Section \ref{sec:data}, we present the details of sample definition, spectral fitting and the measurement of parameters. In Section \ref{sec:results}, we present the SFR, sSFR and stellar surface density profiles of normal main sequence galaxies, as well as the SFR profiles of galaxies that are significantly above and below the SFMS. We further construct a quantitative toy model in Section \ref{sec:toymodel} to explain the observed result. In Section \ref{sec:discussion}, we discuss the implications and speculations of our interpretation for the observational result.  We summarize this work in Section \ref{sec:summary}.  Throughout this paper, we assume a flat cold dark matter cosmology model with $\Omega_m=0.27$, $\Omega_\Lambda=0.73$ and $h=0.7$ when computing distance-dependent parameters.

\section{Observational data}
\label{sec:data}

\subsection{Sample selection and definition}
\label{subsec:2.1}
%A, Pre-selection based on the Wang et al. 2018
%B, Selection SF galaxies ...
As one of the largest on-going IFS surveys, MaNGA aims at obtaining the 2-dimensional spectra for $\sim$10,000 galaxies with redshifts in the range of $0.01<z<0.15$ by the year 2020 \citep{Bundy-15}.  Utilizing the two dual-channel BOSS spectrographs at the 2.5 Sloan Telescope \citep{Gunn-06, Smee-13}, MaNGA is able to cover wavelengths of over 3600-10300\AA\ at R$\sim$2000 and reach a target $r$-band S/N=4-8 (\AA$^{-1}$ per 2 arcsec fiber) at 1-2\re\ with a typical integration time of 3 hr. The MaNGA IFS includes 12 seven-fiber ``mini-bundles'' for flux calibration and 17 science bundles of five different sizes with the corresponding on-sky diameters ranging 12 to 32 arcsec, on the Sloan 3 degree diameter field of view \citep{Drory-15}.  Flux calibration mainly accounts for the flux loss due to atmospheric absorption and instrument response, and is accurate to better than 5\% for more than 89\% of MaNGA's wavelength range \citep{Yan-16}. The typical effective spatial resolution of the MaNGA data cubes can be described by a Gaussian with FWHM$\sim$2.5 arcsec \citep{Law-15, Law-16}.
In the MaNGA sample design and optimization, three samples are defined: the primary, secondary and color-enhanced samples. The primary and secondary samples are selected to have a flat distribution of $i$-band absolute magnitude with the assigned IFS bundles covering 1.5\re\ and 2.5\re, respectively. The color-enhanced sample is selected to increase the fraction of rare populations in the color-magnitude diagram, such as low-mass red galaxies and high-mass blue galaxies.

\begin{figure}
\epsfig{figure=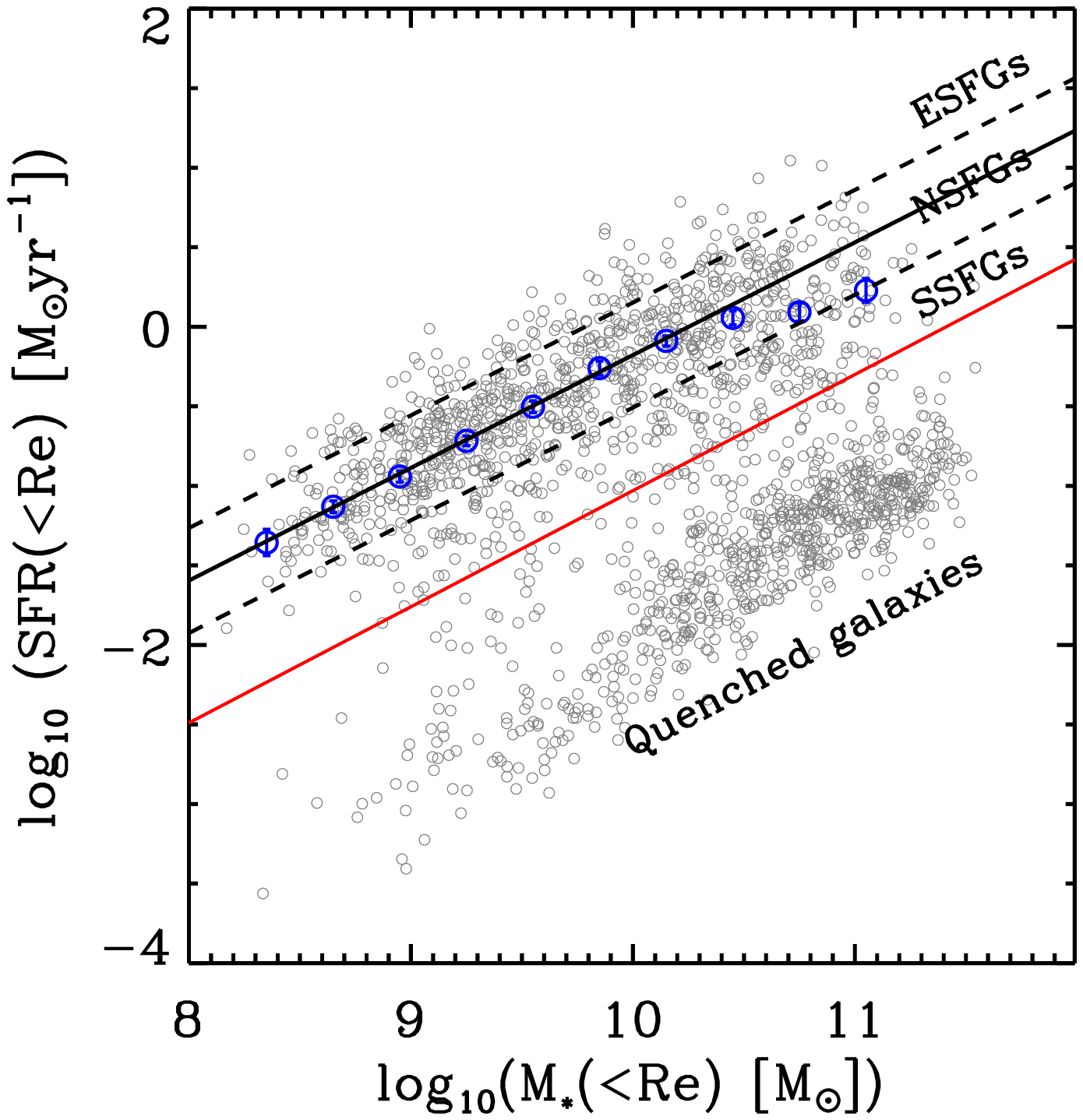,clip=true,width=0.45\textwidth}
\caption{The star formation main sequence for the sample galaxies selected from MaNGA. The red solid line is the demarcation line for SF galaxies and quenched galaxies, which is taken from \cite{Belfiore-18}.  The blue data points show the median \sfr\ of SF galaxies in different stellar mass bins. The black solid line shows the best-fit line to the median \sfr-\mstar\ relation of the sample galaxies with \mstar$<10^{10}$\Msun. The two dashed lines show the typical scatter of the SFMS of 0.33 dex, and are used to define different samples in this work.  }
\label{fig:SFMS}
\end{figure}

The sample of galaxies used in this work are selected from a sample of 1917 MaNGA galaxies in \cite{Wang-18a}, which are originally derived from the MaNGA primary and secondary samples of SDSS Data Release 14 \citep{Abolfathi-18}, excluding the mergers, irregulars and heavily disturbed galaxies ($\sim13\%$).   Based on these 1917 galaxies, we further exclude $\sim11.4\%$ of galaxies in which the median S/Ns of 5500\AA\ continuum are less than 3.0 at the effective radius of the galaxy in question, which results in a set of 1698 galaxies.  Figure \ref{fig:SFMS} shows the relation of stellar mass and SFR measured within \re\ ($M_{\rm *, <Re}$-SFR$_{\rm <Re}$)  for these galaxies. The $M_{\rm *, <Re}$ and SFR$_{\rm <Re}$  are obtained by integrating the stellar mass and SFR of all the spaxels within \re. These quantities within \re\ are used as the representatives of the overall stellar mass and SFR throughout the current work, but they are of course a factor of roughly two less than the actual integrated value.  The determination of the stellar mass and SFR of each spaxel are presented in Section \ref{subsec:2.2} and \ref{subsec:2.3}. 

 As expected, the galaxies are bimodally distributed on the $M_{\rm *, <Re}$-SFR$_{\rm <Re}$ diagram.   The red line separates the populations of galaxies that we define to be ``star-forming" and ``quenched".  For simplicity, this is taken from \cite{Belfiore-18} and is located 1.0 dex below the SFMS of the MaNGA galaxies with SFR and stellar mass determined by integrating within an aperture of 2.5\re.  According to this demarcation line, there are 976 SF galaxies and 722 quenched galaxies.  We note that the quenched population also exhibits a tight correlation in Figure \ref{fig:SFMS}, but this is an artificial effect derived from the methodology used in Section \ref{subsec:2.3}. The SFR measurements for quenched galaxies should not be fully trusted (details see \ref{subsec:2.3}). In the current work, we will only consider the SF galaxies (above the red line) and their \sigsfr\ profiles to investigate the rules governing the elevation and suppression of star formation.  

To define the SFMS, we compute the median $M_{\rm *, <Re}$-SFR$_{\rm <Re}$ relation of SF galaxies as a function of mass. This is shown with blue circles. We fit the median relation at the low mass end ($M_{\rm *, <Re}<10^{10}$\Msun) with a straight line denoted with the black solid line: \lgsfr $= 0.71\times$\lgmstar$-7.27$.  This solid line is nearly parallel to the red line in Figure \ref{fig:SFMS}, indicating that the slope of our SFMS is quite similar to that of \cite{Belfiore-18}.
We refer to this black straight line as the ``nominal SFMS'' in the following analysis of this paper.  

The representative scatter of the SF galaxies on \mstar-\sfr\ diagram is 0.33 dex, indicated by two dashed lines. This is calculated as the median of the scatter at different stellar masses.   Based on Figure \ref{fig:SFMS}, we then define a parameter  $\Delta$SFR to be the deviation of a given galaxy from the straight-line nominal SFMS. In other words this quantifies by how much the {\it overall} star formation in a galaxy is elevated or suppressed with respect to the ``typical'' galaxy at the same stellar mass as defined by the straight solid black line in Figure \ref{fig:SFMS} (and not the curved blue line).  

Furthermore, we then divide galaxies into four subsamples based on their $\Delta$SFR: galaxies with $0.33<\Delta$SFR, $0.0<\Delta$SFR$<0.33$, $-0.33<\Delta$SFR$<0.0$ and $\Delta$SFR$<-0.33$. According to the definition, galaxies with $0.33<\Delta$SFR are the galaxies with a significant elevation of star formation (referred as ESFGs), while galaxies with $\Delta$SFR$<-0.33$ are the galaxies with a significant suppression of star formation (referred as SSFGs). Galaxies within 0.33 dex of the nominal SFMS (i.e. between the black dashed lines) are referred as NSFGs.  

\subsection{Spectral fitting}
\label{subsec:2.2}

The spectral fittings are done to obtain the emission line fluxes spaxel-by-spaxel, following the same method as described in \cite{Li-15} and \cite{Wang-18a}. 
To model the stellar continuum and absorption lines, we use the $\chi^2$ minimization spectral fitting code developed by \cite{Li-05} assuming the Cardelli-Clayton-Mathis (CCM) attenuation curve \citep{Cardelli-Clayton-Mathis-89}. This code is efficient and stable, especially for spectra of low signal-to-noise ratio (S/N). More importantly, it can effectively mask the emission-line regions by iteration during the fitting, which is critical to accurately model the absorption pits and further obtain the SFR based on the emission lines \citep[see examples in][]{Li-15}.  The input templates are a set of nine galactic eigenspectra constructed based on the observed galactic spectra using the technique of principal component analysis. We refer the readers to \cite{Li-05} and \cite{Li-15} for a detailed description of the templates and tests of the spectral fitting results.  We then measure the emission lines based on the stellar component-subtracted spectrum by fitting a Gaussian profile to these lines.   However, the code developed by \cite{Li-05} is unable to provide the stellar mass of the spaxels. Thus, in this work the stellar mass map of the sample galaxies are derived from a public fitting code {\tt STARLIGHT} \citep{CidFernandes-04}. The templates used in the  {\tt STARLIGHT} fitting are a base of 45 single stellar populations from \cite{Bruzual-Charlot-03} with a \cite{Chabrier-03} initial mass function (IMF), which are evenly distributed on an age-metallicity grid with 15 ages ranging from 1 Myr to 13 Gyr and 3 different metallicities (Z = 0.01, 0.02, 0.05). 
In order to obtain the reliable measurements of SFR and stellar mass, only spaxels with S/N greater than 3 in the continuum at 5500\AA\ are considered in this work.  Throughout this work, all the measurements of physical properties, such as stellar mass and SFR, are based self-consistently from the MaNGA data alone, avoiding any potential inconsistencies that could be introduced by considering other data such as the SDSS single-fiber spectra.

\subsection{SFR determination}
\label{subsec:2.3}
\begin{figure*}
\center{
\epsfig{figure=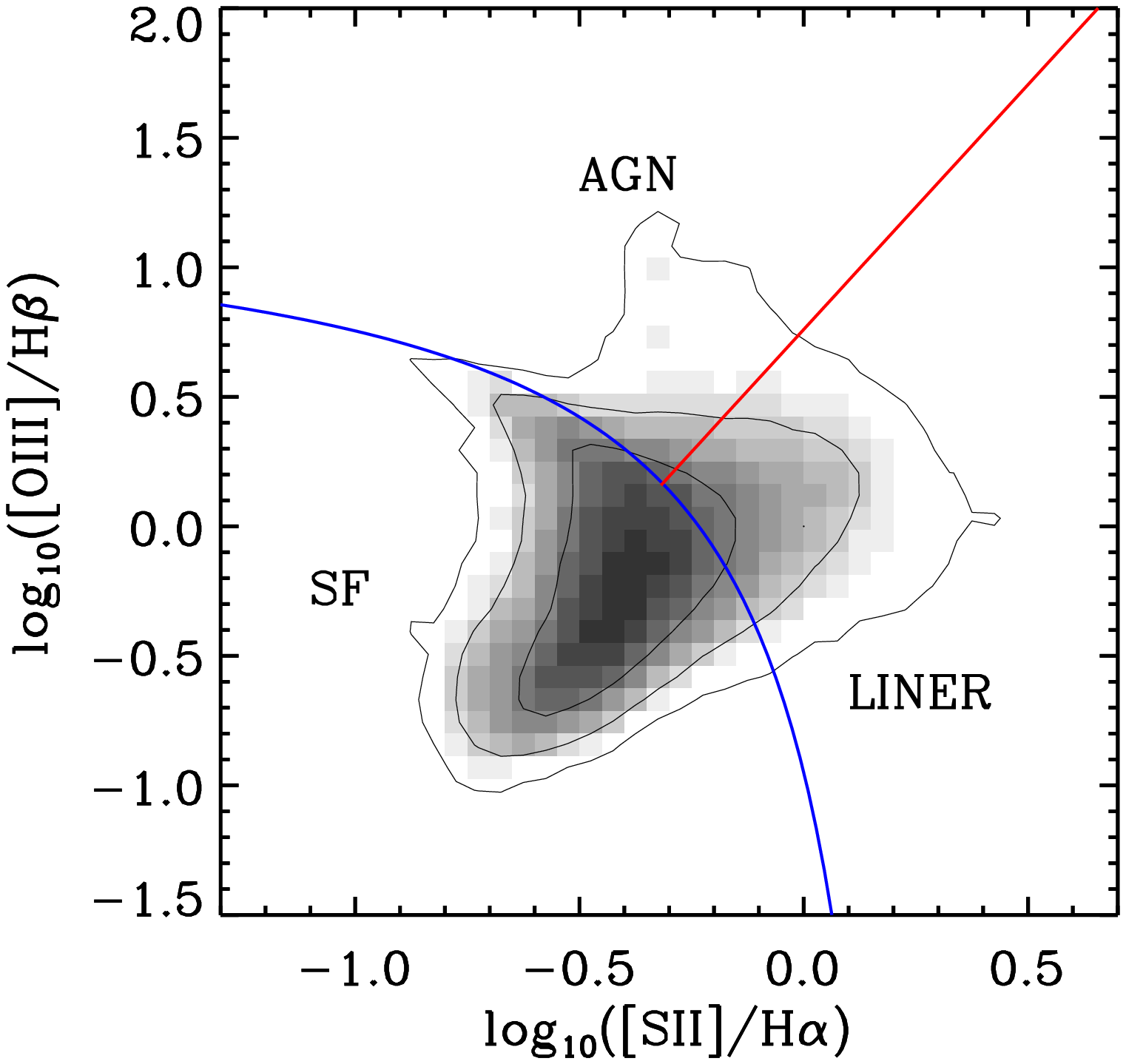,clip=true,width=0.4\textwidth}
\epsfig{figure=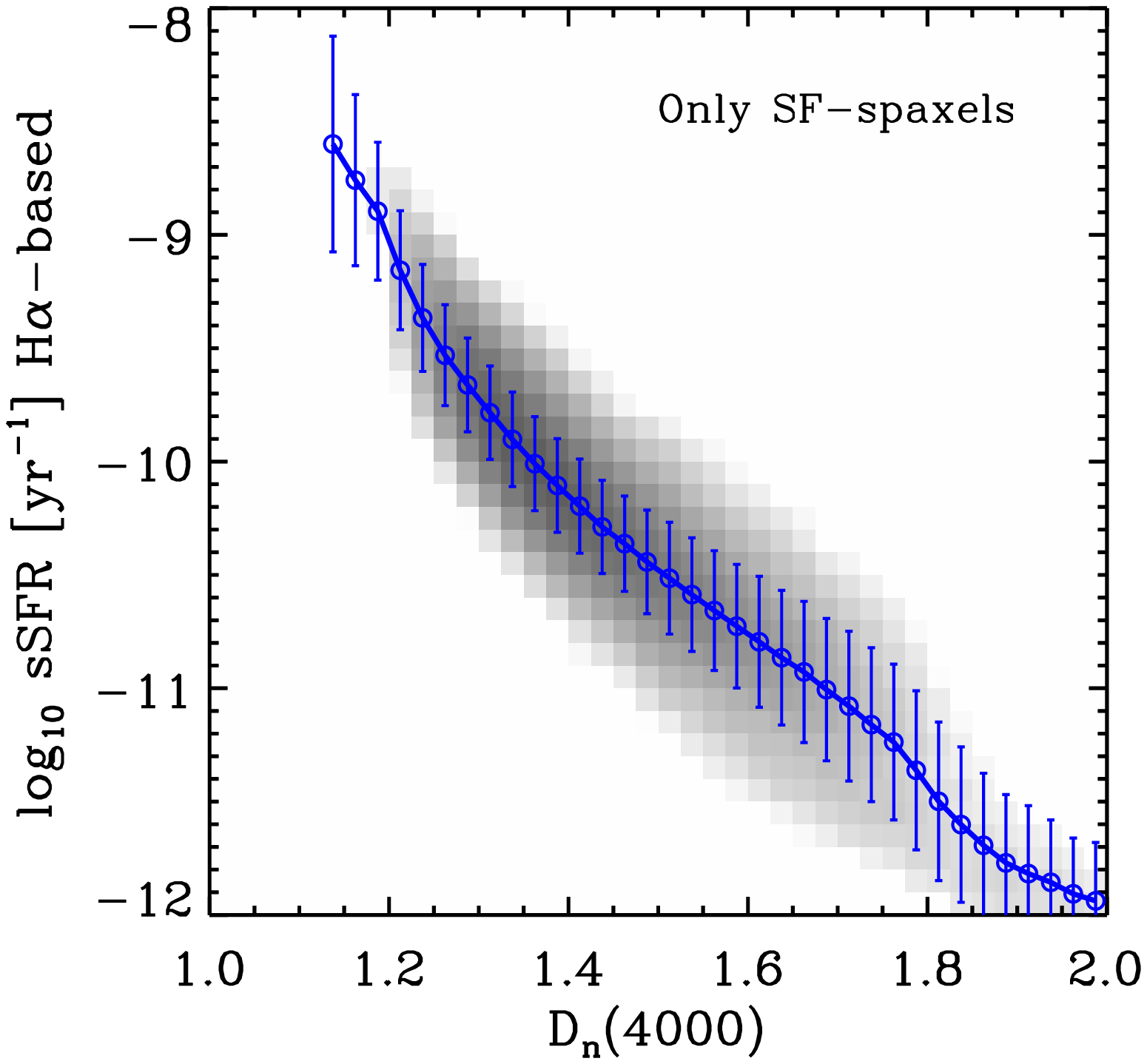,clip=true,width=0.4\textwidth}
}
\caption{ Left panel: the ${\rm \log_{10}([SII]/H\alpha)-\log_{10}([OIII]/H\beta)}$ diagram for all the spaxels.  The gray-scale represents the number density of the spaxels in this diagram in logarithmic space. The blue and red lines are taken from \cite{Kewley-01} and \cite{Kewley-06}  to separate
the spaxels in SF, LINER or AGN regions.  Right panel: the \dindex-sSFR relation for all the SF spaxels. The blue data points show the median sSFR at given \dindex\ bins with the error bars indicating the scatter. 
}
\label{fig:BPT}
\end{figure*}

The H$\alpha$ emission in galaxies can be contributed by many things, such as the star formation, the hot evolved stars on the post-asymptotic giant branch, shocks in interactions or mergers, and AGN activity in the centers of galaxies \citep{Belfiore-16, Zhang-17, Yan-18}.  
Usually for SF spaxels, the H$\alpha$ emission is mainly due to the star formation activities, so that the H$\alpha$ emission can be a good tracer for SFR. However, in case that the H$\alpha$ emission is not dominated by the star formation, simply using the H$\alpha$ to trace star formation would overestimate the SFR. In this work,  we adopt the same method as in \cite{Brinchmann-04} to correct the measurements of SFR for the AGN regions and LINERs\footnote{Low-ionization nuclear emission-line regions}. 

The left panel of Figure \ref{fig:BPT} presents the  Baldwin-Phillips-Terlevich (BPT) diagram \citep[e.g.][]{Baldwin-Phillips-Terlevich-81, Kewley-06} for the whole set of spaxels with the signal-to-noise ratio of the relevant emission lines greater than 3.0.  The demarcation lines from \cite{Kewley-01} and \cite{Kewley-06} separate spaxels into three classes: SF regions, LINERs \citep[or LIERs; see][]{Belfiore-16} and AGN regions.  For spaxels located in the SF regions, the SFR is calculated by the extinction-corrected H$\alpha$ luminosity using the conversion formula from \cite{Kennicutt-98} with a \cite{Chabrier-03} IMF, the same IMF used to derive the stellar mass.  The fluxes of emission lines are corrected for the intrinsic extinction based on the Balmer decrement assuming the CCM dust attenuation curve and the case B recombination with the intrinsic flux ratio of H$\alpha$/H$\beta$=2.86.  
For spaxels located in the LINER and AGN regions, the star formation is no longer well traced by the H$\alpha$ emission. 
We then apply an estimation based on the SF spaxels following the method of \cite{Brinchmann-04}.  The right panel of Figure \ref{fig:BPT} shows the 4000 \AA\ break versus the sSFR for SF spaxels. The 4000 \AA\ break is measured based on the best-fitting stellar spectra by the method of \cite{Li-05}, which exhibits a tight correlation with the sSFR with median scatter of 0.28 dex. For spaxels located in LINERs and AGN regions, we use the median relation of \dindex-sSFR indicated by the blue solid line to estimate the SFR.  We note that using this method would naturally lead to the tight correlation between SFR and stellar mass for quenched galaxies (see Section \ref{subsec:2.1}), since quenched galaxies usually have a small range of \dindex\ and thus a derived small range of sSFR.  

\subsection{Normalization to the sizes of galaxies}
\label{subsec:2.4}

Even at a fixed stellar mass, SF galaxies exhibit a range of size with a scatter of 0.2 dex \citep{Shen-03, Trujillo-06, vanderWel-14}, as parameterized by the effective radius (\re), taken from the NASA-Sloan-Atlas\footnote{http://www.nsatlas.org} \citep[NSA;][]{Blanton-11}. This variation in size may reflect differences in angular momentum etc., but this will not be a focus of this paper.  Rather the variation of size raises issues on how to compute and present radial profiles of surface density quantities in a way that is, as much as possible, independent of the different sizes of the galaxies under study.  

Of course, it is relatively straightforward to plot profiles as a function of the normalized radius $r/$\re.   But in so doing, it is then sensible to also plot a normalized surface density, i.e. a surface density that is computed as the mass (or star-formation rate) in an area that itself scales as \re$^2$.  This ensures that the (visual) integration of a profile on a given surface density-radius diagram reflects the actual integrated quantity in physical terms.  As an example, the profiles of two galaxies that both have (precisely) exponential profiles of \sigsfr\ but which differ by a factor of two in their overall SFR will be displaced by a factor of two in plots of \sigsfr\ vs. $r/$\re, {\it independent} of the sizes \re\ of the two galaxies, if the \sigsfr\ is computed per area scaled as \re$^2$.

In plotting the data, and in considering relative surface density profiles, we therefore always compute the surface density as the mass (or star-formation rate) per area of (0.2\re)$^{2}$.  It should be noted that, when considering, later in the paper, the physical surface density and its role in determining the star-formation efficiency, we will calculate the real physical surface density, i.e. the mass per kpc$^2$.

\section{Results}
\label{sec:results}

In this section, we present the typical SFR profiles for NSFGs (-0.33 $<\Delta {\rm SFR}<$0.33 dex), as well as the SFR profiles for galaxies that lie significantly above and below the SFMS to understand where the boosting, or suppression, of star formation occurs in the galaxy.

\subsection{Typical SFR profiles at different stellar masses}
\label{subsec:3.1}

\begin{figure*}
\center{
\epsfig{figure=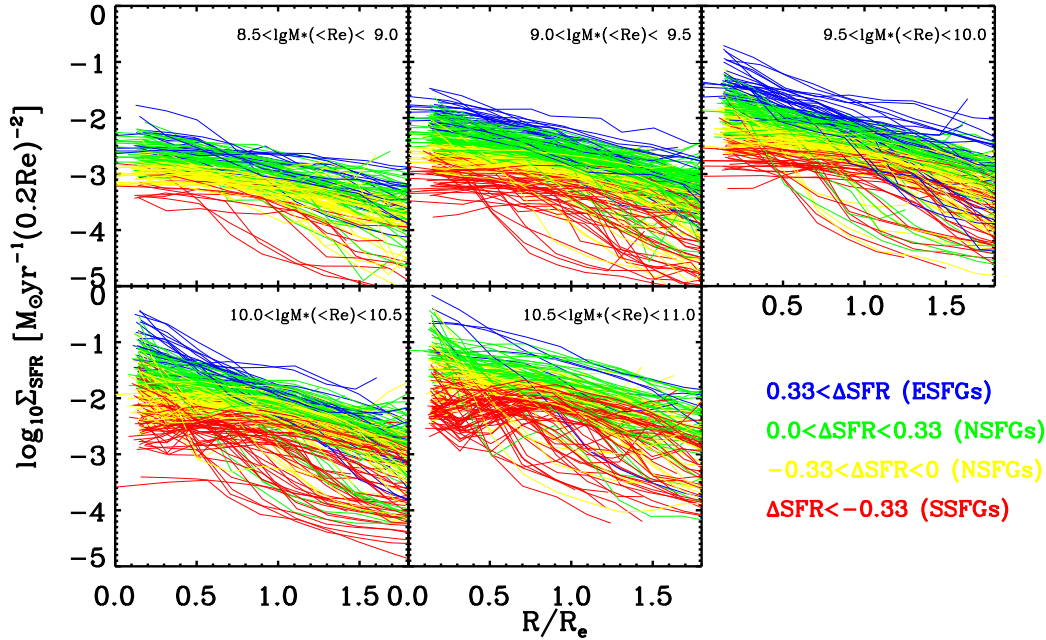,clip=true,width=0.8\textwidth}
}
\caption{The median \sigsfr\ profile for individual galaxies. The sample galaxies are divided into four subsamples according to their locations on the SFMS: $0.33<\Delta$SFR (blue lines), $0.0<\Delta$SFR$<0.33$ (green lines), $-0.33<\Delta$SFR$<0.0$ (yellow lines), and $\Delta$SFR$<-0.33$ (red lines).  For each subsamples,  we also separate galaxies into five stellar mass bins of the same width in logarithmic space, which are 8.5$<$\lgmstar$<$9.0, 9.0$<$\lgmstar$<$9.5, 9.5$<$\lgmstar$<$10.0, 10.0$<$\lgmstar$<$10.5, and 10.5$<$\lgmstar$<$11.0, respectively.  In each panel, the profiles are plotted in a random order to avoid one color dominating. }
\label{fig:individual_sfr}
\end{figure*}

\begin{figure*}
\center{
\epsfig{figure=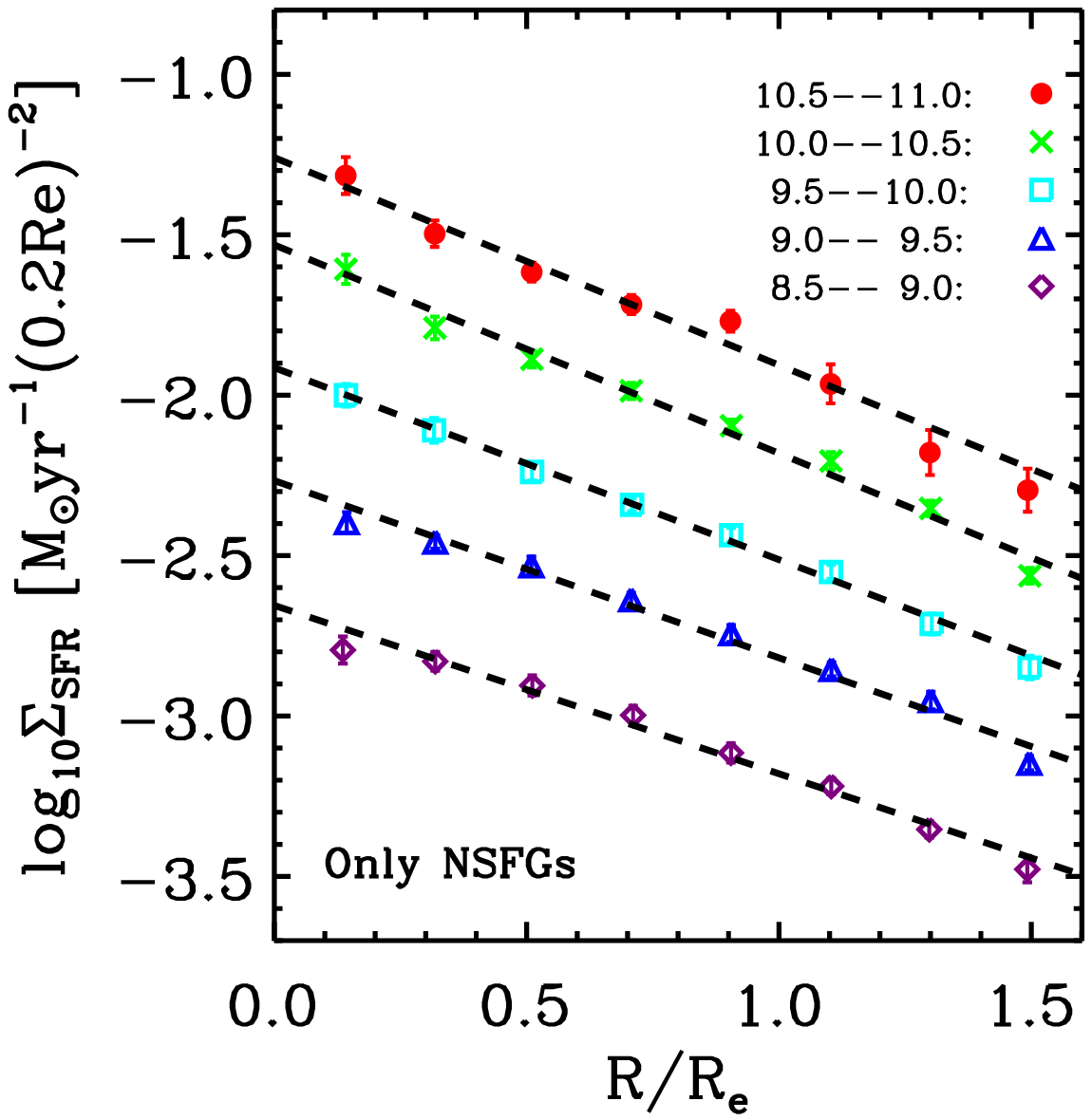,clip=true,width=0.32\textwidth}
\epsfig{figure=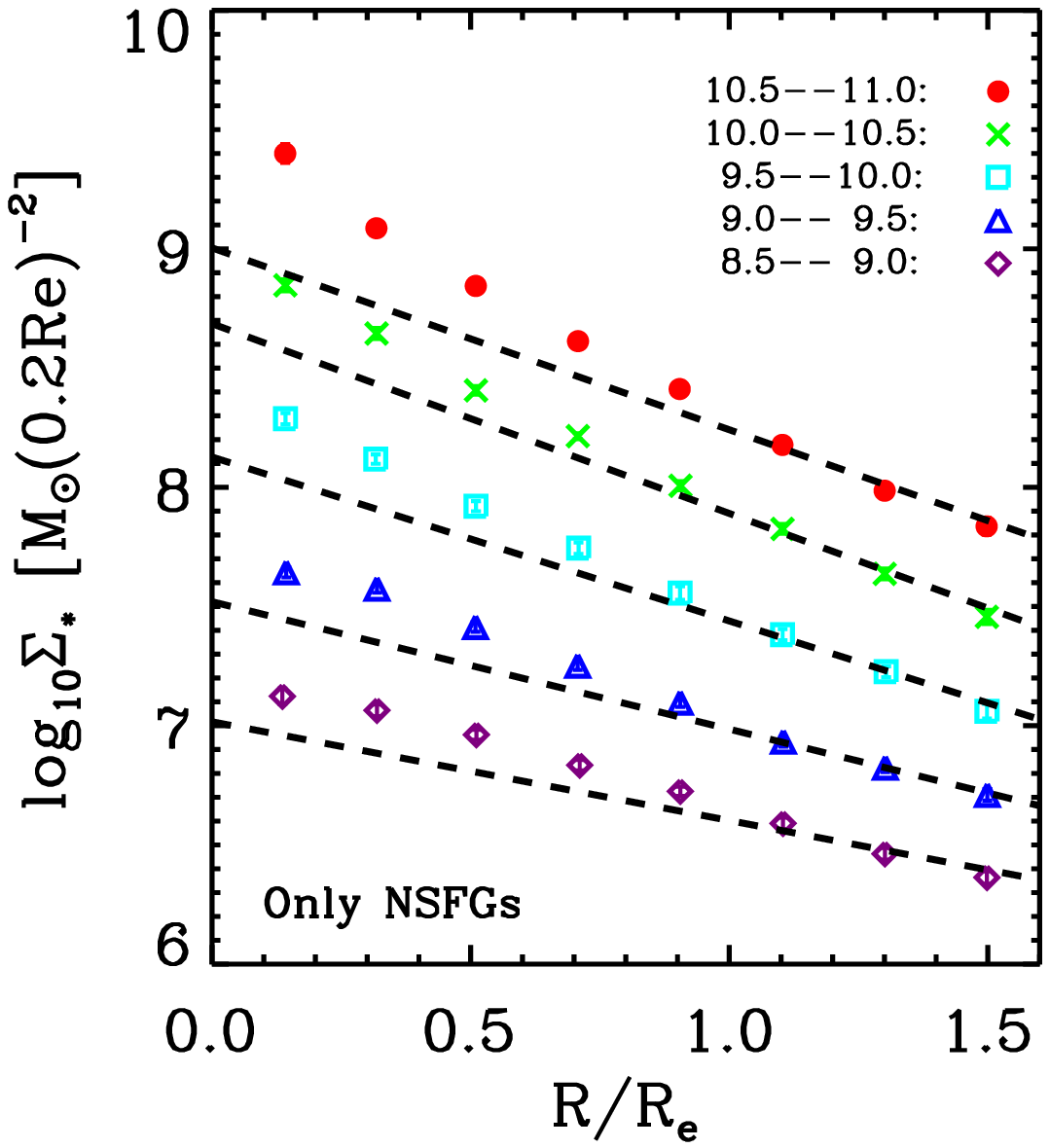,clip=true,width=0.32\textwidth}
\epsfig{figure=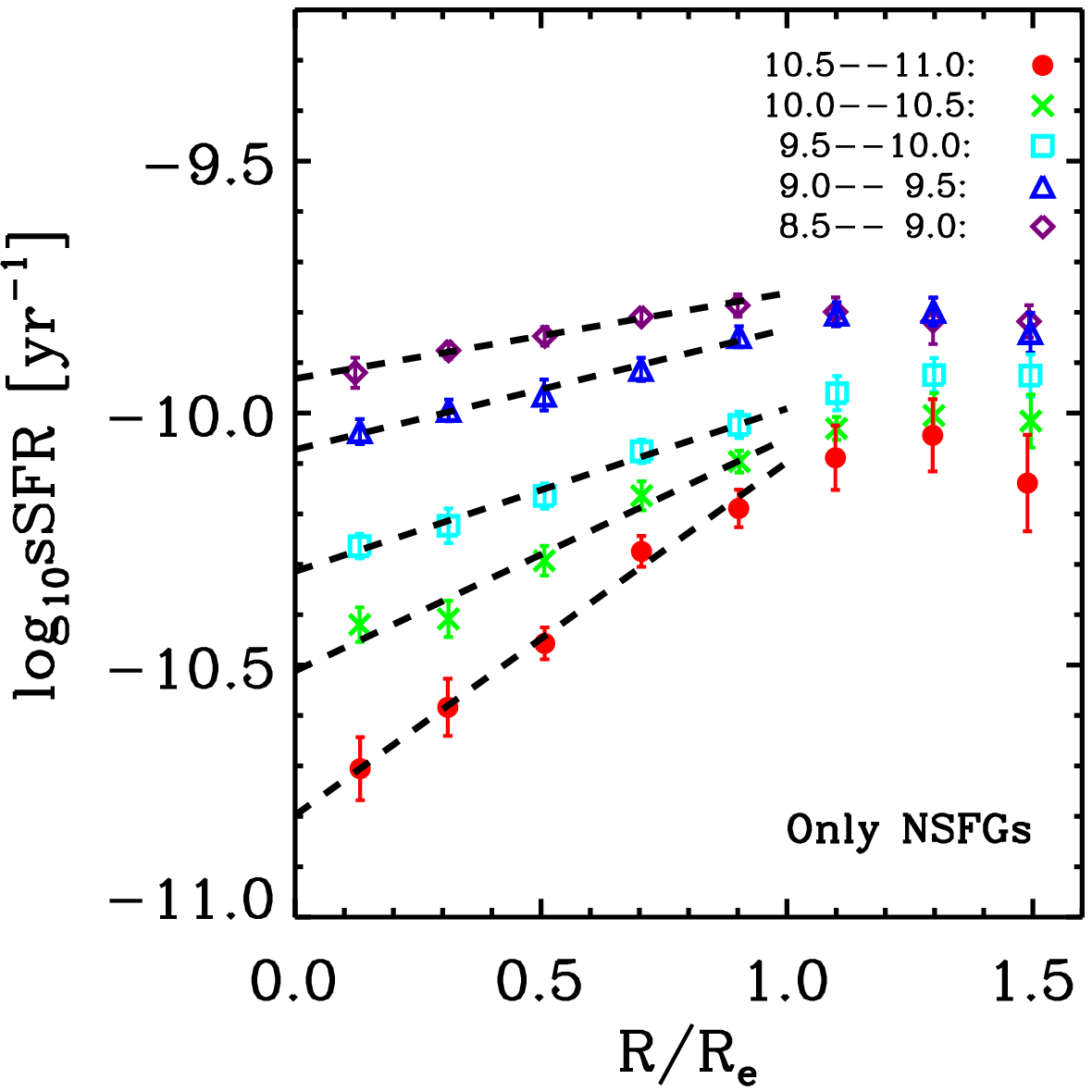,clip=true,width=0.32\textwidth}
}
\caption{The median profiles of \sigsfr\ (left panel), \sigmstar\ (middle panel), and sSFR (right panel) for the sample galaxies within 0.33 dex of the nominal SFMS at the five stellar mass bins, i.e. NSFGs.  Galaxies are separated into the same five stellar mass bins as in Figure \ref{fig:individual_sfr}.  The dashed line are the linear fittings of these profiles at different radius intervals. The \sigsfr\ profiles are fitted within 1.5\re, the \sigmstar\ profiles are fitted within 1.0-2.0\re, and the sSFR profile are fitted within \re.  The fitting results are listed in Table \ref{tab:1}.
 }
\label{fig:typical_sfr}
\end{figure*}

% --------- input Table 1-------------------
%    the fitting result of Figure 3

\begin{table*}[ht]
\renewcommand\arraystretch{1.3}
\begin{center}
\caption{The slope and intercept of \sigsfr, $\Sigma_*$ and $\Sigma_{\rm sSFR}$ profiles shown in Figure \ref{fig:typical_sfr} \label{tab:1}}
\begin{tabular}{@{}lrrrrrrrr@{}}
\tableline
\tableline
%                           &       \sigsfr(r)   $[r<1.5Re]$       & {}    & {}          & $\Sigma_*$(r)   $[r>Re]$     & {}  & {}  &    $\Sigma_{\rm sSFR}$(r)   $[r<Re]$\\
                          &       \sigsfr(r)   & $[r<1.5Re]$      & {}        & $\Sigma_*$(r)   &  $[r>Re]$    & {}  &    $\Sigma_{\rm sSFR}$(r)  &  $[r<Re]$\\                          
                           
%                           &          $[r<1.5Re]$    & {}    & {}          & $[r>Re]$        & {}  & {}  &    $[r<Re]$ \\
\cmidrule{2-3}
\cmidrule{5-6}
\cmidrule{8-9}
\lgmstar/$M_{\odot}$                           & Slope    &  Intercept & {}  & Slope  & Intercept & {}  &  Slope  &   Intercept \\
\tableline
 $[8.5-9.0]$                     &   -0.52      &  -2.66    & {}    &  -0.44     &   7.02      & {}    & 0.17    & -9.93     \\
 $[9.0-9.5]$                     &   -0.55      &  -2.27    & {}    & -0.53      &   7.52      & {}    & 0.24    &  -10.07  \\
 $[9.5-10.0]$                   &   -0.60      &  -1.91    & {}    &  -0.69     &   8.13      & {}    & 0.32    &   -10.31    \\
 $[10.0-10.5]$                 &   -0.65      &  -1.53    & {}    &  -0.80     &   8.68      & {}    & 0.46    &   -10.51    \\
 $[10.5-11.0]$                 &   -0.65      &  -1.26    & {}    &  -0.76     &   9.00      & {}    & 0.70    &    -10.80   \\
\tableline
\tableline
\end{tabular}
\end{center}
\end{table*}

%Why we study the SFR and stellar mass profile? 
The SFR and stellar mass are two basic properties of galaxies, reflecting the instantaneous rate of increase of the stellar mass and how many stars have been formed in the past.  The tight SFMS at both low and high redshift implies that the bulk of star formation in the Universe occurs in a quasi-steady state \citep[e.g.][]{Brinchmann-04, Daddi-07, Noeske-07, Salim-07, Elbaz-11}.  Not only the global SFR and stellar mass are strongly correlated: the local SFR surface density has also been found to tightly correlate with the local stellar mass surface density with a similar scatter of 0.3-0.4 dex as seen in the integrated quantities \citep{Cano-Daz-16, -Akiyama-17, Liu-18}. This suggests the existence of quasi-steady state in governing star formation within galaxies and the scatter reflecting the oscillation of this quasi-steady state.  Motivated by this, we study the \sigsfr\ and \sigmstar\ profiles of galaxies on the SFMS, which may shed light on the basic rule of how galaxies assemble their stellar mass. 

%A, typical SFR profile show no sign of quenching in galactic center. 
%B, the negative sSFR profile for massive galaxies is due to the fact that the bulge become more and more pronounced. 
%C, discussion of ``inside-out'' growth, or ``inside-out'' quenching? 

We first present the median \sigsfr\ profiles for all the individual galaxies. This is shown in Figure \ref{fig:individual_sfr}.  We separate galaxies in five bins of stellar mass and show by different colors their positions relative to the overall nominal SFMS.  Specifically, galaxies are divided into five stellar mass intervals of the same width of 0.5 dex in logarithmic space from \lgmstar/M$_{\odot}=8.5$ to \lgmstar/M$_{\odot}=11.0$. In each stellar mass bin, we separate galaxies into four subsamples according to the distance with respect to the nominal SFMS, $\Delta$SFR (see Section \ref{subsec:2.1}).  The \sigsfr\ profile for each individual galaxy is generated based on the 2-dimensional \sigsfr\ maps, as follows.  For a given galaxy, we compute the deprojected radius from the center of the galaxy based on the minor-to-major axis ratio from the NSA catalog \citep{Blanton-11}.  We also correct the observed mass (or star-formation rate) per spaxel for this same inclination effect.   

We divide the spaxels into a set of non-overlapping radial bins with a constant interval of de-projected radius $\Delta(r/R_{\rm e})=0.2$. The radial profiles of \sigmstar\ and \sigsfr\ are then determined by computing the median value of these quantities in the spaxels falling in each radial bin.  Regions of the galaxy that lie outside of the MANGA field of view are simply ignored.   Spaxels that have a continuum S/N$<$3 at 5500\AA\ are simply considered to lie at the bottom of the distribution in both stellar mass and SFR in computing the median.  However, we ignore completely any radial bins in which more than 30\% of the spaxels have S/N$<$3 at 5500\AA.   For the sSFR profiles, we first compute the sSFR in each individual spaxel based on the SFR and stellar mass in that spaxel.  In computing the median sSFR, we again do not consider spaxels with S/N$<$3 at 5500\AA\, but now compute the median from the remaining spaxels, and not from the full set as above (since sSFR as a relative quantity will be less biassed by the S/N cut).

As shown in Figure \ref{fig:individual_sfr}, the \sigsfr\ profile varies from galaxy to galaxy.    As would be expected, there is a clear trend for the decrease of the \sigsfr\ from the upper envelop to the bottom envelop of the SFMS.   The \sigsfr\ profiles of the sample galaxies however exhibit a large variation, especially for SSFGs lying significantly below the overall main sequence (shown in red).  The NSFGs within 0.33 dex of the nominal SFMS appears to show smaller variation in \sigsfr\ profile than SSFGs, suggesting that the star formation in these NSFGs is likely in a quasi-steady state with rare elevation or suppression of star formation.  This suggests that the NSFGs can serve as a good reference when studying the elevation or suppression of star formation in galaxies lying further above or below the SFMS. 

We now construct a median  \sigsfr\ profile for the NSFGs (within 0.33 dex of the nominal SFMS)  by calculating the median \sigsfr\ for all the NSFGs in this sample that had measurements at these radii.     
Thus, some galaxies may be not included in calculating the median \sigsfr\ profile at large radii ($r>$1.0\re), because of missing measurements of SFR. While these galaxies only account for 6.6\% of the whole population,  we have examined that our basic result does not change with excluding these galaxies. 
  Figure \ref{fig:typical_sfr} shows the median \sigsfr, \sigmstar\ and sSFR profiles for these NSFGs (-0.33$<\Delta$SFR$<0.33$ dex) at the same five different stellar mass bins, denoted in the top right corner of each panel.  
%Here we do not include the ESFGs and SSFGs because these galaxies are suffering from star formation elevation or suppression, which have very different features in \sigsfr\ profiles with normal SFMS galaxies (see Figure \ref{fig:individual_sfr} and Section \ref{subsec:3.2}). 
%In generating the median \sigsfr\ profile for each subsample, we first produce the median \sigsfr\ for each galaxy based on the 2-dimensional \sigsfr\ maps.  In practice, for a given galaxy, we take the spaxels with a continuum S/N$>$3 at 5500\AA\ and divide them into a set of non-overlapping radial bins with a constant interval of $\Delta(R/Re)=0.2$. The radial profile of \sigsfr\ is then determined by the median value of the spaxels falling in each radial bin.  During the process, we have corrected the inclination effect for each spaxel based on the minor-to-major axis ratio from the NASA-Sloan-Atlas\footnote{http://www.nsatlas.org} \citep{Blanton-11}. 

As shown in the left panel of Figure \ref{fig:typical_sfr}, the median SFR profiles for the NSFGs in the five stellar mass intervals can be well fitted by a straight line out to 1.5\re, indicating a pure exponential profile. With increasing stellar mass, the slope of the \sigsfr\ profile (in the left panel) becomes slightly steeper from -0.52 to -0.65.  These different slopes in Figure \ref{fig:typical_sfr} reflect small differences  in the ratio of the exponential scale length of the H$\alpha$ emission (used to trace the \sigsfr) to the effective radius of the $r$-band light that was used to normalize the radii.  If we had normalized the radii to the H$\alpha$ scale length, then these profiles would be exactly parallel, with a slope of -0.43.  

However, the \sigmstar\ profiles (in the middle panel) can clearly not be simply characterized by an exponential function. With increasing stellar mass, the inner slope of \sigmstar\ profiles are obviously becoming steeper due to the presence of a central stellar core, or bulge.  This leads to increasingly positive sSFR gradients for the galaxies of higher stellar mass in the right panel of Figure \ref{fig:typical_sfr}. 

This makes clear that the positive sSFR gradients in the inner regions (within \re) of massive galaxies may have nothing to do with the suppression of star formation or quenching, since the galaxies used to generate the Figure \ref{fig:typical_sfr} are NSFGs, and there is no evidence for star formation suppression in the sense that the \sigsfr\ profiles are pure exponential into the inner regions of massive galaxies, as seen from the left panel of Figure \ref{fig:typical_sfr}.  In addition, it can be seen that, for all the mass bins, the slopes of the median \sigmstar\ profiles at radii greater than \re\ are comparable to the slopes of the median \sigsfr\ profiles, resulting in nearly flat sSFR profiles at radii greater than \re, as shown in the right panel. 

We conclude that the positive sSFR profiles for NSFGs, especially at high masses, do not provide any information about star formation quenching, but reflect the combination of an exponential SF disk plus a pronounced stellar core or bulge in massive galaxies. The more and more positive gradients of sSFR with increasing stellar mass only indicates the greater prominence of the stellar core or bulge in more massive NSFGs.  
The presence of a substantial older population of stars in galaxies is a natural consequence of inside-out assembly \citep[e.g.][]{Larson-76, Chiappini-Matteucci-Romano-01, Pezzulli-15}.
In agreement with this, many papers based on the resolved spectroscopic analysis of galaxies demonstrates that the stellar population is much older in the inner regions than in the outer regions for massive SF galaxies with stellar mass greater than $10^{10}$\Msun \citep[e.g.][]{Perez-13, Li-15, Ibarra-Medel-16, Wang-17, Wang-18a}.  \cite{Lilly-Carollo-16} showed that significant sSFR gradients were achieved in a toy model based on the observed evolution of the size-mass relation of star-forming galaxies.  The positive sSFR gradients may well have nothing whatsoever to do with star formation quenching. 

Rather, a less confusing way to study star formation elevation or suppression within galaxies is to directly examine the \sigsfr\ profiles rather than the sSFR gradients of galaxies. We turn to this in the next subsection. 
 
\subsection{The variations of \sigsfr\ profile at different stellar masses}
\label{subsec:3.2}

\begin{figure*}
\center{
\epsfig{figure=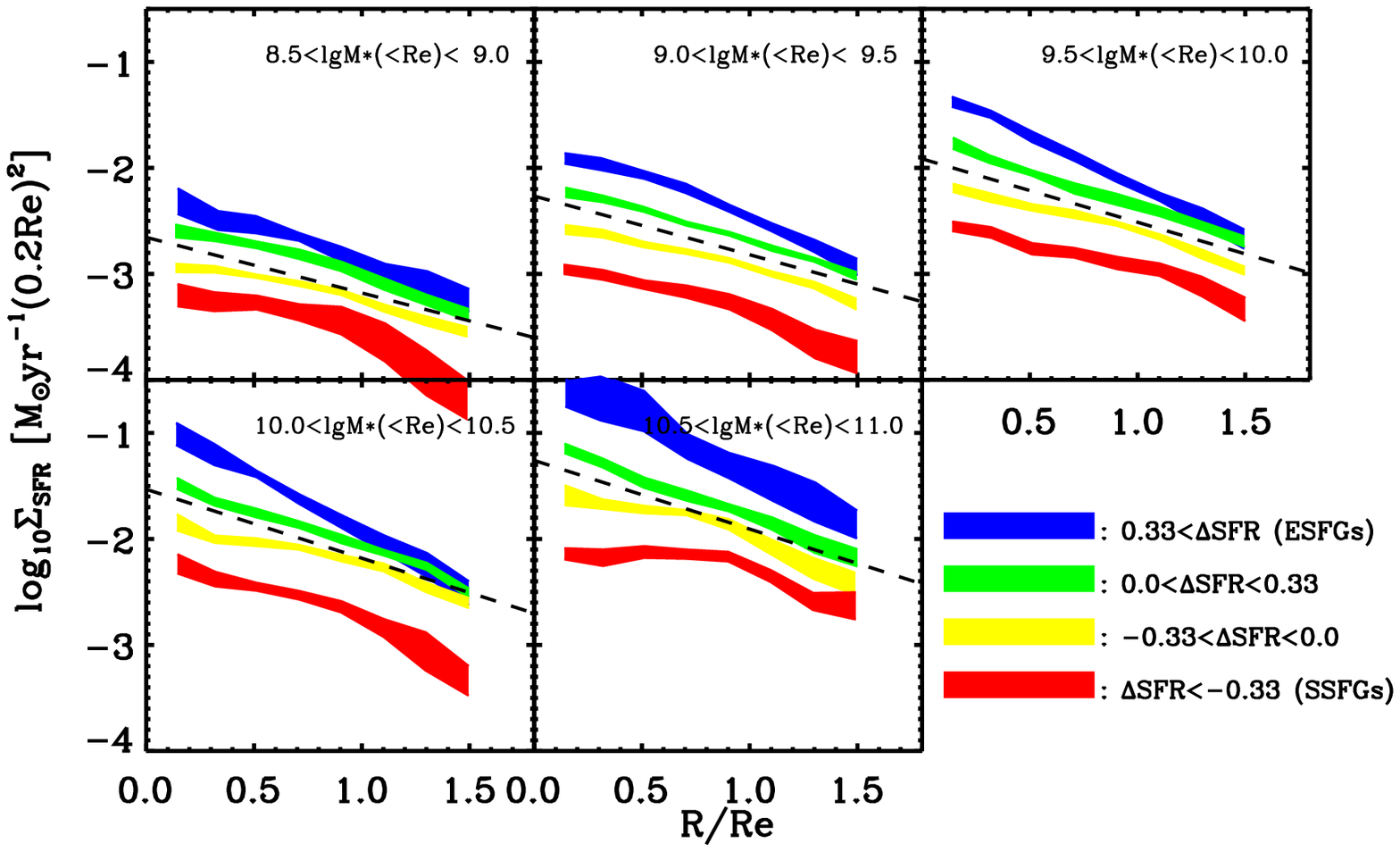,clip=true,width=0.8\textwidth}
\epsfig{figure=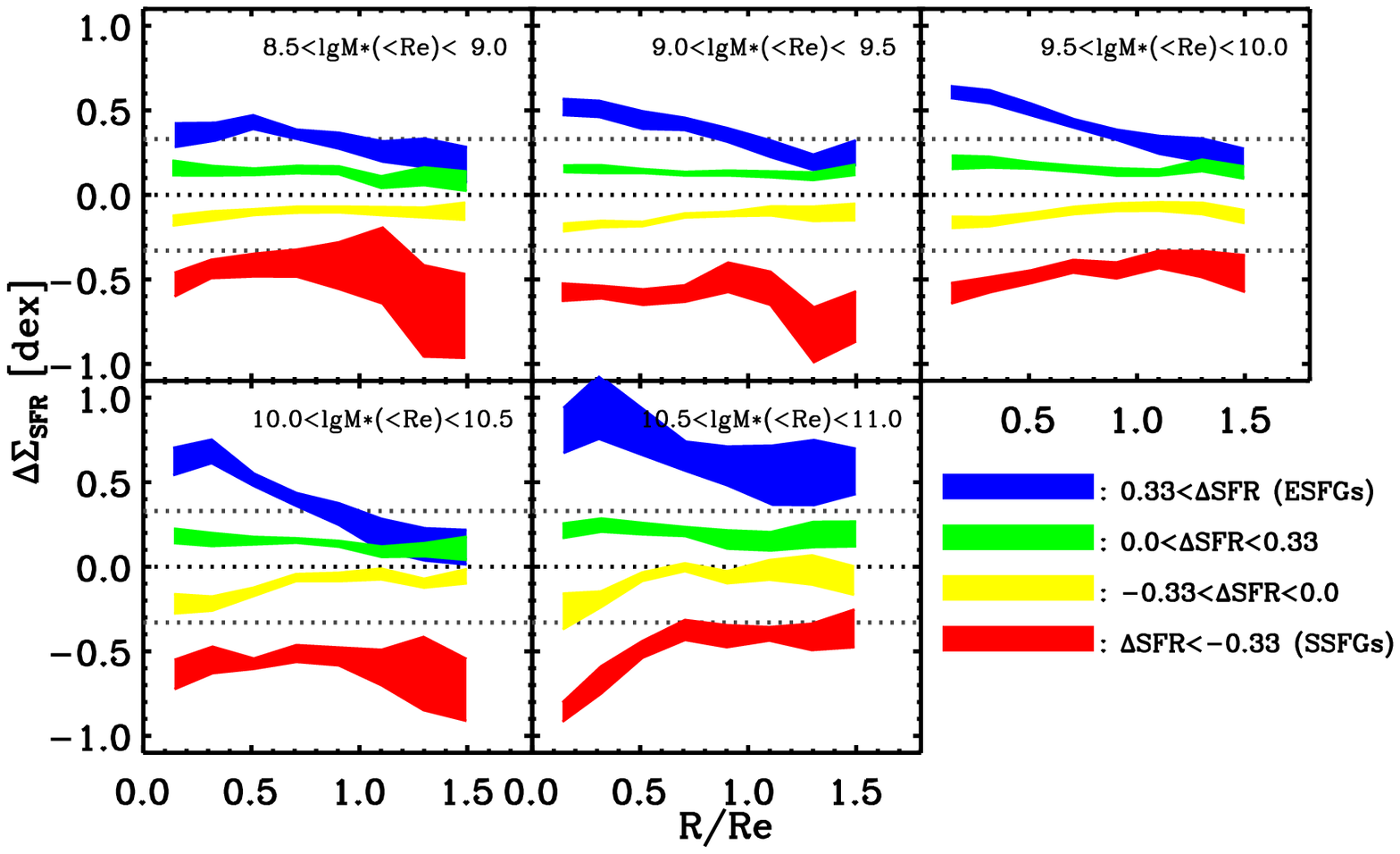,clip=true,width=0.8\textwidth}
}
\caption{Top group of panels: The median \sigsfr\ profile for galaxies at given stellar mass and given $\Delta$SFR. The dashed lines in each panel are taken from the left panel of Figure \ref{fig:typical_sfr}.  Bottom group of panels: The median \dsigsfr\ profile for galaxies at given stellar mass and given $\Delta$SFR. The NSFGs are divided into two subsamples according to $\Delta {\rm SFR}$. In each panel, the dotted horizontal lines represent \dsigsfr$=$0, -0.33 and 0.33 dex, respectively. The stellar mass intervals are denoted at the top right corner of each panel. }
\label{fig:dsfr}
\end{figure*}

%A, galaxies above the SFMS show elevation of star formation everywhere, especially in the center. 
%B, galaxies below the SFMS show suppression of star formation everywhere, especially in the center. This are usually referred as ``inside-out'' quenching in many literature. 
%C,  we do not find  the evidence of compaction (the shrinking of galaxy size) purely by the star formation, or even the gas compaction.    
%D, We interpret this in a different way. The symmetric of boost and quenching is likely produced by more general process, rather than compaction and ``inside-out'' quenching. 

In the previous subsection, we have presented the typical \sigsfr\ profiles for NSFGs at different stellar mass bins.  In this subsection we try to answer where in the galaxy the elevation or suppression of star-formation occurs for those galaxies lying further above or below the SFMS, by investigating the \sigsfr\ profiles for the ESFGs ($\Delta$SFR $>$ 0.33 dex) and the SSFGs ($\Delta$SFR $<-0.33$ dex).  The top group of plots in Figure \ref{fig:dsfr} show the median \sigsfr\ profiles for ESFGs, SSFGs compared with NSFGs in the usual five stellar mass intervals. The width of each profile is calculated using the boot-strap method from the sample in question. The dashed lines are taken from the left panel of Figure \ref{fig:typical_sfr} as reference.  

As shown, a first order result is that the median ESFG profiles exhibit enhanced star formation levels over the whole range of galactocentric distance (at least up to 1.5\re\ as we investigated here) and over the whole stellar mass range we considered.  Similarly,  SSFGs exhibit suppressed star formation levels over the whole range of galactocentric distance and over the whole stellar mass range.   

There is an additional effect in that ESFGs or SSFGs show {\it more} enhanced or suppressed star formation (respectively) in the central regions with respect to the outer regions. This effect clearly becomes more and more prominent with increasing stellar mass. 

This can be seen more clearly by subtracting from each profile the \sigsfr\ profile that would be expected for a completely typical NSFG of the same mass.  For each individual galaxy, the \dsigsfr\ profile is defined as the deviation from the median \sigsfr\ profile of the NSFGs at the same stellar mass (See Figure \ref{fig:typical_sfr}). Since the \sigsfr\ profile strongly depends on the stellar mass (see the left panel of Figure \ref{fig:typical_sfr}), we calculate for each galaxy the reference \sigsfr\ profile by interpolating (in stellar mass) the \sigsfr\ profiles shown in the left panel of Figure \ref{fig:typical_sfr}.  The result of this is shown in the bottom group of panels in Figure \ref{fig:dsfr}, where the median \dsigsfr\ profiles are plotted for ESFGs, SSFGs and NSFGs in the five stellar mass bins.   The dotted lines represent offsets of 0.33 dex in \dsigsfr\ corresponding to the offsets in the {\it overall} SFR associated with the definition of the four categories of star-forming galaxies.
  
As can again be seen, galaxies that are significantly above the nominal SFMS show enhanced star formation over the whole range of galactocentric distance, while galaxies that are significantly below the nominal SFMS show suppressed star formation again over the whole galactocentric distance.  Now it is more clearly visible that, with increasing stellar mass, the star formation in the central regions of ESFGs is relatively more enhanced, and that of SSFGs is relatively more suppressed, relative to the enhancement or suppression in the outer regions.  

It is quite striking in Figure \ref{fig:dsfr} that the inner elevation and suppression of star formation in the ESFG and SSFGs is quite symmetric, especially in the highest mass bin.  In the two lowest stellar mass bins, the outer regions of SSFGs also show strong suppression of star formation, which is likely due to the environmental effects \citep{Medling-18}, and we will show below that, after removing the low-mass satellites, a better symmetry is shown also at these lower masses (see Figure \ref{fig:dsfr_tau}). 

This result is broadly consistent with the work of \cite{Ellison-18}, who found that the SFR is enhanced throughout the galaxies especially in the center for galaxies above the global main sequence, although they have not examined the dependence on stellar mass. However, for galaxies below the SFMS, they found that only galaxies that lie at least 1 dex below the normal SFMS exhibit more suppressed star formation in the center than in the outer regions.  They also did not find the  symmetry of star formation elevation and suppression in galaxies for galaxies below and above the SFMS. 
This difference with our results may be due to the fact that 1) they did not split by galaxy mass, 2) they defined the global SFMS using the SFR from MPA-JHU\footnote{https://wwwmpa.mpa-garching.mpg.de/SDSS/DR7} catalog originally based on the SDSS 3-arcsec fiber spectra, and 3) they did not consider the contribution of LINERs in converting the H$\alpha$ emission to SFR.   

\begin{figure*}
\center{
\epsfig{figure=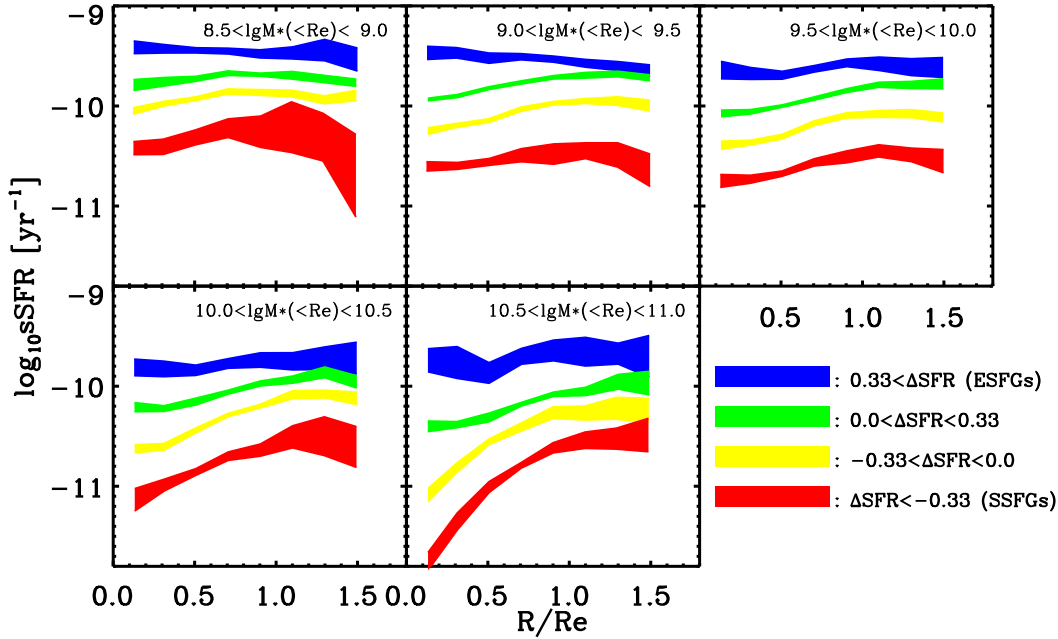,clip=true,width=0.8\textwidth}
}
\caption{The median sSFR profile for galaxies at given stellar mass and given $\Delta$SFR (the deviation from the nominal SFMS). Colors and lines are the same as Figure \ref{fig:dsfr}. }
\label{fig:ssfr_mass}
\end{figure*}

Figure \ref{fig:ssfr_mass} presents the median sSFR profiles for ESFGs, SSFGs and NSFGs in the five stellar mass bins. As shown, for galaxies with \lgmstar/M$_{\odot}<10.0$, the sSFR profiles exhibit flat or positive gradients for all the subsamples. While for galaxies with \lgmstar/M$_{\odot}>10.0$, galaxies show increasingly positive radial gradients in sSFR when moving from the upper envelop of the SFMS (ESFGs) down to the lower envelop of the SFMS (SSFGs).  It is noticeable that we do not find significant negative sSFR gradients (in these median sSFR profiles) in ESFGs over the whole stellar mass range we considered (except for an extremely weak negative gradient in the ESFGs with 9.0$<$\lgmstar/M$_{\odot}<$9.5).    We have also examined the sSFR profiles for individual galaxies and find that galaxies with negative sSFR gradients are a minor population \citep[c.f.][]{Pezzulli-15}. In total 72\% of the sample galaxies show positive sSFR gradients, and 58\% of ESFGs show positive sSFR gradients. 

It is an obvious statement that galaxies with positive sSFR gradients must be increasing their mass-weighted sizes (e.g. their half-mass radii), and vice versa, neglecting any subsequent stellar migration.   We conclude that despite their clearly elevated \sigsfr\  in their inner regions (see Figure \ref{fig:dsfr}) even the ESFGs are nevertheless still increasing their mass-weighted sizes as a whole, i.e. they are growing ``inside-out".

\cite{Ellison-18} have interpreted the centrally concentrated star formation in ESFGs as a ``galaxy compaction'' phase. This has been originally proposed as the contraction of cold gas along with central starburst activities at high redshift, triggered by violent disk instabilities \citep{Dekel-Burkert-14, Zolotov-15}, mergers and accretion of counter-rotating streams \citep{Danovich-15}.  Such ``wet compaction'' in high-redshift galaxies usually leads to the shrinking of their size during the rapid build-up of their stellar cores or bulges \citep{Zolotov-15, Tacchella-16a}. 
However, the word ``compaction'' may not be suitable at low redshift, because the increase in the inner stellar surface density of low redshift galaxies does not lead to the overall shrinking of their size due to a typical positive sSFR gradient (see Figure \ref{fig:ssfr_mass}).

%\cite{Ellison-18} interpret their results as the ``compaction'' scenario in which central starburst drives the build-up of the bulge and may ultimately precede galactic quenching from the inside-out.  
% the motivation of the new explanation ...  

\subsection{The dependence of \sigsfr\ fluctuation on gas depletion time}
\label{subsec:3.3}
%  why we see in this way?  and why we using gas depletion time ... 

\begin{figure}
\center{
\epsfig{figure=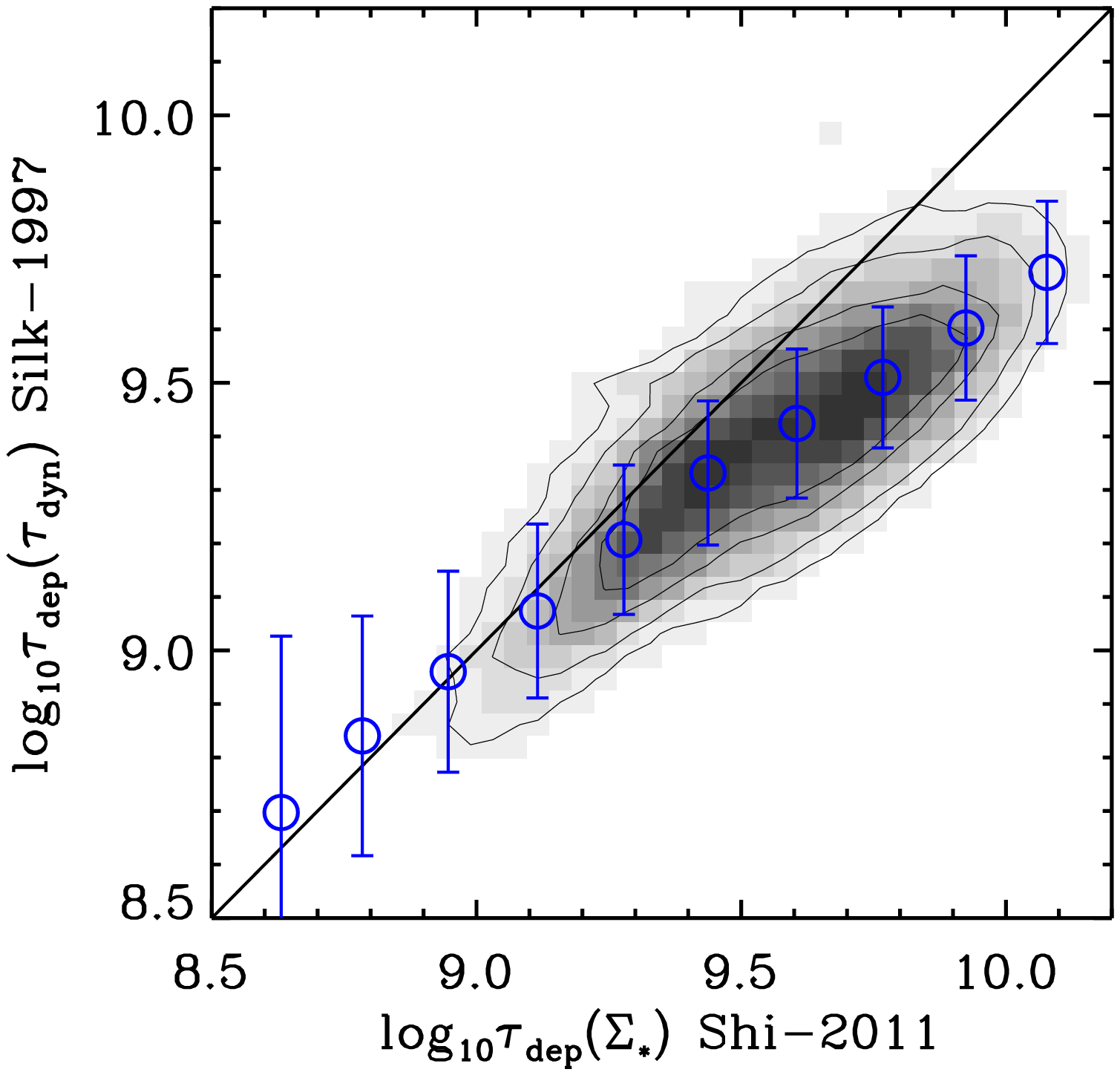,clip=true,width=0.45\textwidth}
}
\caption{ The gas depletion time derived from the extended Schmidt law \citep{Shi-11, Shi-18} versus the gas depletion time derived from the Silk-Elmegreen law \citep{Elmegreen-97, Silk-97, Krumholz-Dekel-McKee-12}. The grayscale represents the number density of spaxels of the sample galaxies. 
The blue data points show the median relation between $\tau_{\rm dep}(\Sigma_*)$ and $\tau_{\rm dep}(\tau_{\rm dyn})$, with the error bars representing the scatter of  $\tau_{\rm dep}(\tau_{\rm dyn})$ at given $\tau_{\rm dep}(\Sigma_*)$ bins.  }
\label{fig:com_tdep}
\end{figure}

\begin{figure*}
\center{
\epsfig{figure=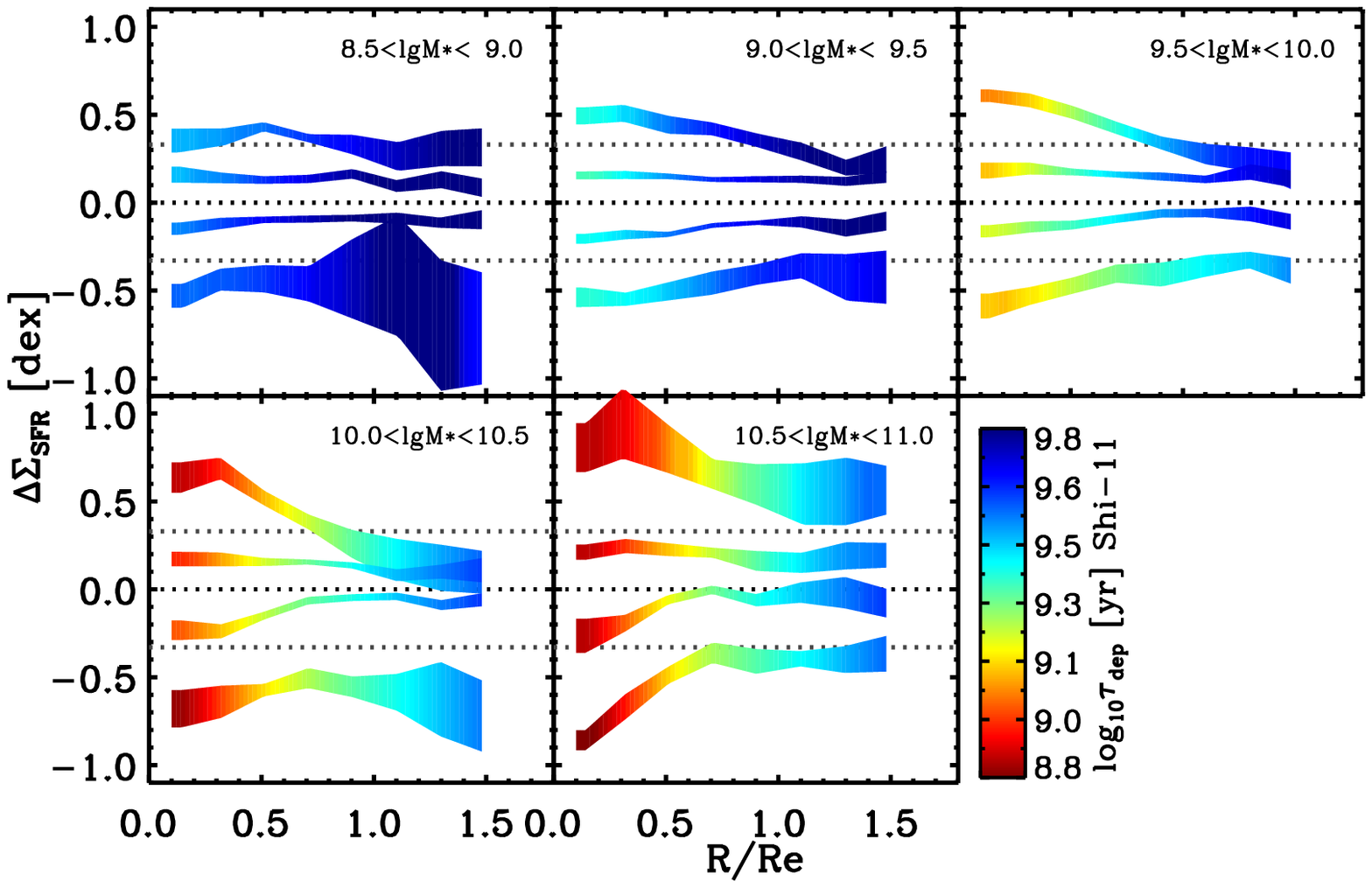,clip=true,width=0.8\textwidth}
\epsfig{figure=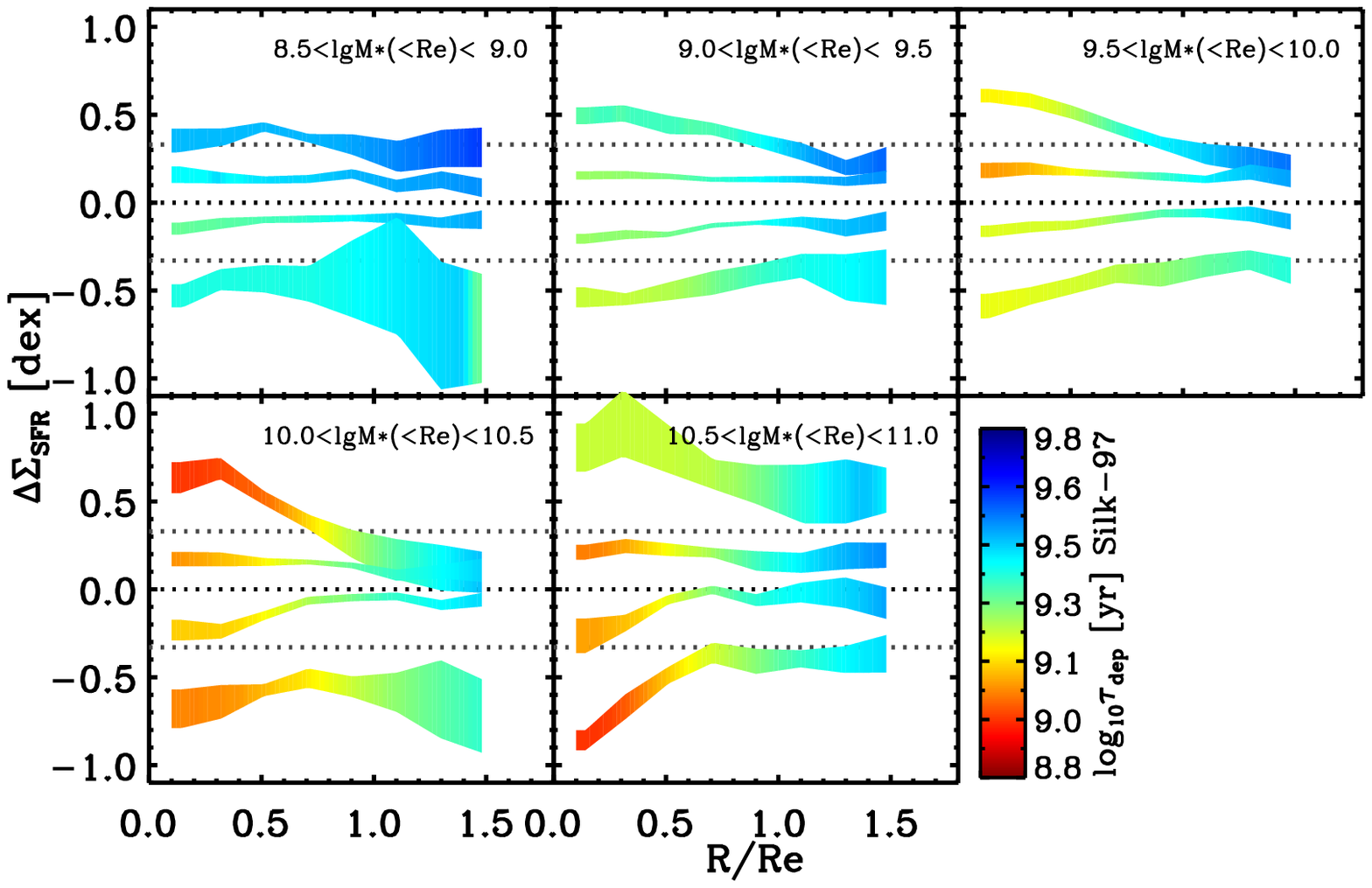,clip=true,width=0.8\textwidth}
}
\caption{The same as the bottom panels of Figure \ref{fig:dsfr} but now with color-coding the median gas depletion time derived from each galaxy as a function of radius.  Top group of panels are color-coded with the $\tau_{\rm dep}(\Sigma_*)$ derived from the extended Schmidt law, while the bottom group of panels are color-coded with the $\tau_{\rm dep}(\tau_{\rm dyn})$ derived from the Silk-Elmegreen law. 
Derived from Figure \ref{fig:dsfr}, in generating this figure, we additionally exclude 135 satellite galaxies with stellar mass less than one fifth of its centrals to reduce the environmental effects \citep{Wang-18b, Wang-18c}. }
\label{fig:dsfr_tau}
\end{figure*}

By investigating the spatially resolved star formation of 1494 MaNGA galaxies, \cite{Spindler-18} have claimed that there is a population of centrally-suppressed galaxies revealed by a bimodal distribution of the sSFR profiles, and that high mass galaxies are more likely to be centrally suppressed than low mass ones.  We also find that massive SSFGs show large positive sSFR gradients as seen in Figure \ref{fig:ssfr_mass}, but argue that the large positive sSFR gradients are the combination effect of the more suppressed SFR in galactic center and the early-formed stellar cores. 

Indeed, as shown in the previous section, when we look at the \sigsfr\ profiles for galaxies above and below the SFMS, we do not find that the centrally {\it suppressed} star formation in SSFGs is any more striking than the centrally {\it elevated} star formation in ESFGs. The nice {\it symmetry} of the elevation and suppression of star formation in galaxies (see Figure \ref{fig:dsfr}) gives us a new perspective to understand star formation: are the elevation {\it and} suppression of star formation both governed by the same physical principles?    If so, it might be dangerous to associate these with the one-way quenching process. 

We now address the variation of \sigsfr\ within and across galaxies in the context of a simple gas regulator model.
\cite{Lilly-13} have proposed a simple gas regulator model, which links the evolution of the cosmic specific star formation rate, the metallicity-mass-SFR relation and the stellar content of halos.  The gas-regulator model can account for the metallicity-mass-SFR relation in the local universe, the evolving metallicity of SF galaxies at redshift of 2, and even the existence of the FMR \citep[][]{Lara-Lopez-10, Mannucci-10}.  In this model, the formation of stars is instantaneously regulated by the mass of gas in the galaxy.  In a steady state, the gas regulator sets the sSFR of the galaxy to the specific accretion rate of gas onto the galaxy.  Variations in the infall rate produces variations in star-formation rate by adjustments of the gas reservoir.  According to this simple physical basis,  the Figure 3 of \cite{Lilly-13} presents examples of the response of the gas regulator to sudden changes in the accretion rate at different SFEs, defined as the ratio of SFR to the cold gas mass (the SFE is the inverse of gas depletion time, $\tau_{\rm dep}$). Sudden increases or decreases in the specific inflow rate of cold gas \citep[sMIR$_{\rm B}$; see Equation 16 in][]{Lilly-13}, produces a reaction of the sSFR on a timescale that is related to the gas depletion time.  If changes in the inflow are very gradual, the sSFR will track the change of sMIR$_{\rm B}$.  But sudden changes of specific inflow rate, on timescales that are short compared with the gas depletion time scale will produce delayed responses of sSFR.    A prediction of this generic model is therefore that the variation of SFR in galaxies, in response to variations in the inflow rate, should depend on the ratio of the gas depletion time and the timescale for the variation in inflow. This is a very general expectation. 

We therefore calculate the gas depletion time (the inverse of the SFE) of individual galaxies and investigate the links of this with the variation of \sigsfr\ within and across the population. We use two independent methods to calculate the SFE of each individual spaxel in the sample galaxies: the extended Schmidt law \citep{Shi-11, Shi-18} and the Silk-Elmegreen law \citep{Silk-97, Elmegreen-97}.  Both give very similar results.

The extended Schmidt law, proposed by \cite{Shi-11, Shi-18},  is based on integrated observations of individual galaxies over five orders of magnitude in stellar mass density, but was shown to also be applicable for spiral galaxies at sub-kiloparsec resolutions and low-surface-brightness regions.  The extended Schmidt law has the form 
\begin{equation}
\frac{\rm SFE}{\rm yr^{-1}} = 10^{-10.28} \left(\frac{\Sigma_*}{\rm M_{\odot}pc^{-2}}\right)^{0.48}, 
\label{eq:1}
\end{equation} 
which shows that the SFE, defined as the SFR per unit cold gas mass, depends only on the stellar mass surface density. In contrast to the classical Kennicutt-Schmidt law \citep[e.g.][]{Schmidt-59,Kennicutt-98}, the extended Schmidt law shows the role of existing stars in controlling the star-formation efficiency, which can be interpreted in terms of the free fall time of cold gas collapse in a stellar-dominated potential \citep{Talbot-Arnett-75, Dopita-Ryder-94}.  As \cite{Shi-11} pointed out, the \sigsfr\ can be written as: \sigsfr$=\eta \Sigma_{\rm gas}/\tau_{\rm ff}$, where $\eta$ is the fraction of the cold gas that collapses into stars and $\tau_{\rm ff}$  is the free-fall timescale of gas collapse. In a stellar-dominated potential, the $\tau_{\rm ff}$ can be expressed as: 
\begin{equation}
\begin{aligned}
\tau_{\rm ff} &= \frac{1}{4}\left(  \frac{3\pi}{2G\rho_{*}}  \right)^{0.5}
                    &= \frac{1}{4}\left(  \frac{3\pi h_*}{G} \right)^{0.5}\cdot \Sigma_*^{-0.5}, 
\end{aligned}
\label{eq:ex_sk}
\end{equation}
where $\rho_*$ is the stellar mass volume density, and $h_*$ is the stellar scale height. Assuming the $h_*$ is constant, Equation \ref{eq:ex_sk} naturally leads to SFE $\propto$\sigmstar$^{0.5}$.  Based on the de-projected stellar mass surface density map, the SFE (or gas depletion time) map of each individual galaxies can be easily obtained using Equation \ref{eq:1}.  

Unlike the extended Schmidt law, the Silk-Elmegreen law invokes the dynamical orbital timescale ($\tau_{\rm dyn}$) as \sigsfr$\propto\Sigma_{\rm gas}/\tau_{\rm dyn}$ \citep{Elmegreen-97, Silk-97}. In this law, the SFE is linearly proportional to $\tau_{\rm dyn}^{-1}$.  We therefore re-compute the SFE map for each individual galaxies based on the orbital timescale, adopting \sigsfr$=0.1\Sigma_{\rm gas}/\tau_{\rm dyn}$ as in \cite{Krumholz-Dekel-McKee-12}.  We first obtain the rotation curve of the gas disk based on the gas velocity map from the H$\alpha$ emission lines. For each galaxy, the rotation curve is determined by the velocity of spaxels within a long slit of 3-arcsec width aligned along the major axis.  Then we correct for the inclination effect on the rotation curve by dividing $(1-b^2/a^2)^{1/2}$, where $b/a$ is the minor-to-major axis ratio.  Then the dynamical time of each spaxel is calculated as the orbital time assuming a circular orbit. 

\begin{figure*}
\center{
\epsfig{figure=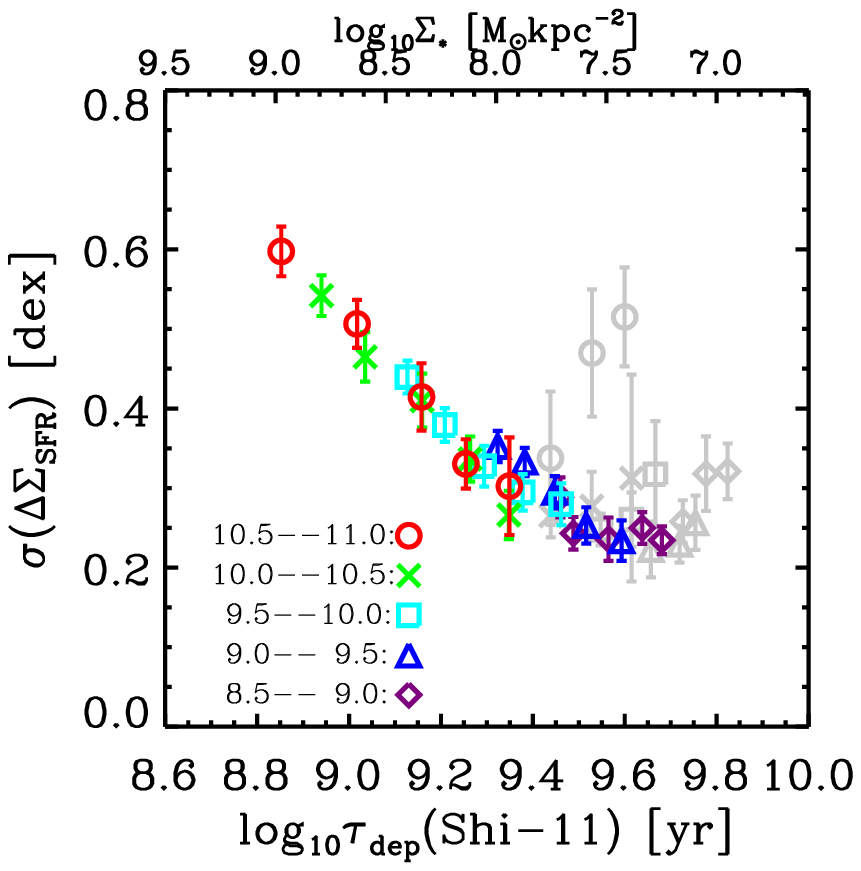,clip=true,width=0.48\textwidth}
\epsfig{figure=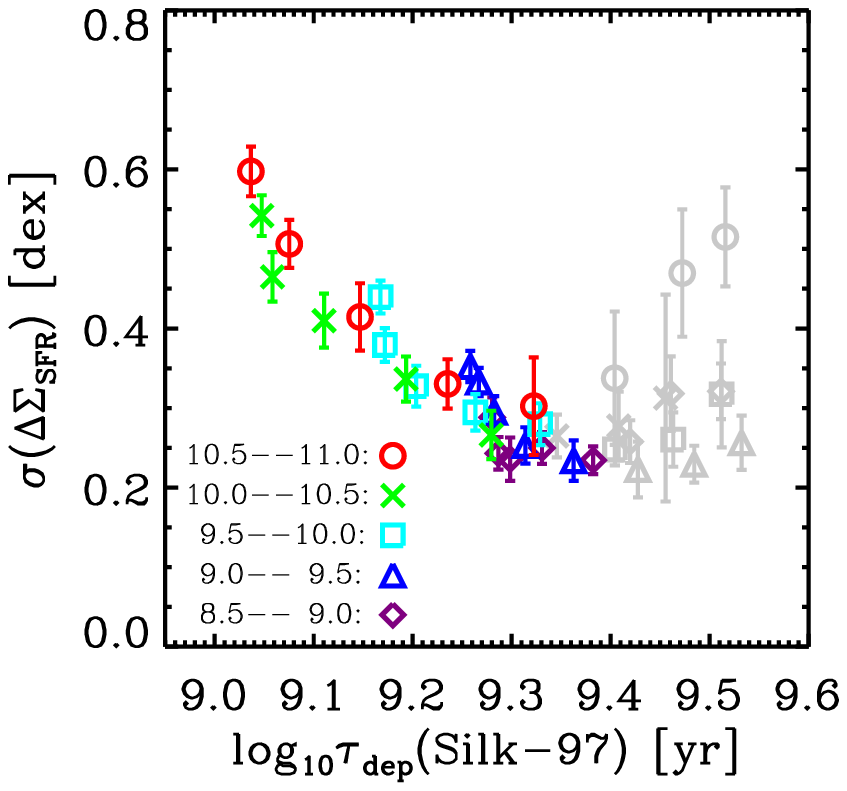,clip=true,width=0.48\textwidth}
}
\caption{Left panel: The scatter of \dsigsfr\ as a function of $\tau_{\rm dep}(\Sigma_*)$ derived using the extended Schmidt law. Right panel: The scatter of \dsigsfr\ as a function of $\tau_{\rm dep}(\tau_{\rm dyn})$ derived using the Silk-Elmegreen law.   The different colors represent different stellar mass bins, as denoted in the bottom-left corner.  Data points with the radius larger than \re, are indicated in gray.  Note that the x-axis of the left-hand panel is exactly equivalent to the stellar surface density, as indicated in the top of the plot.}
\label{fig:am_tau}
\end{figure*}

We thus obtain, for each individual galaxy, two maps of the gas depletion time (or SFE) based on the above two approaches. 
It is important to note that the gas depletion times in Figure \ref{fig:com_tdep}-\ref{fig:am_tau} are derived from the assumed SFE laws and are not based on the measurements of SFR and gas density (the latter being not available). 
One may worry that using the \dindex-sSFR relation to estimate the SFR in Section \ref{subsec:2.3} would introduce systematic bias for regions of high and low SFEs.  We therefore examine the dependence of \dindex-sSFR relation on SFE, and find that the SFR estimated by \dindex-sSFR relation would not introduce systematic bias for spaxels of low and high SFEs.  

The gas depletion time derived from the extended Schmidt law is referred as $\tau_{\rm dep}(\Sigma_*)$, and the gas depletion time derived from the Silk-Elmegreen law is referred as $\tau_{\rm dep}(\tau_{\rm dyn})$. 
   Figure \ref{fig:com_tdep} compares the gas depletion timescale of all individual spaxels derived by these two star formation laws, with the grayscale representing the number density of spaxels at each point. The blue circles represent the median relation of $\tau_{\rm dep}(\Sigma_*)$ and $\tau_{\rm dep}(\tau_{\rm dyn})$ with the error bars set to be the scatter of $\tau_{\rm dep}(\tau_{\rm dyn})$ at a given $\tau_{\rm dep}(\Sigma_*)$.  
As shown, $\tau_{\rm dep}(\Sigma_*)$ is broadly consistent with $\tau_{\rm dep}(\tau_{\rm dyn})$.  At high $\tau_{\rm dep}(\Sigma_*)$ end (corresponding to the low $\Sigma_*$),  the gas depletion times based on \cite{Shi-11} are larger than the measurements based on the Silk-Elmegreen law.  This is not surprising, because the extended Schmidt law is proposed to improve the Kennicutt-Schmidt law for the galaxies or the regions where the gravity is weak, such as the low-surface-brightness galaxies. As pointed out in \cite{Shi-11} and \cite{Shi-18}, the outer regions of dwarf galaxies, which are outliers of both the Kennicutt-Schmidt and Silk-Elmegreen laws, also follow the extended Schmidt law.   Here we do not present more discussion for the similarities and differences of these star formation laws \citep[see e.g.][and references therein]{Bacchini-18}, since it is beyond the scope of this work. We emphasize that our basic result does not depend on which star formation law is adopted (see Figure \ref{fig:dsfr_tau} and Figure \ref{fig:am_tau} below). 

We can then replot the bottom group of panels of Figure \ref{fig:dsfr} with a color-coding to represent the median gas depletion time. This is shown in Figure \ref{fig:dsfr_tau}. The top group of panels are color-coded with $\tau_{\rm dep}(\Sigma_*)$, and the bottom group of panels are color-coded with $\tau_{\rm dep}(\tau_{\rm tdy})$. Relative to Figure \ref{fig:dsfr}, we take the opportunity to exclude those galaxies that are satellites with stellar mass less than one fifth\footnote{The dependence of the quenched fraction on halo-centric distance is significant only for galaxies with stellar mass less than one fifth of the masses of their centrals, suggesting that the environmental processes are probably not the main drivers for quenching these massive satellites \citep{Wang-18b, Wang-18c}. } of their centrals \citep{Yang-07} to reduce the importance of environmental effects \citep{Wang-18b, Wang-18c}, such as tidal stripping etc.  In addition it should be noted that \cite{Medling-18} have also found that galaxies in denser environments show a suppressed sSFR from the outside in, probably caused by environmental effects.  This is in agreement with what we find in Figure \ref{fig:individual_sfr} that some SSFGs show extremely suppressed star formation at outer regions.  Comparing the locus of curves in Figure \ref{fig:dsfr_tau} with those in the original Figure \ref{fig:dsfr} suggests that indeed the impression of symmetry is strengthened by this modification.

As shown in Figure \ref{fig:dsfr_tau}, the {\it variation} in \sigsfr\ across the star-forming galaxy population appears to be strongly correlated with the derived gas depletion times, in all the stellar mass bins, at all radii, regardless of which star formation law we used.  Regions with the smaller $\tau_{\rm dep}$ have the largest variation in \sigsfr\ across the population.    

This agrees well with the prediction of the gas regulator model that the sSFR should track the change of specific accretion rate on a shorter timescale at higher SFE, assuming that the variation of \sigsfr\ in galaxies is determined by the variation in the accretion rate.  We explore this further in the next section.

To quantify the variation of \sigsfr, we next calculate the scatter of  $\Delta$\sigsfr\  in a given radial bin for each of the five stellar mass intervals. In order to reduce the contribution from outliers (see Figure \ref{fig:individual_sfr}),  we compute the {\it scatter} of $\Delta$\sigsfr, $\sigma(\Delta\Sigma_{\rm SFR})$ ,  with a sigma-clipping algorithm by iteratively rejecting points beyond 2.5$\sigma$.  
Figure \ref{fig:am_tau} presents the scatter of $\Delta$\sigsfr\ as a function of the median gas depletion time for each of the five stellar mass bins.  We note that each data point represents the variation of $\Delta$\sigsfr\ in a radial bin for galaxies at given stellar mass, and the higher $\tau_{\rm dep}$ usually corresponds to the larger radii.  
For each data point,  the $\sigma(\Delta\Sigma_{\rm SFR})$ is calculated as the scatter of the median $\Delta$\sigsfr\  at given radial bin of all galaxies located in that stellar mass bin. For the x-axis of Figure \ref{fig:am_tau}, we first calculate the median $\tau_{\rm dep}$ of all spaxels in a given radial bin of each individual galaxy, and then x-axis of each data point in Figure \ref{fig:am_tau} is the median of all galaxies in a given radial bin and a given stellar mass bin. 

Figure \ref{fig:am_tau} shows the scatter of $\Delta\Sigma_{\rm SFR}$ in different radial bins in different mass galaxies. It is therefore directly comparable to Figure \ref{fig:dsfr_tau}, and uses the same information (specifically the $\Sigma_{\rm SFR}$ profiles) but presented in a different way.  
It is therefore closely connected to the previous results in Section \ref{subsec:3.1} and \ref{subsec:3.2}.  We also comment that the $\sigma(\Delta \Sigma_{\rm SFR})$ in Figure \ref{fig:am_tau} is not exactly equivalent to the scatter of $\Sigma_{\rm SFR}$, because the ``typical" $\Sigma_{\rm SFR}$ profile is calculated continuously in stellar mass, by interpolation, according to the right-hand panel of Figure \ref{fig:typical_sfr}.

As can be seen, the $\Sigma_{\rm SFR}$ variation in galaxies is strongly anti-correlated with the gas depletion time, regardless of which of the two methods are used to calculate the gas depletion time.  This anti-correlation is especially seen at radii within \re, shown by the coloured points in Figure \ref{fig:am_tau}.   
To be clear, the ``star-formation efficiencies" or ``gas depletion times" used in the current analysis are derived from formulae in the literature rather than from direct measurements of the gas content in these particular galaxies.  It will be important to check, or test, the result in Figure \ref{fig:am_tau} using resolved measurements of cold gas in galaxies.   For instance, on the basis of Figure \ref{fig:am_tau}, we would predict that the variation in the (normalised) surface density of cold gas should decrease with increasing gas depletion time ($[M_{\rm HI}+M_{\rm H_2}]/{\rm SFR}$). This could be examined in future spatial-resolved CO and HI observations of these or other galaxies. 

At radii greater than \re\ (the grey points in Figure \ref{fig:am_tau}) the relation becomes flat or even increases, especially for the most massive galaxies. 
This may reflect the possible role of other processes in suppressing or elevating star formation in the outer regions of galaxies \citep{Medling-18}, leading to a larger variation of SFR at larger radius. It may also be due to the fact that it needs too long time to reach the quasi-steady state of star formation for the outer regions with a $\tau_{\rm dep}\sim3$ Gyr.  
   
 According to the basic assumption of gas regulator, the variation of SFR should depend on both the change of cold gas accretion rate and the timescale of the reaction to this change.  
 The result in Figure \ref{fig:am_tau} can therefore be interpreted in terms of the different response times of different regions of galaxies to changes in the  accretion rate.  In the inner regions, with higher SFE, the SFR reacts quicker, leading to larger changes in the SFR.  In regions with lower SFE, the response time is longer, smoothing out variations in accretion rate and producing a lower amplitude of variations in the SFR.   This is the most important new result of this paper. %According to this explanation, one would expect that the scatter of $\Sigma_{\rm SFR}$ increases with decreasing gas depletion time, or increasing with $\Sigma_*$.  Thus, we examine the scatter of $\Sigma_{\rm SFR}$ for all the spaxels as a function of $\Sigma_*$, and find that the scatter indeed increases with increasing stellar surface density as a whole, especially for the two most massive bins. This is consistent with the result of Figure \ref{fig:am_tau}. 
 
This simple interpretation suggests that {\it both} the elevation and suppression of SFR in the centers of galaxies may simply be caused by their more rapid response to increases or decreases in the rate of accretion from the halo. This is quite different from the previous interpretations, such as negative AGN feedback, the role of galactic bars, and ``compaction'' and quenching scenario  \citep{Ellison-18, Guo-19}.  In Section \ref{sec:toymodel}, we will illustrate how this model could work in a more quantitative way.

\subsection{The $\Sigma_{\rm SFR}-\Sigma_*$ resolved main sequence}
\label{subsec:3.4}

\begin{figure*}
\center{
\epsfig{figure=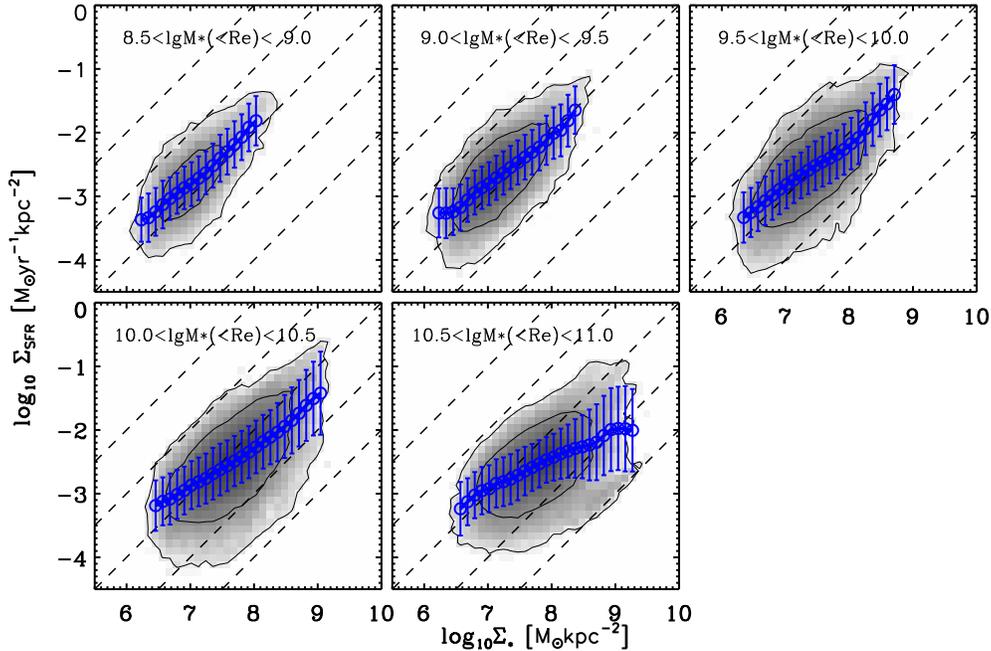,clip=true,width=0.8\textwidth}
}
\caption{The $\Sigma_{\rm SFR}-\Sigma_*$ resolved main sequence for the sample galaxies of the five stellar mass bins.  In each panel, the grayscale shows the number density of the spaxels, and the blue data points show the median $\Sigma_{\rm SFR}-\Sigma_*$ relation with the error bars indicating the scatter of $\Sigma_{\rm SFR}$ at given $\Sigma_*$.  Note that, different from the previous subsections, both $\Sigma_{\rm SFR}$ and $\Sigma_*$ surface densities are in physical units, without scaling by (0.2\re)$^2$. In each panel, the dashed lines from top-left to bottom-right are for the constant sSFR of $10^{-8}$, $10^{-9}$, $10^{-10}$, $10^{-11}$ and $10^{-12}$  ${\rm yr}^{-1}$, repectively.   }
\label{fig:resolved_ms_mstar}
\end{figure*}

\begin{figure}
\center{
\epsfig{figure=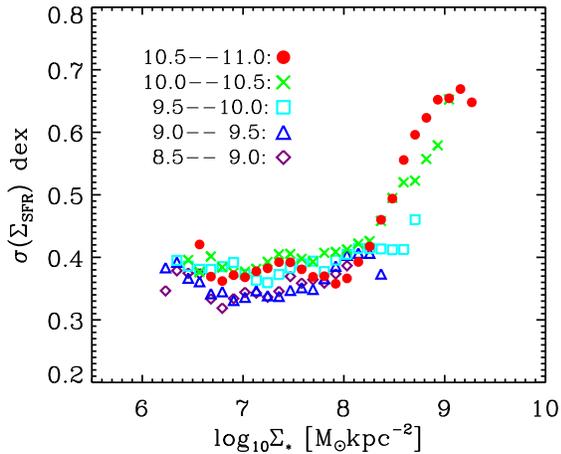,clip=true,width=0.47\textwidth}
}
\caption{The scatter of $\Sigma_{\rm SFR}$ as a function of $\Sigma_*$ for the sample galaxies of the five stellar mass bins.  As in Figure \ref{fig:resolved_ms_mstar}, both $\Sigma_{\rm SFR}$ and $\Sigma_*$ surface densities are in physical units, without scaling by (0.2\re)$^2$.  }
\label{fig:resolved_ms_sig}
\end{figure}

The idea presented in Section \ref{subsec:3.3} is that the observational results based on the $\Sigma_{\rm SFR}$ profiles (including Figure \ref{fig:am_tau}) can be interpreted in terms of the response of a gas regulator system to changes in the accretion rate. We will explore this further in the following Section \ref{sec:toymodel}.  This gas-regulator model is itself based on the idea that the SFR in galaxies is set by the accretion of gas from the surrounding dark matter halo.
In this context, the global SFMS is believed to be more fundamental than the resolved $\Sigma_*-\Sigma_{\rm SFR}$ relation.   A $\Sigma_*-\Sigma_{\rm SFR}$ relation would then naturally follow if the radial profile of how new gas was added to the galaxy was more or less stable over time.
However, some authors have argued for the opposite, i.e. that the global SFMS is the result of a more fundamental resolved $\Sigma_{\rm SFR}-\Sigma_*$ main sequence \citep[see][and references therein]{Cano-Daz-16}.  Although it is not clear to us whether this is correct, it is still interesting to examine our earlier result in the context of the resolved SFMS.

If the resolved SFMS is fundamental, a natural prediction of the idea that was presented in Section \ref{subsec:3.3} is that the scatter in the resolved SFMS, $\Sigma_{\rm SFR}$ at a given $\Sigma_*$, should increase with increasing $\Sigma_*$. This is because, in the extended Schmidt law formulation of star-formation efficiencies, the efficiency is derived purely from $\Sigma_*$.  We examine this in Figure \ref{fig:resolved_ms_mstar}, which shows the $\Sigma_{\rm SFR}-\Sigma_*$ relation for all the spaxels of the sample galaxies, split into the five stellar mass bins.  It is important to note that, unlike the $\Sigma_{\rm SFR}$ and $\Sigma_*$ in the previous subsections, we in this subsection use physical units in Figure \ref{fig:resolved_ms_mstar} for both $\Sigma_{\rm SFR}$ and $\Sigma_*$ {\it without} scaling with (0.2\re)$^2$. 

The $\Sigma_{\rm SFR}-\Sigma_*$  relation depends strongly on the stellar mass, in the sense that the slope of the $\Sigma_{\rm SFR}-\Sigma_*$  relation becomes flatter with increasing stellar mass. This probably reflects the fact that the bulge component becomes more pronounced with increasing stellar mass (see Figure \ref{fig:typical_sfr}), leading to a higher $\Sigma_*$ at the centers of galaxies that may be unrelated to the star formation rate in the disks.

More importantly, Figure \ref{fig:resolved_ms_mstar} shows that the scatter of $\Sigma_{\rm SFR}$ does indeed increase with increasing $\Sigma_*$, for the two highest stellar mass bins.  The result can be seen more clearly in Figure \ref{fig:resolved_ms_sig}, where the scatter of $\Sigma_{\rm SFR}$ as a function of $\Sigma_*$ for the five stellar mass bins is directly compared.   At $\Sigma_*$ greater than $\sim10^8$\Msun kpc$^{-2}$, the scatter of $\Sigma_{\rm SFR}$ strongly increases with $\Sigma_*$ (or the gas depletion time), which is in good agreement with the the idea in Section \ref{subsec:3.3}.   At low $\Sigma_*$, the scatter of $\Sigma_{\rm SFR}$ levels out, and depends at most weakly on the $\Sigma_*$. 

Overall, the similarity between Figures \ref{fig:am_tau} and \ref{fig:resolved_ms_sig} is reassuring.  It should be noted that, although they present similar results, the two plots are not completely equivalent. The $\sigma(\Delta\Sigma_{\rm SFR})$ in Figure \ref{fig:am_tau} is calculated with rescaled $\Sigma_{\rm SFR}$ (see Section \ref{subsec:2.4}), which takes out the effect of galaxy  size (or galaxy structure). 
As noted above, the scatter in Figure \ref{fig:resolved_ms_sig} may also reflect the effects of  galaxy structure,  in the sense that, the more pronounced bulge may lead to a flatter relation of $\Sigma_{\rm SFR}-\Sigma_*$. In this case, galaxies of different bulge fraction may broaden the scatter of $\Sigma_{\rm SFR}$ at given $\Sigma_*$. 
Furthermore, on a spaxel-by-spaxel basis,  one also needs to worry about the effect of scale, when calculating the scatter of $\Sigma_{\rm SFR}$, especially at the low-$\Sigma_*$ end,  since one would expect to obtain a larger scatter of $\Sigma_{\rm SFR}$ on a smaller scale.  
However, this effect should not be important in Figure \ref{fig:am_tau}, because the scatter in Figure \ref{fig:am_tau} is computed (across the population) based on the median value of $\Delta\Sigma_{\rm SFR}$ in the pixels within a given radial bin within each galaxy, whereas in Figure \ref{fig:resolved_ms_sig} the scatter is computed using simply all pixels in the sample with a particular $\Sigma_*$.
These may explain the fact that the scatter of $\Sigma_{\rm SFR}$ in Figure  \ref{fig:resolved_ms_sig}  is overall larger than the $\sigma(\Delta\Sigma_{\rm SFR})$ in Figure \ref{fig:am_tau}. 

In addition, it is possible that the resolved SFMS is not after all fundamental. If so, we could expect additional scatter in Figure \ref{fig:resolved_ms_sig}.  Actually, it is clear on Figure \ref{fig:resolved_ms_mstar} that the locus of the resolved main sequence varies strongly with the total stellar mass of the galaxy. Furthermore, we have examined the dependence of the $\Sigma_{\rm SFR}-\Sigma_*$  locus on the displacement of the overall SFR from the nominal {\it integrated} SFMS (i.e. $\Delta$SFR), and find that the  $\Sigma_{\rm SFR}-\Sigma_*$  relation also strongly depends on $\Delta$SFR: galaxies with higher $\Delta$SFR (i.e. higher overall SFR at a given mass) have higher $\Sigma_{\rm SFR}$ than low-$\Delta$SFR galaxies at the same $\Sigma_*$.  

Although the question of whether the global or the resolved SF main sequence is the more fundamental is beyond the scope of this paper, the important point for us is that the result in Figure \ref{fig:resolved_ms_sig} is consistent with the result in Figure \ref{fig:am_tau}.   We will explore this further below.

\section{A quantitative toy-model}

\label{sec:toymodel}
 
At the end of previous section, we argued that the striking inverse correlation between the amplitude of the variations in \sigsfr\ across the population (at a given radius and for a given galaxy mass bin) with the median gas depletion timescale (at this same radius and galaxy mass) could be qualitatively understood in terms of the different reaction timescales to changes in the accretion rate in the simple gas-regulator picture.
In this section, we construct a few heuristic toy models to explore how this could work, and also whether the variations in SFR could reflect variations in other parameters such as the SFE.

If the formation of stars is instantaneously regulated by the mass of available cold gas in the reservoir, and if there is an instantaneous mass loss scaling as the star-formation rate,  then the change in the mass of cold gas in terms of the gas inflow rate ($\Phi$) is given by the simple continuity equation: 
\begin{equation}
\begin{aligned}
\frac{d{M_{\rm gas}}(t)}{dt} = \Phi(t) - {\rm SFE(t)}\cdot(1+\lambda)\cdot M_{\rm gas}(t), 
\end{aligned}
\label{eq:resp}
\end{equation}
where $M_{\rm gas}$ is the mass of cold gas, and $\lambda$ is the mass-loading factor of stellar wind \citep{Lilly-13}.  As noted earlier, the $dM_{\rm gas}(t)/dt$ is set to zero in the ``bath-tub'' model \citep[e.g.][]{Bouche-10, Dave-Finlator-Oppenheimer-12}.  It is however this term that is the key to understanding the effects presented above in terms of the dynamic response of the gas mass in the gas regulator model \citep[see also][]{Lilly-13}. 
This differential equation can be solved, yielding $M_{\rm gas}(t)$ in the form 
\begin{equation}
\begin{aligned}
M_{\rm gas}(t) = e^{\rm -P(t)} \cdot \left({\rm Const}+\int {\Phi(t)}\cdot e^{\rm P(t)}dt\right),  
\end{aligned}
\label{eq:ans}
\end{equation}
where  ${\rm P(t)}=\int{\rm SFE(t)}\cdot(1+\lambda) dt$.   It should be noted that the evolution of SFR(t) is assumed to  follow the combined evolution of the cold gas mass and star formation efficiency, i.e. SFR(t) = SFE(t)$\cdot M_{\rm gas}(t)$.  According to Equation \ref{eq:ans}, the variation of SFR therefore only depends on the gas inflow rate, $\Phi(t)$, the mass-loading factor $\lambda$ and the SFE(t).

\subsection{Simple variations in inflow rate with constant SFE}
\label{subsec:model_A}

We first look at the case that the SFE is independent of time, and explore the behaviour of SFR(t) with given inflow flow rate, $\Phi(t)$, at a series of different SFEs. To see the effects of different SFE (or gas depletion time) more clearly, we consider the simple case where the inflow rate $\Phi(t)$ is oscillating sinusoidally with time. For simplicity, we will assume two simple forms of the inflow rate: a sinusoidal function in {\it linear} space and one in {\it logarithmic} space.  We refer to the sinusoidal inflow rate in linear space as Model A1, and in logarithmic space as Model A2. 

The inflow rate of Model A1 can be written as 
\begin{equation}
\Phi(t) = \Phi_{\rm 0} +\Phi_{\rm t}\cdot {\rm sin}({\rm 2\pi} t/T_{\rm p}), 
\label{eq:sin_phi}
\end{equation}
We then look for solutions of $M_{\rm gas}$ in the form of
\begin{equation}
M_{\rm gas} = M_{\rm 0} + M_{\rm t}\cdot {\rm sin}(2\pi t/T_{\rm p}-\delta), 
\end{equation}  
This can be substituted into Equation \ref{eq:resp} and the various time-dependent terms equated in the usual way.   It is convenient to define a parameter $\xi$ to be the ratio of the effective gas-depletion timescale to the timescale of variation of inflow rate $T_{\rm p} (2\pi)^{-1}$, i.e. $\xi=2\pi \tau_{\rm dep, eff}/T_{\rm p}$ and to define an effective depletion time as $\tau_{\rm dep, eff}$, i.e. including the outflow driven by stellar wind, i.e.  
\begin{equation}
\tau_{\rm dep, eff}={\rm SFE}^{-1}(1+\lambda)^{-1}.  
\label{eq:teff}
\end{equation}
Note that because of uncertainties in $\lambda$, we did not use $\tau_{\rm dep, eff}$ in Figure \ref{fig:dsfr_tau} and \ref{fig:am_tau}.

We then find that 
\begin{equation}
\begin{aligned}
&M_{\rm 0}= \Phi_{0}\tau_{\rm dep,eff},  \\
&\delta=  {\rm arctan}(\xi),  \\
&M_{\rm t} = \frac{\Phi_{t} \tau_{\rm dep,eff}}{[1+\xi^2]^{1/2}}.
\end{aligned}
\label{eq:solution}
\end{equation}
  
The resulting SFR(t) is therefore also a sinusoidal function with a phase delay $\delta$.  The ratio of the relative oscillation amplitude of the star formation rate (SFR$_{\rm t}$) to the relative amplitude in inflow $\Phi(t)$ can be written as: 
\begin{equation}
\frac{{\rm SFR_t}}{{\rm SFR_0}} = \frac{1}{[\xi^2+1]^{1/2}}\frac{\Phi_{\rm t}}{\Phi_{\rm 0}},
\end{equation}
where SFR$_{\rm 0}$ is the steady-unperturbed star formation rate, i.e. SFR$_{\rm 0}$=SFE$\cdot M_{\rm 0}$= $\Phi_{\rm 0} (1+\lambda)^{-1}$.

To summarize, when inputing a simple sinusoidally varying inflow rate, the SFR(t) responds in a sinusoidal way with a phase delay of $\rm arctan (\xi)$ and a reduction of the {\it relative} oscillation amplitude by a factor of $1/[\xi^2+1]^{1/2}$, where $\xi$ is given by the relative timescales of the variation in inflow and (effective) gas depletion.    Because the SFE is constant in time and independent of gas mass in this model, the variation in the gas content of the system will scale exactly as the SFR.

\begin{figure*}
\center{
\epsfig{figure=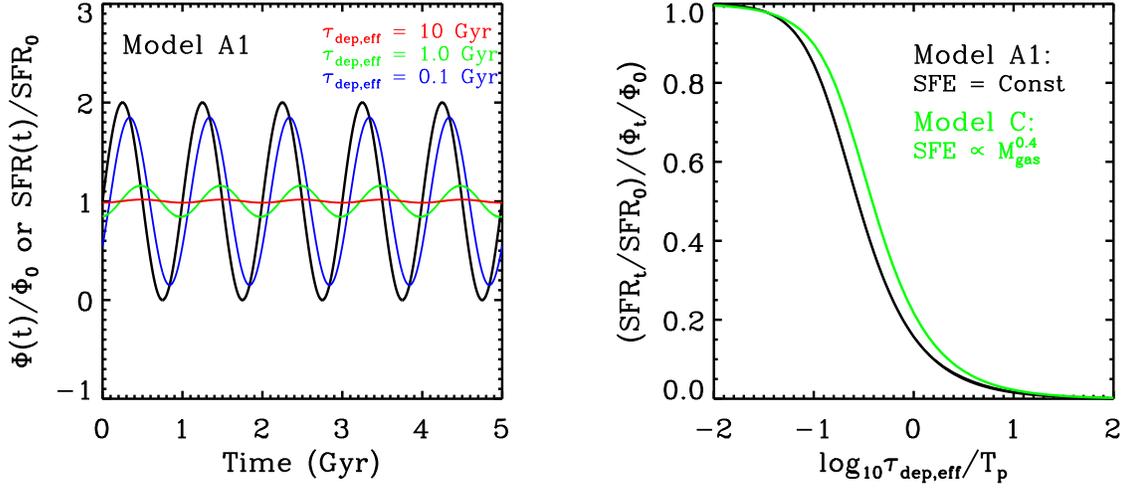,clip=true,width=0.9\textwidth}
}
\caption{Illustration of the toy Model A1 assuming that the inflow rate oscillates in a sinusoidal function in linear space. Left panel: The input gas inflow rate and the reaction of SFR as a function of time for Model A1 under the circumstances of three different effective gas depletion times: $\tau_{\rm dep,eff}=$0.1, 1.0, and 10 Gyr, as denoted in the top right corner.  
Right panel: The ratio of reaction amplitude of SFR to the input oscillation amplitude of $\Phi$  as a function of  $\tau_{\rm dep, eff}/T_{\rm p}$ for Model A1. 
In addition, we also present the reaction amplitude of SFR(t) of Model C as a function of  $\tau_{\rm dep, eff}/T_{\rm p}$ in the right panel, shown in green solid lines, with assuming the same oscillation of inflow rate as in Model A1, but the SFE varies in response to variations in the gas content of the system according to the Kennicutt-Schmidt law, i.e. SFE $\propto M_{\rm gas}^{0.4}$. }
\label{fig:toy_model_A}
\end{figure*}

\begin{figure*}
\center{
\epsfig{figure=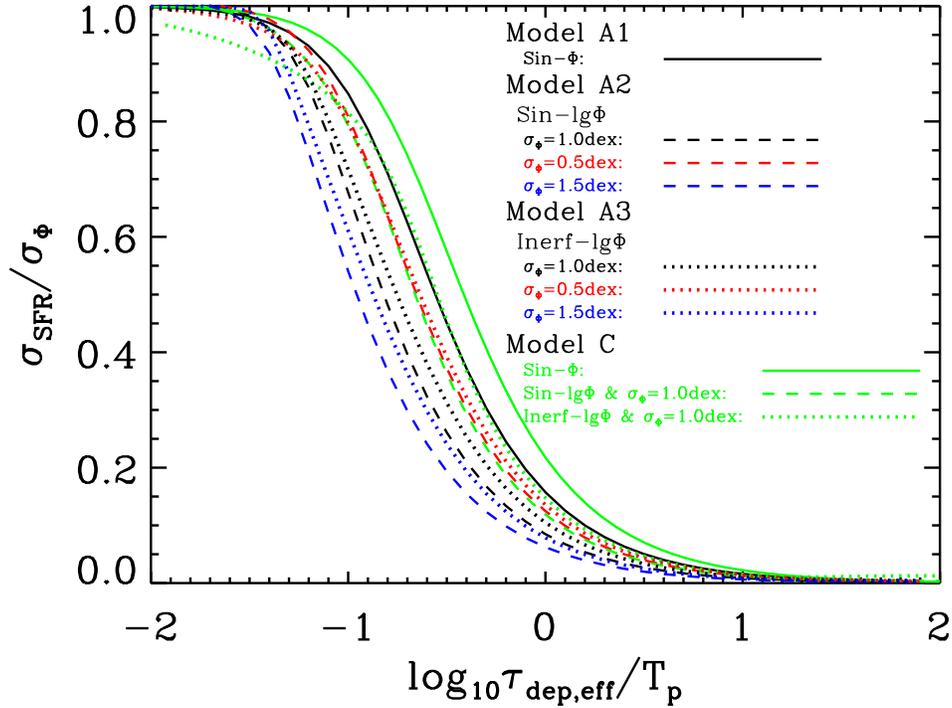,clip=true,width=0.70\textwidth}
}
\caption{ The ratio of reaction variation of SFR(t) to the input variation of $\Phi$ as a function of $\tau_{\rm dep, eff}/T_{\rm p}$ for a series of different models of inflow rate.  The models of inflow rate are labeled in the top right corner in the figure.  In addition, we also consider the case of Model C that the SFE varies in response to variations in the gas content of the system according to the Kennicutt-Schmidt law. The response  curves for Model C are shown in green curves with different forms of inflow rate.  }
\label{fig:toy_model_curve}
\end{figure*}

Figure \ref{fig:toy_model_A} summarize the behaviour of SFR(t) in model A1 as the period of oscillation of the inflow is varied relative to the effective gas depletion timescale. The left panel of Figure \ref{fig:toy_model_A} presents the reaction of SFR as a function of time under three different effective gas depletion times:  $\tau_{\rm dep,eff}=$0.1, 1.0, and 10 Gyr, when the inflow varies with $T_{\rm p}=1$ Gyr and $\Phi_{\rm t}=\Phi_{\rm 0}$.   As expected, the resulting SFR oscillates sinusoidally with time, with a phase delay with respect to the inflow rate (see Equation \ref{eq:solution}) that increases as the depletion time increases.    The amplitude of oscillations in the SFR similarly shows a reduction with respect to the oscillation amplitude of the input $\Phi(t)$.  The longer the gas depletion time, the more there is a reduction of the amplitude of variations in the gas content as it is unable to quickly respond to the varying inflow rate.  The right panel of Figure \ref{fig:toy_model_A} presents the analytic form of the ratio of SFR$_{\rm t}$ to $\Phi_{\rm t}$ as a function of the ratio of effective gas depletion time to the period of $\Phi(t)$ (confirmed by our numerical calculations).  When $\tau_{\rm dep, eff}$ is a tenth of $T_{\rm p}$, the oscillation amplitude of SFR is reduced by 15\% with respect to $\Phi_{\rm t}$. However, when $\tau_{\rm dep, eff}$ is comparable to $T_{\rm p}$, the oscillation amplitude of SFR is reduced by 84\% with respect to $\Phi_{\rm t}$.  The oscillation amplitude of SFR is reduced by half with respect to $\Phi_{\rm t}$ at $\tau_{\rm dep, eff}/T_{\rm p}$= 0.28.

According to Equation \ref{eq:sin_phi}, in the case of the linear Model A1, the peak-to-peak variation of $\Phi(t)$ is limited to be less than a factor of two with respect to $\Phi_{\rm 0}$, since the inflow rate should always be greater than zero. However, the inflow rate can be much more complicated in a real case, and the amplitude of variations in inflow could be larger than the steady state flow, since the variation of \sigsfr\  is up to 0.6 dex as seen in Figure \ref{fig:am_tau}.  

Therefore we also consider in Model A2 an inflow rate that oscillates sinusoidally in logarithmic space (Sin-lg$\Phi$).  However, both these sinusoidal inflow rates would lead to a double-horned distribution of inflow rate, and likely also a double-horned distribution of SFR of the galaxy population, which is clearly inconsistent with the roughly Gaussian distribution of SFR on the SFMS \citep[e.g.][]{Davies-19}.  Thus, we also consider a form of inflow rate in Model A3 that oscillates following the periodic inverse error function in logarithmic space (Inerf-lg$\Phi$), which produces a Gaussian distribution of the inflow rate.  In a similar way as Model A1, we solve the Equation \ref{eq:resp} for Models A2 and A3, but do so numerically.

Figure \ref{fig:toy_model_curve} presents the ratio of the variation of SFR(t) to the variation of $\Phi(t)$ as a function of  $\tau_{\rm dep, eff}/T_{\rm p}$ under the circumstance of these different forms of the inflow rate.  With the logarithmic variation of the Models A2 and A3, the reduction in variation amplitude not surprisingly depends on the (logarithmic) amplitude of the variations in inflow. This can be clearly seen in Figure \ref{fig:toy_model_curve} that offsets in the response curves are found for different $\sigma_{\Phi}$, even with the same form of inflow rate.

Although the oscillation amplitude of the output SFR(t) depends on the effective gas depletion time, the period of inflow rate, the oscillation amplitude, and even the functional form of the changes in the inflow rate, we find that all three models (A1, A2 and A3) qualitatively reproduce the trend that the variation of SFR is strongly anti-correlated with the effective gas depletion time, as shown observationally in Figure \ref{fig:am_tau}.  

In this, we naively assume that the amplitude and timescale of the changes in the inflow rate do not  vary with radius or galaxy mass, which may not be realistic.  However, with this caveat in mind, the variation in \sigsfr\ within and across galaxies in Figure \ref{fig:am_tau} can be interpreted as representing the dynamic reaction of the regulator, at different gas depletion timescales, to changes in the inflow rate.

\begin{figure}
\center{
\epsfig{figure=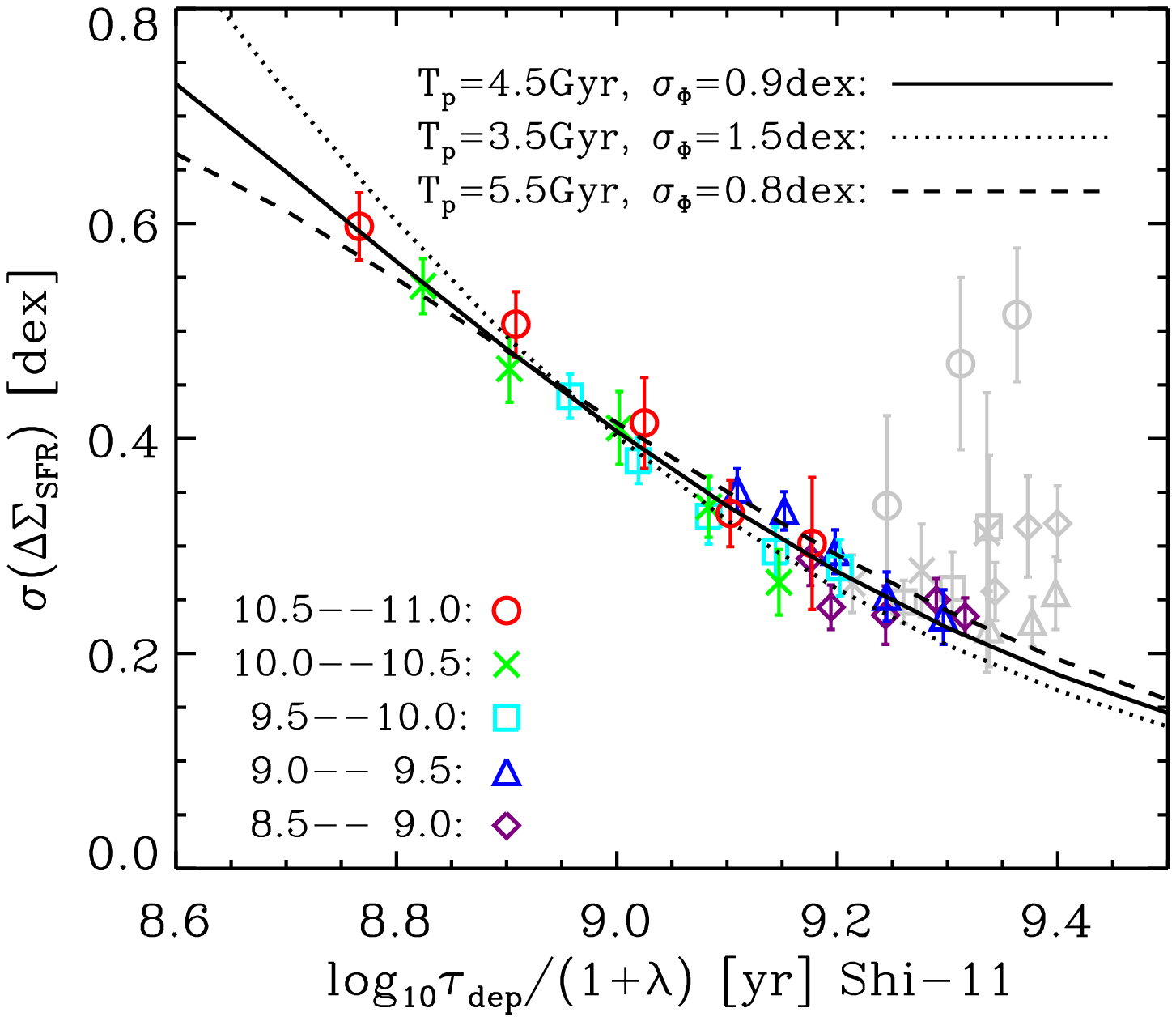,clip=true,width=0.45\textwidth}
}
\caption{ The same as the left panel of Figure \ref{fig:am_tau}, but overlapped with the best-fit model (black solid line), and the fitting models (dotted and dashed lines) at $T_{\rm p}=$3.5 Gyr and 5.5 Gyr, as denoted in the top right corner. We note that the x-axis is converted to the effective gas depletion time with respect to the Figure \ref{fig:am_tau}. }
\label{fig:model_fit}
\end{figure}

With this assumption, we can try to match the model with the observational result in Figure \ref{fig:am_tau}. For the purpose of illustration, we only present the model fitting for the left panel of Figure \ref{fig:am_tau} (based on the extended Schmidt law), and only consider the data points within \re\ for the five stellar mass bins.  Specifically, we adopt the Inerf-$\Phi$ of Model A3, because it is more realistic than the other forms of inflow rate we considered.   Since the model contains the mass-loading factor,  it is necessary to first estimate the $\lambda$.  Here we determine the mass-loading factor in a similar way as \cite{Lilly-13}.  We assume the mass-loading factor to be in the form of $\lambda_{\rm 0}(\Sigma_*/10^9[{\rm M}_{\odot}{\rm kpc}^{-2}])^\alpha$. The $\lambda_{\rm 0}$ is determined by fitting the gas-regulator model to match the resolved $\Sigma_*$-sSFR-metallicity relation of all the spaxels for galaxies in the five stellar mass bins (E. Wang et al., in preparation). During the fits, we fix the value of $\alpha$ to be -1/3, because the fitted value of $\alpha$ is close to -1/3 for all the five stellar mass bins when setting it free in the fittings. 

After obtaining the mass-loading factor, we then convert the $\tau_{\rm dep}$ of the left panel of Figure \ref{fig:am_tau} into the effective gas depletion time, $\tau_{\rm dep, eff}$ plotting this quantity in Figure \ref{fig:model_fit}. The best fitting model is shown in black solid line,  giving $T_{\rm p}$=4.5 Gyr and $\sigma_{\Phi}$=0.9 dex. However, the $T_{\rm p}$ and $\sigma_{\Phi}$ are highly degenerate according to Figure \ref{fig:toy_model_A} and Figure \ref{fig:toy_model_curve}. We also present the best-fitting models with fixing $T_{\rm p}$=3.5 Gyr and 5.5 Gyr, shown in dotted and dashed lines in Figure \ref{fig:model_fit}. All the three fitting models match the data points reasonably well. 

%We do not consider it appropriate to try to quantitatively constrain the variations in the inflow rate based on the observational result in Figure \ref{fig:am_tau}.  
Although the models match the data points well, we regard this as a heuristic exercise, and the derived values should be treated with a great deal of caution. This is because 1) the variation of SFR depends on the form of inflow rate and the real inflow rate is unlikely to be any of the forms we assumed, 2) the variations in inflow rate may well depend on radius and on galaxy mass, 3) the mass-loading factor determined by fitting the $\Sigma_*$-sSFR-metallicity relation may carry large uncertainties, and finally 4) the oscillation amplitude and period of inflow rate are highly degenerate as shown in Figure \ref{fig:toy_model_curve}. 

Nevertheless, despite these caveats, the toy model suggests that this idea of dynamic response to changes in accretion is a useful one in interpreting radial variations of star-formation rates in galaxies.  This is different from the previous mechanisms proposed to explain the elevation or suppression of star formation \citep[e.g.][]{Ellison-18, Guo-19, Spindler-18}.   Previously the enhanced star formation in galactic centers has been interpreted as the result of disk instabilities, galaxy-galaxy interactions or the existence of bar, and suppressed star formation in galactic center has been interpreted as ``inside-out'' quenching by AGN feedback or morphological quenching. However, based on our toy model, the symmetric elevation and suppression in star formation is simply due to the fact that the inner regions of massive SF galaxies with high SFE respond more rapidly to any change in the accretion from the host halo than the outer regions with low SFE.  Our result suggests that the variation of SFR within and across galaxies may be primarily governed by the different responses to variations of the cold gas inflow, due to different SFEs at different galactocentric distances.  

In addition to the variation of SFR within and across the galaxies, the gas-regulator model can also explain the elevation (or suppression) of star formation at all galactocentric radii for ESFGs (or SSFGs). Assuming an overall increasing or decreasing inflow rate for a galaxy, both the inner and outer regions would have enhanced (or suppressed) star formation.  An interesting question is whether the phase lag, evident in Figure \ref{fig:toy_model_A} and Equation \ref{eq:solution}, would produce noticeable effects within a given galaxy, e.g. by causing the elevation or suppression of star-formation at different radii to be out of phase.  The phase delay depends on the effective gas depletion time.  According to the Equation \ref{eq:solution}, the phase delay is always less than $\pi$/2. 
In practice, the phase delay will be considerably smaller, because the change of the effective gas depletion times from the inner regions to the outer regions is always less than 0.5 dex (see Figure \ref{fig:model_fit}).   According to Equation \ref{eq:solution}, and the actual range of $\tau_{\rm dep,eff}/T_{\rm p}$ in Figure \ref{fig:model_fit}, we would expect that the largest phase delay to be around $\delta$ = 0.16$\pi$, only a third of the maximum $\pi$/2.   Given this, our very naive and heuristic model would predict a more or less coherent elevation/suppression of SFR at all radii for ESFGs/SSFGs, as observed (see Figure \ref{fig:dsfr}). 

\subsection {Variations in SFE with constant inflow rate}
\label{subsec:model_B}

Looking at the continuity Equation \ref{eq:resp}, we could also imagine that the variations in \sigsfr\ are due to imposed variations in the SFE, rather than the inflow rate.  It is not clear how these variations could be imposed on a galaxy, but for completeness we consider this scenario as a Model B. We consider only the linear case (equivalent to Model A1 above) and simply impose that the SFE varies as 
\begin{equation}
{\rm SFE(t) = SFE_0 +SFE_t}\cdot {\rm sin}({\rm 2\pi} t/T_{\rm p}), 
\label{eq:model_B}
\end{equation}
where SFE$_{\rm 0}$ is the median SFE(t), SFE$_{\rm t}$ is the oscillation amplitude of SFE(t), and $T_{\rm p}$ is the period of SFE(t). 

\begin{figure}
\center{
\epsfig{figure=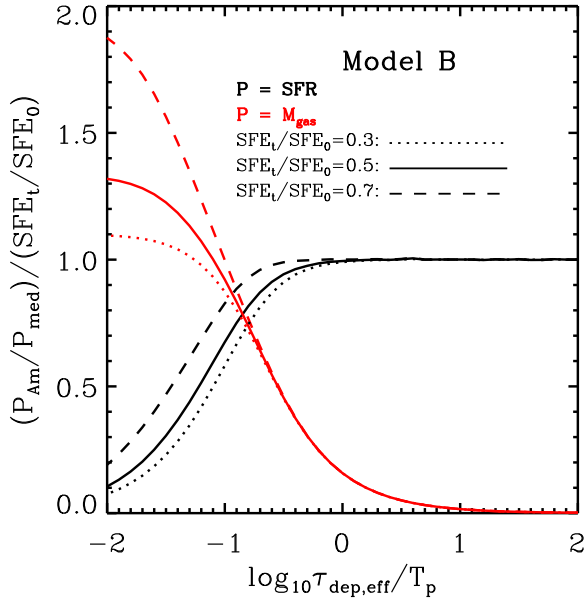,clip=true,width=0.45\textwidth}
}
\caption{Illustration of toy Model B with assuming that the SFE oscillates in a sinusoidal function in linear space with fixing the inflow rate.  The behavior of SFR(t) and $M_{\rm gas}(t)$ in Model B are shown as the oscillation period of the SFE is varied relative to the median effective gas depletion timescale. 
The black (or red) lines represent the oscillation amplitude of SFR(t) (or $M_{\rm gas}(t)$) to median SFR(t) (or $M_{\rm gas}(t)$) relative to SFE$_{\rm t}$/SFE$_{\rm 0}$ as a function of $\tau_{\rm dep,eff}/T_{\rm p}$ at three different oscillation amplitude of SFE: SFE$_{\rm t}$/SFE$_{\rm 0}$= 0.3, 0.5, and 0.7.  }
\label{fig:toy_model_B}
\end{figure}

Substituting Equation \ref{eq:model_B} into Equation \ref{eq:resp} and fixing the inflow rate,  we again determine solutions numerically.  The behavior of  SFR(t) and $M_{\rm gas}(t)$ of Model B is shown in Figure \ref{fig:toy_model_B}. The Figure \ref{fig:toy_model_B} shows the relative amplitude of oscillations in star-formation rate (or mass of gas), i.e. the oscillation amplitude of the SFR(t)  (or $M_{\rm gas}(t)$) relative to the median values, relative to the variation in star-formation efficiency, SFE$_{\rm t}$/SFE$_{\rm 0}$, as a function of the ratio of the median effective gas depletion timescale $\tau_{\rm dep,eff}$ relative to the driving period $T_{\rm p}$, for three different oscillation amplitudes of the SFE: SFE$_{\rm t}$/SFE$_{\rm 0}$= 0.3, 0.5, and 0.7.

Interestingly, the relative amplitude of variations in star-formation rate appears to {\it increase} monotonically with the median $\tau_{\rm dep,eff}/T_{\rm p}$, while the relative amplitude of variations in $M_{\rm gas}(t)$ decreases.  When the SFE varies on a very short timescale with respect to the effective gas depletion timescale ($\log_{10}\tau_{\rm dep,eff}/T_{\rm p}>$0),  the relative oscillation amplitude of SFR(t) exactly follows the input relative oscillation amplitude of SFE, and the mass of gas hardly varies at all.  With decreasing the $\tau_{\rm dep,eff}/T_{\rm p}$, the oscillation of SFR(t) becomes less and less significant, while the oscillation of $M_{\rm gas}(t)$ becomes larger and larger. We note that while the amplitude of the variations in SFE has some effect, the basic qualitative trends of the relative oscillation amplitude of SFR(t) and $M_{\rm gas}(t)$ are similar at different SFE$_{\rm t}$/SFE$_{\rm 0}$ (see different line styles in Figure \ref{fig:toy_model_B}). At fixed $\tau_{\rm dep,eff}/T_{\rm p}$ (at low $\tau_{\rm dep,eff}/T_{\rm p}$), the relative oscillation amplitude of SFR(t) and $M_{\rm gas}(t)$ is higher at higher SFE$_{\rm t}$/SFE$_{\rm 0}$. 

Apart from the difficulty in imagining how the SFE could be driven in this way, we note that this Model B produces the opposite trend in the amplitude of SFR variations as a function of $\tau_{\rm dep,eff}/T_{\rm p}$ as seen in the observational data (see Figure \ref{fig:am_tau}) and as successfully reproduced by Model A1 and A2.  This suggests to us that variations in SFE with a more or less constant inflow rate are unlikely to be the primary factor in determining the variation of SFR in galaxies, and we do not consider Model B further.  In this next subsection we explore a hybrid situation in which variations in SFE are consequent to variations in the inflow rate via variations in the gas mass.

\subsection {Variations in inflow rate with consequently varying SFE}
\label{subsec:model_C}

As a third option, we therefore consider a toy model in which variations in SFR are driven by variations in the inflow rate, as in Model A1-A3, but in which the SFE varies in response to variations in the gas content of the system.  This would be appropriate if the SFE was given by a pure Kennicutt-Schmidt type relation, rather than the constant SFE used in Section \ref{subsec:model_A}, and is more in line with e.g. the numerical simulations of \cite{Tacchella-16b} who emphasized the role of varying $\tau_{\rm dep}$ in the oscillation of galaxies above and below the SFMS.  We refer this model as Model C. 

In the (steady-state) gas regulator model \citep{Lilly-13}, the value of the SFE has no effect on the sSFR, which in the ``ideal regulator'' case is always set by the regulator to be exactly equal to the specific accretion rate, independent of $\tau_{\rm dep}$.   Although in that paper, an SFE independent of gas mass was assumed, it was commented that an SFE that increased with gas mass would not affect this basic result, since it would only affect the reaction timescale of the regulator $\tau_{\rm dep}$ and the equilibrium gas fraction and not the resulting equilibrium level of star-formation.   We would not therefore expect major changes to the results of Subsection \ref{subsec:model_A}.

To verify this, we re-ran Model A1-A3, but now introduced an SFE that varied as the mass of gas to the 0.4 power, i.e. a pure Kennicutt-Schmidt Law $\Sigma_{\rm SFR} \propto \Sigma_{\rm gas}^{1.4}$.  This star formation law also indicates a radially decreasing SFE from the center outward as the extended Schmidt law and Silk-Elmegreen law \citep[c.f.][]{Utomo-17},  since the star formation and the gas surface density are almost always peaked at the center of galaxies (see Figure \ref{fig:dsfr}). 
  As we would expect, this change produces only a small change in the amplitude of the response of the SFR to changes in the inflow relative to the results of Models A1-A3.  This is shown in the right-hand panel of Figure \ref{fig:toy_model_A} and Figure \ref{fig:toy_model_curve} (green lines).  For a positive dependence of SFE on gas mass, a slightly larger amplitude of SFR for a given amplitude of variation in inflow is produced because, as the gas mass increases, the increase in star-formation rate is even stronger with this SFE law.  However the effect is as expected, quite small.  This suggests that the variation of SFE implied by a pure Kennicutt-Schmidt type relation plays only a secondary role in determining the variation of SFR on the SFMS with respect to the variation of inflow rate. 

By using 26 simulated galaxies from their zoom-in hydrocosmological simulations, \cite{Tacchella-16b} found that both the gas mass and SFE play a role in determining where galaxies located on the SFMS: galaxies above the SFMS have higher cold gas mass and higher SFE than galaxies below the SFMS.  However, the dependence of $\Delta$SFR on SFE in their result is not surprising, since the SFE in their simulation was set to mimic the empirical Kennicutt-Schmidt law. 

Whether the SFE depends on the cold gas mass (or cold mass surface density) or not, and whether the overall SFE is different above and below the SFMS,  are still open questions observationally. The extended Schmidt law used earlier in this paper suggests no direct link between gas mass and SFE.  Observationally, \cite{Genzel-15} have found a positive correlation between the SFE (inverse gas depletion time) and the $\Delta$SFR with a power law index of -0.46.   However, since the observed star-formation rate enters into both $\Delta$SFR and the SFE, uncertainties in the observational determination of the SFR may produce a spurious correlation between $\Delta$SFR and the SFE.  

\subsection{Discussion of the toy-model}
\label{subsec:toymodeldiscussion}

%In interpreting qualitatively Figure \ref{fig:am_tau} in terms of the response of a regulator system to variations in the inflow rate there is an implicit assumption that these variations are, in terms of both amplitude and timescale, more or less independent of both the mass of the galaxy and of position within the galaxy.  Both of these are unlikely to be true in detail, which is why we refrained from attempting to quantitatively interpret the Figure \ref{fig:am_tau} to try to isolate these quantities. 

It is nevertheless interesting that the data points in Figure \ref{fig:am_tau} lie in a narrow sequence on the $\sigma(\Delta\Sigma_{\rm SFR})$-$\tau_{\rm dep}$ relation, for all the stellar mass bins and at least for all the radii less than \re, regardless of which star formation law is adopted.  %This may imply the existence of a uniform inflow rate for galaxies of different stellar mass bins and different galactocentric radii. 
Although the assumption of a uniform inflow rate within and across galaxies in our model can account for the observational result in Figure \ref{fig:am_tau}, it is also possible that a different oscillation amplitude combined with a different period of $\Phi(t)$ make the data points form a narrow sequence on the $\sigma(\Delta\Sigma_{\rm SFR})$-$\tau_{\rm dep}$  diagram  by coincidence, because the oscillation amplitude and period of inflow rate are highly degenerate according to our toy model.  The mass-loading factor is another factor, which certainly introduces uncertainties.  It is generally assumed that the mass-loading factor is anti-correlated with stellar mass, but how the mass-loading factor should vary within galaxies at different radii is not well understood.  

One obvious issue with our interpretation is that, while our toy model provides a very nice explanation for the elevation and suppression of star formation within and across galaxies,  the variation of gas inflow rate should in principle always be {\it larger} than the observed variation of \sigsfr\ (see Figure \ref{fig:toy_model_A} and Figure \ref{fig:toy_model_curve}), since the sluggish response of the regulator is only to {\it decrease} the amplitude of variation. According to Figure \ref{fig:am_tau} and Figure \ref{fig:toy_model_curve}, the variation of specific gas inflow rate should therefore be larger than 0.6 dex, at least for the inner region of the most massive bin.  This is much larger than the scatter of the overall SFMS at both low and high redshift which is comparable also to the dispersion in overall halo specific accretion rates averaged within 20 per cent of the Hubble time \citep{Rodrguez-Puebla-16}.  

We have not explored this in detail, but there are a number of possibilities that could account for this.  Depending on whether the accretion occurs as a steady stream or associated with discrete events (like minor mergers), any theoretical estimate of the variation of specific accretion rates would itself require a time average over some interval of time.  Indeed, one could envisage that there will be a frequency spectrum of accretion which will produce, via the response curves shown in Figure \ref{fig:toy_model_curve}, a frequency spectrum of star-formation.  It is not surprising that the variation of the overall halo specific accretion  would be much larger than 0.3 dex if averaged within a much shorter timescale than that in \cite{Rodrguez-Puebla-16}. One could also imagine that the mass inflow onto the central galaxies out of the halo showed larger temporal variations than the accretion of gas onto the larger halo, perhaps due to instabilities within the halo itself. 

Determining the frequency spectrum of star formation, whether in simulations \citep{Matthee-Schaye-19} or observationally \citep{Caplar-Tacchella-19} will be interesting. We would expect this quantity to be given by the frequency spectrum of accretion (at a given radius and mass) multiplied by the response curves in Figure \ref{fig:toy_model_curve}. 

\section{Other Discussion}
\label{sec:discussion}
\subsection{Relation to the variation in integrated SFR}

\begin{figure}
\center{
\epsfig{figure=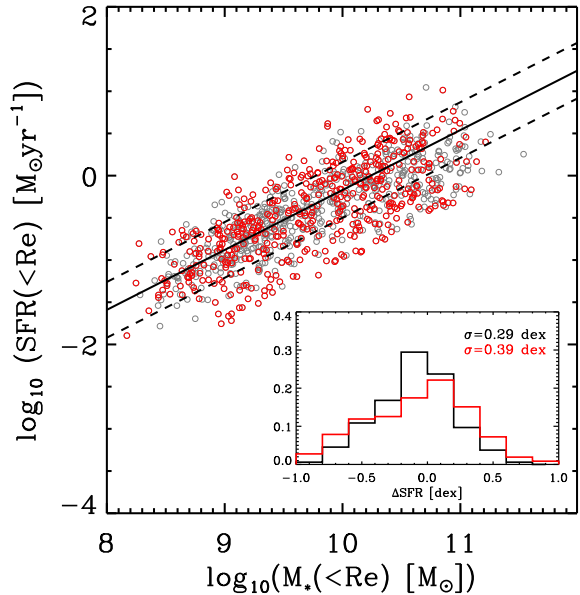,clip=true,width=0.45\textwidth}
}
\caption{The SFMS for the compact SF galaxies (red circles) and extended SF ones (gray circles). Enclosed plot shows the distribution of $\Delta$SFR for compact SF (red histogram) and extended SF galaxies (gray histogram). The scatter of $\Delta$SFR for the two populations are also denoted. }
\label{fig:sfms_1}
\end{figure}

%A, a natural prediction of our model: the compact SF galaxies are more scatteredly distributed on the SFMS than extended SF galaxies, because compact SF have higher SFE, and can be better following the fluctuation of sMIR than extended SF galaxies. 
%B, the data is in good agreement with the prediction, help us to understand the scatter of SFMS. Also see Wang et al. 2017, and Wang et al. 2018. 
%(To do in the future:  to check the SFMS at higher redshift, or check the upper envelope of SFMS, whether the compact galaxies are more scatteredly distributed on the MS.  ) 

The observational results in Section \ref{sec:results} above was based on spatially resolved measurements within galaxies, but the principles should apply also to galaxies as a whole.  Based on both of our adopted SFE laws, we would expect that, at a given stellar mass, compact galaxies would have higher SFE than extended galaxies.  Evidence for this is given in the analysis of  \cite{Wang-Kong-Pan-18} in which it was shown that the sample of compact SF galaxies as a whole have half the \HI\ gas content as that of extended SF galaxies.   \cite{Wang-Kong-Pan-18} interpreted this deficiency of gas as a signature of imminent quenching.   But the difference in HI content can be equally well interpreted as compact galaxies having twice the SFE as those of extended SF galaxies, which is actually in good qualitative agreement with the extended Schmidt law \citep{Shi-11} used in this paper.  

If this is the case, then we would expect within the picture that we have developed in this paper, that the shorter SFE of the compact galaxies would lead to quicker responses to changes in the accretion rate and therefore to larger variations in the SFR.  All other things being equal, we would predict that compact SF galaxies would have the {\it same} mean SFR as extended ones, but a larger {\it dispersion} around this mean, assuming that the halo accretion properties of the two sets of galaxies would be the same. 

These predictions are both in good agreement with the results found in the SDSS analysis in \cite{Wang-Kong-Pan-18}. The compact SF galaxies indeed show a very similar median SFR to the extended SF galaxies, while the dispersion (standard deviation) of the global SFR for compact SF galaxies is indeed larger than those of extended SF galaxies for $\sim30\%$ over the whole stellar mass range.  

We therefore also examine the distribution of compact and extended SF galaxies on the SFMS for our MaNGA sample examined here, for which we do not of course have HI data.  Figure \ref{fig:sfms_1} presents the SFMS dividing the MaNGA galaxies into compact and extended SF galaxies according to their mass-size relation.   In practice, we first fit a straight line for the \mstar-\re\ relation, then separate galaxies into two more or less equal classes, compact SF and extended SF galaxies by this straight line.  As shown in Figure \ref{fig:sfms_1}, the compact and extended SF galaxies have similar mean SFR with respect to the SFMS. However, the dispersion in the $\Delta$SFR of the two populations is again significantly different, in the sense that compact galaxies show a 34\% broader range of $\Delta$SFR than the extended SF galaxies.  We perform a Kolmogorov-Smimov test to quantify the significance of this difference and obtain a probability of 0.002,  suggesting a 99.8\% significance for the null hypothesis of the two distributions being drawn from the same parent distribution to be rejected.  We have also examined this result by broadly separating galaxies into two stellar mass bins, and find that compact SF galaxies show a larger scatter on the SFMS than extended SF galaxies for both low and high stellar mass bins.

A natural prediction of our model is that the scatter of the SFMS should increase with decreasing effective gas depletion time. Specifically, it is expected that massive galaxies have a larger scatter on the SFMS than less massive galaxies because more massive galaxies usually have higher stellar surface density, and therefore shorter gas depletion time.  This prediction is consistent with previous studies that the scatter of the SFMS broadens with increasing stellar mass at stellar mass greater than $\sim10^9$\Msun   \citep[e.g.][]{Guo-15, Willett-15, Davies-19}.   However, at the low mass end, the scatter of the SFMS appears to increase again with decreasing stellar mass \citep{Davies-19}.  This does not necessarily argue against our model. Actually, at the low mass end, the wind mass-loading factors driven by stellar feedback may become very large, resulting in short {\it effective} depletion times.  If the mass-loading factor increases more quickly than the star formation efficiency declines, then the effective gas depletion time would actually decrease with decreasing stellar mass. In this case, our model would predict an increase of the scatter of SFMS at the low mass end. However, we note that it is still rather unclear observationally what the mass-loading factors actually are for the inflow and outflow of galaxies, especially for low-mass galaxies. 

Here we can also comment that the difference between $\tau_{\rm dep}$ and $\tau_{\rm dep,eff}$ may also account for some of the flattening in Figures \ref{fig:am_tau} and \ref{fig:resolved_ms_sig}, in which $\tau_{\rm dep}$ was plotted.  In particular, on Figure \ref{fig:resolved_ms_sig}, it is possible that the $\tau_{\rm dep,eff}$ for low $\Sigma_*$ are substantially shorter than the nominal $\tau_{\rm dep}$ computed from $\Sigma_*$, producing a similar response in the regulator to higher values of $\Sigma_*$ with lower wind mass-loading factors.

As noted above, a gas regulator fed by some steady specific accretion rate will set the sSFR to be exactly this specific accretion \citep{Lilly-13}, quite {\it independent} of the SFE.  As shown in this paper, however, the response to variations in the specific accretion rate will however depend on the SFE.  The quicker response of systems with high SFE produces larger variations in sSFR for a given variation in specific accretion rate.   The observational result in Figure \ref{fig:sfms_1} is therefore in very good agreement with the prediction of gas regulator combined with the extended Schmidt law to determine the SFE. 
 
The analysis in this subsection therefore supports the scenario in which the movement of SF galaxies going up and down on the SFMS is primarily governed by the variations in cold gas inflow rate, while the scatter of the SFMS depends on both the oscillation amplitude of the gas inflow rate {\it and} by the reaction of the system to the oscillation which is set by the parameter $\xi$. 

The continuity Equation \ref{eq:resp} can be trivially re-written \citep[c.f. ][]{Tacchella-16b} in terms of the depletion and replenishment timescales

\begin{equation}
\begin{aligned}
\frac{d {\ln M_{\rm gas}}(t)}{dt} = \Phi(t)/M_{\rm gas} - {\rm SFE}\cdot(1+\lambda),
\end{aligned}
\label{eq:resp1}
\end{equation}

where $\Phi(t)/M_{\rm gas}$ is an inverse gas replenishment timescale, $t_{\rm rep}$ and SFE$(1+\lambda)$ is the inverse gas depletion timescale, $t_{\rm dep,eff}$.   Clearly if $t_{\rm rep} < t_{\rm dep,eff}$, the gas mass will increase and the galaxy will move up in sSFR, and vice versa.    \cite{Tacchella-16b} associated the replenishment timescale $t_{\rm rep}$ with the inverse specific accretion rate of the halo, but we believe that this is not strictly correct.  These quantities differ by a potentially large factor of $M_{\rm int}/M_{\rm gas}$, where $M_{\rm int}$ is the time integral of all previous inflow onto the galaxy, not the remaining gas in the galaxy (see \cite{Lilly-13} for a discussion).    This distinction becomes more and more important at later times in the Universe.

\subsection{Perspectives on the inside-out growth and the inside-out quenching of galaxies}

%A,  the exponential SF disk of MS galaxies, the stellar mass surface density profile indicates the formation of stellar core or bugle at early time, suggesting the ``inside-out'' growth picture for MS galaxies. 

As shown in Figure \ref{fig:typical_sfr}, the median \sigsfr\ profiles of NSFGs can be well expressed as an exponential function, indicating that these NSFGs have an exponential SF disk. However, their sSFR profiles show positive gradients especially for massive galaxies at galactic center, due to the contribution of the pronounced stellar core in galactic center.  These old cores are the result of the ``inside-out'' build up of the galaxy and are likely to have nothing to do with star formation quenching.  This is seen in the stellar age profiles \citep{Perez-13, Ibarra-Medel-16, Goddard-17, Wang-18a}.   

%B, the ``inside-out'' quenching can naturally  obtained if the accretion of cold gas in inefficient. There is no need to introduce extra mechanism to quench galactic center, such as the black hole feedback, and the morphological quenching. 
Quite distinct from the inside-out growth scenario,  we find, as have others before, that galaxies significantly below the nominal SFMS have suppressed star-formation at all radii, with the inner SFR more suppressed than in their outer regions. This finding has usually been interpreted as evidence for inside-out quenching  \citep[e.g.][]{Tacchella-15, Belfiore-18, Ellison-18, Medling-18, Sanchez-18}. 
In this picture, star formation quenching in galaxies first occurs in the center and then propagates out to larger galactocentric radii, which implies that there must be some physical processes happened in the galactic center to quench star formation.  Thus, the feedback of central massive black holes and the possible quenching effects of galaxy morphology (e.g. the presence of bars or bulges) have been invoked to account for this proposed inside-out quenching. 
 However, the morphological quenching is unable to simultaneously explain the fact that galaxies above the SFMS show more enhanced star formation in the center than in outer regions as we have shown in this paper. It is possible that the positive and negative effects of AGN feedback may be able to explain both the elevation and suppression of star formation in ESFGs and SSFGs, although it is still unclear whether, and how in detail, they actually work.

In this work we interpret the centrally-suppressed star formation in SSFGs to be due to a quicker response of these regions to a reduction in the overall gas inflow onto galaxies, rather than some quenching mechanisms operating initially at the center.   If the gas inflow rate recovers or becomes higher again, then the center would likewise quickly recover and have an elevated star-formation rate.  This naturally explains the symmetric elevation and suppression of star formation in galaxies.  
%Our toy model explains the variation of \sigsfr\ within galaxies and across galaxies. According to the model, the {\it apparent} inside-out quenching in galaxies can be naturally obtained with suppressing the global cold gas inflow rate without invoking any mechanisms to quench galactic center, since central regions response to this change more rapidly than the outer regions.  

%C, the compact SF galaxies quench their star formation in a shorter timescale than extended SF galaxies at given stellar mass if the cold gas replenishment is inefficient. This can explain the link between structural properties and quiescence of galaxies, even for satellite galaxies.   
Our scenario indicates that the SF galaxies go up and down across the SFMS during their lifetime \citep[e.g.][]{Tacchella-16b, Matthee-Schaye-19}.  What, if anything, has this got to do with the ``quenching'' of galaxies, i.e. the cessation of star-formation and the transformation of the galaxy into a passively evolving system.   We could imagine that quenching could occur when the inflow rate or the SFR falls below a certain threshold due to the cessation of cold gas inflow, for reasons that are poorly understood.  Assuming quenching begins with a sudden cutoff of cold gas inflow (regardless of the cause of such an event), the central region responds more rapidly than the outer region to the change of inflow, naturally resulting in a more suppressed star formation in the center than in the outer region during the quenching without invoking any specific physical mechanisms to preferentially quench the galactic center.  This is in good agreement with the observed more suppressed star formation in the center than in the outer regions for green valley galaxies \citep{Belfiore-18}.  
We also infer that compact SF galaxies would be transformed into passive systems in a shorter timescale with respect to  extended SF galaxies, since the compact galaxies with higher $\Sigma_*$ have higher SFE  and response quicker to the sudden cessation of cold gas inflow than the extended galaxies.

\section{Summary and Conclusion}
\label{sec:summary}

In this paper, we examined a sample of 976 SF galaxies selected from the SDSS-MaNGA DR14 in order to investigate the variation of star formation within galaxies.  We excluded mergers, irregulars and heavily disturbed galaxies \citep{Wang-18a}.  For SF regions, the SFRs were calculated based on the H$\alpha$ emission with correcting the intrinsic dust extinction, while the SFRs for LINER/AGN regions are determined by the \dindex-sSFR relation based on the SF spaxels \citep{Brinchmann-04}. We thereby generated maps and profiles for \sigsfr\ and \sigmstar\ for each individual galaxy and then also constructed median profiles for sets of galaxies selected according to their overall stellar mass and their location relative to the Main Sequence of star-forming galaxies. The median profiles of SFR (or mass) were normalized to the observed effective radius, i.e. we computed in units of SFR (or mass) per (0.2\re)$^2$ area. 
Our main results are summarized as follows. 

\begin{itemize}

\item The typical \sigsfr\ profiles of NSFGs (-0.33$<\Delta {\rm SFR}<0.33$ dex) can be well described by an exponential function at both low and high stellar mass bins. The sSFR profiles show increasingly positive gradients with increasing stellar mass.  These two facts suggest that the sSFR gradients are due to the increased prominence of central stellar cores or bulges, rather than being the result of any kind of star formation quenching.  Thus one cannot simply interpret the positive sSFR profile as evidence for ``inside-out'' quenching (see Section \ref{subsec:3.1}). 

\item The median sSFR profiles show little evidence for systematic negative gradients in all SFR bins over the whole stellar mass range, suggesting that essentially galaxies are increasing in mass-weighted size (see Figure \ref{fig:ssfr_mass}), and are not shrinking in size, i.e. galaxies are continuous to grow ``inside-out'', at least in the local universe only due to star formation (and without considering mergers and radial migration of stars). 

\item As a whole, galaxies above the SFMS exhibit an elevation of star formation everywhere, especially in the inner region. Similarly, galaxies  below the SFMS exhibit a global suppression of star formation everywhere, especially in the inner region. This enhanced elevation, and suppression, of star-formation in the inner regions of galaxies becomes progressively more pronounced as the stellar mass increases.   Interestingly, the pattern of radial elevation and suppression of star formation in galaxies with overall elevated, or suppressed, star-formation rates is strikingly symmetric over the whole stellar mass range (see Figure \ref{fig:dsfr} and Figure \ref{fig:dsfr_tau}). 

\item  In the main new result of the paper, we show that the amplitude of these variations in \sigsfr\, at a given radius in galaxies of a given mass, is well-correlated with the gas depletion times for the galaxies that can be derived from either the extended Schmidt law or Silk-Elmegreen law (see Figure \ref{fig:am_tau}). This result is also confirmed by a spaxel-by-spaxel analysis that shows that the scatter of the resolved Main Sequence relation, i.e. $\sigma(\Sigma_{\rm SFR})$ at a given $\Sigma_*$, increases with $\Sigma_*$ (see Figure \ref{fig:resolved_ms_mstar} and \ref{fig:resolved_ms_sig}). 

\item We interpret this result as reflecting the different dynamical response of a regulator system to changes in the accretion rate of gas onto the system.  In regions of high SFE (short gas depletion time), the SFR is able to follow changes in the accretion rate, while at lower SFE, the changing of SFR is damped with respect to changes in the accretion rate, thereby reducing the amplitude of changes in the SFR (see Figure \ref{fig:toy_model_A} and \ref{fig:toy_model_curve}). 

\end{itemize}

We constructed an idealized gas regulator model with an oscillating inflow in order to show how a tight correlation between the variation of \sigsfr\ and gas depletion time could arise. This was based on the simple physics in the gas regulator model \citep{Lilly-13} in which the formation of stars in galaxies is instantaneously regulated by the mass of gas in the reservoir.
In the model, the inflow rate is assumed to periodically vary in either linear space or in logarithmic space, with some period.  In both cases, the resulting variation of the SFR is found to strongly anti-correlate with the gas depletion time, simply because shorter gas depletion timescales (higher SFE) allow a quicker response to variations in accretion rate. This is parameterized via $\xi$, the ratio of effective gas depletion time to the oscillation period of inflow rate. 
This model therefore explains the observed variation of SFR within and across galaxies, but also it provides some other interesting implications.  

According to this picture, compact SF galaxies should have larger fluctuation of SFR on the SFMS than the extended SF galaxies of the same mass, but have the same average SFR.  Both of these are shown in Figure \ref{fig:sfms_1} \citep[see also][]{Wang-Kong-Pan-18}. 

We conclude that the radial elevation and suppression of SFR within galaxies, and the variation from galaxy to galaxy, are likely due to the simple physics of the gas regulator model and the importance in this model of the gas depletion timescale.  Several mechanisms have been proposed to account for the star formation elevation and suppression in massive galaxies, including the existence of bar-like structures, positive and negative AGN feedback, and the morphological stabilization of gas disk against star formation \citep[e.g.][]{Nelson-16, Spindler-18, Ellison-18, Guo-19}.  
Different from the interpretations of previous works, we suggest a more general and simple explanation for the symmetric fluctuation of \sigsfr\ that accounts for the symmetric nature of the excursions above and below ``normal'' SF galaxies.  Previously, the suppressed star formation in the centers of massive galaxies has been taken as evidence for ``inside-out'' quenching, suggesting that there must be some physical processes to quench galaxies from the center outwards. However, according to our explanation, this is due to the quicker response to decreases in accretion rate in the central regions, and is mirrored by the quicker response of the central regions also to increases in the accretion rate.  There is no necessity of any specific quenching processes preferentially operating at the centers of galaxies, according to the oscillating gas-regulator model.

\acknowledgments
We are grateful to the anonymous referee for their thoughtful and constructive review of the paper and their several suggestions (including the analysis of Section \ref{subsec:3.4}), which have improved the paper. This research has been supported by the Swiss National Science Foundation.

Funding for the Sloan Digital Sky Survey IV has been provided by
the Alfred P. Sloan Foundation, the U.S. Department of Energy Office of
Science, and the Participating Institutions. SDSS-IV acknowledges
support and resources from the Center for High-Performance Computing at
the University of Utah. The SDSS web site is www.sdss.org.

SDSS-IV is managed by the Astrophysical Research Consortium for the
Participating Institutions of the SDSS Collaboration including the
Brazilian Participation Group, the Carnegie Institution for Science,
Carnegie Mellon University, the Chilean Participation Group, the French Participation Group, 
Harvard-Smithsonian Center for Astrophysics,
Instituto de Astrof\'isica de Canarias, The Johns Hopkins University,
Kavli Institute for the Physics and Mathematics of the Universe (IPMU) /
University of Tokyo, Lawrence Berkeley National Laboratory,
Leibniz Institut f\"ur Astrophysik Potsdam (AIP),
Max-Planck-Institut f\"ur Astronomie (MPIA Heidelberg),
Max-Planck-Institut f\"ur Astrophysik (MPA Garching),
Max-Planck-Institut f\"ur Extraterrestrische Physik (MPE),
National Astronomical Observatory of China, New Mexico State University,
New York University, University of Notre Dame,
Observat\'ario Nacional / MCTI, The Ohio State University,
Pennsylvania State University, Shanghai Astronomical Observatory,
United Kingdom Participation Group,
Universidad Nacional Aut\'onoma de M\'exico, University of Arizona,
University of Colorado Boulder, University of Oxford, University of Portsmouth,
University of Utah, University of Virginia, University of Washington, University of Wisconsin,
Vanderbilt University, and Yale University.

\bibliography{rewritebib.bib}

\begin{thebibliography}{121}
\expandafter\ifx\csname natexlab\endcsname\relax\def\natexlab#1{#1}\fi

\bibitem[{{Abadi} {et~al.}(1999){Abadi}, {Moore}, \&
  {Bower}}]{Abadi-Moore-Bower-99}
{Abadi}, M.~G., {Moore}, B., \& {Bower}, R.~G. 1999, \mnras, 308, 947

\bibitem[{{Abdurro'uf} \& {Akiyama}(2017)}]{-Akiyama-17}
{Abdurro'uf}, \& {Akiyama}, M. 2017, \mnras, 469, 2806

\bibitem[{{Abolfathi} {et~al.}(2018){Abolfathi}, {Aguado}, {Aguilar}, {Allende
  Prieto}, {Almeida}, {Ananna}, {Anders}, {Anderson}, \& {et
  al.}}]{Abolfathi-18}
{Abolfathi}, B., {Aguado}, D.~S., {Aguilar}, G., {et~al.} 2018, \apjs, 235, 42

\bibitem[{{Bacchini} {et~al.}(2018){Bacchini}, {Fraternali}, {Iorio}, \&
  {Pezzulli}}]{Bacchini-18}
{Bacchini}, C., {Fraternali}, F., {Iorio}, G., \& {Pezzulli}, G. 2018, arXiv
  e-prints

\bibitem[{{Baldwin} {et~al.}(1981){Baldwin}, {Phillips}, \&
  {Terlevich}}]{Baldwin-Phillips-Terlevich-81}
{Baldwin}, J.~A., {Phillips}, M.~M., \& {Terlevich}, R. 1981, \pasp, 93, 5

\bibitem[{{Barro} {et~al.}(2017){Barro}, {Faber}, {Koo}, {Dekel}, {Fang},
  {Trump}, {P{\'e}rez-Gonz{\'a}lez}, {Pacifici}, \& {et al.}}]{Barro-17}
{Barro}, G., {Faber}, S.~M., {Koo}, D.~C., {et~al.} 2017, \apj, 840, 47

\bibitem[{{Belfiore} {et~al.}(2016){Belfiore}, {Maiolino}, {Maraston},
  {Emsellem}, {Bershady}, {Masters}, {Yan}, {Bizyaev}, \& {et
  al.}}]{Belfiore-16}
{Belfiore}, F., {Maiolino}, R., {Maraston}, C., {et~al.} 2016, \mnras, 461,
  3111

\bibitem[{{Belfiore} {et~al.}(2018){Belfiore}, {Maiolino}, {Bundy}, {Masters},
  {Bershady}, {Oyarz{\'u}n}, {Lin}, {Cano-Diaz}, \& {et al.}}]{Belfiore-18}
{Belfiore}, F., {Maiolino}, R., {Bundy}, K., {et~al.} 2018, \mnras, 477, 3014

\bibitem[{{Bell} {et~al.}(2012){Bell}, {van der Wel}, {Papovich}, {Kocevski},
  {Lotz}, {McIntosh}, {Kartaltepe}, {Faber}, \& {et al.}}]{Bell-12}
{Bell}, E.~F., {van der Wel}, A., {Papovich}, C., {et~al.} 2012, \apj, 753, 167

\bibitem[{{Blanton} {et~al.}(2011){Blanton}, {Kazin}, {Muna}, {Weaver}, \&
  {Price-Whelan}}]{Blanton-11}
{Blanton}, M.~R., {Kazin}, E., {Muna}, D., {Weaver}, B.~A., \& {Price-Whelan},
  A. 2011, \aj, 142, 31

\bibitem[{{Blanton} {et~al.}(2017){Blanton}, {Bershady}, {Abolfathi},
  {Albareti}, {Allende Prieto}, {Almeida}, {Alonso-Garc{\'{\i}}a}, {Anders}, \&
  {et al.}}]{Blanton-17}
{Blanton}, M.~R., {Bershady}, M.~A., {Abolfathi}, B., {et~al.} 2017, \aj, 154,
  28

\bibitem[{{Bouch{\'e}} {et~al.}(2010){Bouch{\'e}}, {Dekel}, {Genzel}, {Genel},
  {Cresci}, {F{\"o}rster Schreiber}, {Shapiro}, {Davies}, \& {et
  al.}}]{Bouche-10}
{Bouch{\'e}}, N., {Dekel}, A., {Genzel}, R., {et~al.} 2010, \apj, 718, 1001

\bibitem[{{Brinchmann} {et~al.}(2004){Brinchmann}, {Charlot}, {White},
  {Tremonti}, {Kauffmann}, {Heckman}, \& {Brinkmann}}]{Brinchmann-04}
{Brinchmann}, J., {Charlot}, S., {White}, S.~D.~M., {et~al.} 2004, \mnras, 351,
  1151

\bibitem[{{Bruzual} \& {Charlot}(2003)}]{Bruzual-Charlot-03}
{Bruzual}, G., \& {Charlot}, S. 2003, \mnras, 344, 1000

\bibitem[{{Bundy} {et~al.}(2015){Bundy}, {Bershady}, {Law}, {Yan}, {Drory},
  {MacDonald}, {Wake}, {Cherinka}, \& {et al.}}]{Bundy-15}
{Bundy}, K., {Bershady}, M.~A., {Law}, D.~R., {et~al.} 2015, \apj, 798, 7

\bibitem[{{Cano-D{\'{\i}}az} {et~al.}(2016){Cano-D{\'{\i}}az}, {S{\'a}nchez},
  {Zibetti}, {Ascasibar}, {Bland-Hawthorn}, {Ziegler}, {Gonz{\'a}lez Delgado},
  {Walcher}, \& {et al.}}]{Cano-Daz-16}
{Cano-D{\'{\i}}az}, M., {S{\'a}nchez}, S.~F., {Zibetti}, S., {et~al.} 2016,
  \apjl, 821, L26

\bibitem[{{Caplar} \& {Tacchella}(2019)}]{Caplar-Tacchella-19}
{Caplar}, N., \& {Tacchella}, S. 2019, arXiv e-prints

\bibitem[{{Cardelli} {et~al.}(1989){Cardelli}, {Clayton}, \&
  {Mathis}}]{Cardelli-Clayton-Mathis-89}
{Cardelli}, J.~A., {Clayton}, G.~C., \& {Mathis}, J.~S. 1989, \apj, 345, 245

\bibitem[{{Ceverino} \& {Klypin}(2009)}]{Ceverino-Klypin-09}
{Ceverino}, D., \& {Klypin}, A. 2009, \apj, 695, 292

\bibitem[{{Chabrier}(2003)}]{Chabrier-03}
{Chabrier}, G. 2003, \pasp, 115, 763

\bibitem[{{Chiappini} {et~al.}(2001){Chiappini}, {Matteucci}, \&
  {Romano}}]{Chiappini-Matteucci-Romano-01}
{Chiappini}, C., {Matteucci}, F., \& {Romano}, D. 2001, \apj, 554, 1044

\bibitem[{{Chown} {et~al.}(2018){Chown}, {Li}, {Li}, {Athanassoula}, {Wilson},
  {Lin}, {Mo}, {Parker}, \& {et al.}}]{Chown-18}
{Chown}, R., {Li}, C., {Li}, N., {et~al.} 2018, arXiv e-prints

\bibitem[{{Cicone} {et~al.}(2014){Cicone}, {Maiolino}, {Sturm},
  {Graci{\'a}-Carpio}, {Feruglio}, {Neri}, {Aalto}, {Davies}, \& {et
  al.}}]{Cicone-14}
{Cicone}, C., {Maiolino}, R., {Sturm}, E., {et~al.} 2014, \aap, 562, A21

\bibitem[{{Cid Fernandes} {et~al.}(2004){Cid Fernandes}, {Gu}, {Melnick},
  {Terlevich}, {Terlevich}, {Kunth}, {Rodrigues Lacerda}, \&
  {Joguet}}]{CidFernandes-04}
{Cid Fernandes}, R., {Gu}, Q., {Melnick}, J., {et~al.} 2004, \mnras, 355, 273

\bibitem[{{Conselice} {et~al.}(2003){Conselice}, {Chapman}, \&
  {Windhorst}}]{Conselice-Chapman-Windhorst-03}
{Conselice}, C.~J., {Chapman}, S.~C., \& {Windhorst}, R.~A. 2003, \apjl, 596,
  L5

\bibitem[{{Cox} {et~al.}(2006){Cox}, {Jonsson}, {Primack}, \&
  {Somerville}}]{Cox-06}
{Cox}, T.~J., {Jonsson}, P., {Primack}, J.~R., \& {Somerville}, R.~S. 2006,
  \mnras, 373, 1013

\bibitem[{{Croom} {et~al.}(2012){Croom}, {Lawrence}, {Bland-Hawthorn},
  {Bryant}, {Fogarty}, {Richards}, {Goodwin}, {Farrell}, \& {et
  al.}}]{Croom-12}
{Croom}, S.~M., {Lawrence}, J.~S., {Bland-Hawthorn}, J., {et~al.} 2012, \mnras,
  421, 872

\bibitem[{{Daddi} {et~al.}(2007){Daddi}, {Dickinson}, {Morrison}, {Chary},
  {Cimatti}, {Elbaz}, {Frayer}, {Renzini}, \& {et al.}}]{Daddi-07}
{Daddi}, E., {Dickinson}, M., {Morrison}, G., {et~al.} 2007, \apj, 670, 156

\bibitem[{{Danovich} {et~al.}(2015){Danovich}, {Dekel}, {Hahn}, {Ceverino}, \&
  {Primack}}]{Danovich-15}
{Danovich}, M., {Dekel}, A., {Hahn}, O., {Ceverino}, D., \& {Primack}, J. 2015,
  \mnras, 449, 2087

\bibitem[{{Dav{\'e}} {et~al.}(2012){Dav{\'e}}, {Finlator}, \&
  {Oppenheimer}}]{Dave-Finlator-Oppenheimer-12}
{Dav{\'e}}, R., {Finlator}, K., \& {Oppenheimer}, B.~D. 2012, \mnras, 421, 98

\bibitem[{{Davies} {et~al.}(2019){Davies}, {Lagos}, {Katsianis}, {Robotham},
  {Cortese}, {Driver}, {Bremer}, {Brown}, \& {et al.}}]{Davies-19}
{Davies}, L.~J.~M., {Lagos}, C.~d.~P., {Katsianis}, A., {et~al.} 2019, \mnras,
  483, 1881

\bibitem[{{Dekel} \& {Burkert}(2014)}]{Dekel-Burkert-14}
{Dekel}, A., \& {Burkert}, A. 2014, \mnras, 438, 1870

\bibitem[{{Dopita} \& {Ryder}(1994)}]{Dopita-Ryder-94}
{Dopita}, M.~A., \& {Ryder}, S.~D. 1994, \apj, 430, 163

\bibitem[{{Drory} {et~al.}(2015){Drory}, {MacDonald}, {Bershady}, {Bundy},
  {Gunn}, {Law}, {Smith}, {Stoll}, \& {et al.}}]{Drory-15}
{Drory}, N., {MacDonald}, N., {Bershady}, M.~A., {et~al.} 2015, \aj, 149, 77

\bibitem[{{El-Badry} {et~al.}(2016){El-Badry}, {Wetzel}, {Geha}, {Hopkins},
  {Kere{\v s}}, {Chan}, \& {Faucher-Gigu{\`e}re}}]{El-Badry-16}
{El-Badry}, K., {Wetzel}, A., {Geha}, M., {et~al.} 2016, \apj, 820, 131

\bibitem[{{Elbaz} {et~al.}(2011){Elbaz}, {Dickinson}, {Hwang},
  {D{\'{\i}}az-Santos}, {Magdis}, {Magnelli}, {Le Borgne}, {Galliano}, \& {et
  al.}}]{Elbaz-11}
{Elbaz}, D., {Dickinson}, M., {Hwang}, H.~S., {et~al.} 2011, \aap, 533, A119

\bibitem[{{Ellison} {et~al.}(2018){Ellison}, {S{\'a}nchez}, {Ibarra-Medel},
  {Antonio}, {Mendel}, \& {Barrera-Ballesteros}}]{Ellison-18}
{Ellison}, S.~L., {S{\'a}nchez}, S.~F., {Ibarra-Medel}, H., {et~al.} 2018,
  \mnras, 474, 2039

\bibitem[{{Elmegreen}(1997)}]{Elmegreen-97}
{Elmegreen}, B.~G. 1997, in Revista Mexicana de Astronomia y Astrofisica
  Conference Series, ed. J.~{Franco}, R.~{Terlevich}, \& A.~{Serrano}, Vol.~6,
  165

\bibitem[{{Fabian}(2012)}]{Fabian-12}
{Fabian}, A.~C. 2012, \araa, 50, 455

\bibitem[{{Fang} {et~al.}(2013){Fang}, {Faber}, {Koo}, \& {Dekel}}]{Fang-13}
{Fang}, J.~J., {Faber}, S.~M., {Koo}, D.~C., \& {Dekel}, A. 2013, \apj, 776, 63

\bibitem[{{Genzel} {et~al.}(2015){Genzel}, {Tacconi}, {Lutz}, {Saintonge},
  {Berta}, {Magnelli}, {Combes}, {Garc{\'{\i}}a-Burillo}, \& {et
  al.}}]{Genzel-15}
{Genzel}, R., {Tacconi}, L.~J., {Lutz}, D., {et~al.} 2015, \apj, 800, 20

\bibitem[{{Goddard} {et~al.}(2017){Goddard}, {Thomas}, {Maraston}, {Westfall},
  {Etherington}, {Riffel}, {Mallmann}, {Zheng}, \& {et al.}}]{Goddard-17}
{Goddard}, D., {Thomas}, D., {Maraston}, C., {et~al.} 2017, \mnras, 466, 4731

\bibitem[{{Gonz{\'a}lez Delgado} {et~al.}(2016){Gonz{\'a}lez Delgado}, {Cid
  Fernandes}, {P{\'e}rez}, {Garc{\'{\i}}a-Benito}, {L{\'o}pez Fern{\'a}ndez},
  {Lacerda}, {Cortijo-Ferrero}, {de Amorim}, \& {et al.}}]{GonzalezDelgado-16}
{Gonz{\'a}lez Delgado}, R.~M., {Cid Fernandes}, R., {P{\'e}rez}, E., {et~al.}
  2016, \aap, 590, A44

\bibitem[{{Gunn} \& {Gott}(1972)}]{Gunn-Gott-72}
{Gunn}, J.~E., \& {Gott}, III, J.~R. 1972, \apj, 176, 1

\bibitem[{{Gunn} {et~al.}(2006){Gunn}, {Siegmund}, {Mannery}, {Owen}, {Hull},
  {Leger}, {Carey}, {Knapp}, \& {et al.}}]{Gunn-06}
{Gunn}, J.~E., {Siegmund}, W.~A., {Mannery}, E.~J., {et~al.} 2006, \aj, 131,
  2332

\bibitem[{{Guo} {et~al.}(2015){Guo}, {Zheng}, {Wang}, \& {Fu}}]{Guo-15}
{Guo}, K., {Zheng}, X.~Z., {Wang}, T., \& {Fu}, H. 2015, \apjl, 808, L49

\bibitem[{{Guo} {et~al.}(2019){Guo}, {Peng}, {Shao}, {Fu}, {Catinella},
  {Cortese}, {Yuan}, {Yan}, \& {et al.}}]{Guo-19}
{Guo}, K., {Peng}, Y., {Shao}, L., {et~al.} 2019, \apj, 870, 19

\bibitem[{{Ibarra-Medel} {et~al.}(2016){Ibarra-Medel}, {S{\'a}nchez},
  {Avila-Reese}, {Hern{\'a}ndez-Toledo}, {Gonz{\'a}lez}, {Drory}, {Bundy},
  {Bizyaev}, \& {et al.}}]{Ibarra-Medel-16}
{Ibarra-Medel}, H.~J., {S{\'a}nchez}, S.~F., {Avila-Reese}, V., {et~al.} 2016,
  \mnras, 463, 2799

\bibitem[{{Kalfountzou} {et~al.}(2017){Kalfountzou}, {Stevens}, {Jarvis},
  {Hardcastle}, {Wilner}, {Elvis}, {Page}, {Trichas}, \& {et
  al.}}]{Kalfountzou-17}
{Kalfountzou}, E., {Stevens}, J.~A., {Jarvis}, M.~J., {et~al.} 2017, \mnras,
  471, 28

\bibitem[{{Kennicutt}(1998)}]{Kennicutt-98}
{Kennicutt}, Jr., R.~C. 1998, \araa, 36, 189

\bibitem[{{Kewley} {et~al.}(2001){Kewley}, {Dopita}, {Sutherland}, {Heisler},
  \& {Trevena}}]{Kewley-01}
{Kewley}, L.~J., {Dopita}, M.~A., {Sutherland}, R.~S., {Heisler}, C.~A., \&
  {Trevena}, J. 2001, \apj, 556, 121

\bibitem[{{Kewley} {et~al.}(2006){Kewley}, {Groves}, {Kauffmann}, \&
  {Heckman}}]{Kewley-06}
{Kewley}, L.~J., {Groves}, B., {Kauffmann}, G., \& {Heckman}, T. 2006, \mnras,
  372, 961

\bibitem[{{Knobel} {et~al.}(2015){Knobel}, {Lilly}, {Woo}, \& {Kova{\v
  c}}}]{Knobel-15}
{Knobel}, C., {Lilly}, S.~J., {Woo}, J., \& {Kova{\v c}}, K. 2015, \apj, 800,
  24

\bibitem[{{Krumholz} {et~al.}(2012){Krumholz}, {Dekel}, \&
  {McKee}}]{Krumholz-Dekel-McKee-12}
{Krumholz}, M.~R., {Dekel}, A., \& {McKee}, C.~F. 2012, \apj, 745, 69

\bibitem[{{Lara-L{\'o}pez} {et~al.}(2010){Lara-L{\'o}pez}, {Cepa},
  {Bongiovanni}, {P{\'e}rez Garc{\'{\i}}a}, {Ederoclite}, {Casta{\~n}eda},
  {Fern{\'a}ndez Lorenzo}, {Povi{\'c}}, \& {et al.}}]{Lara-Lopez-10}
{Lara-L{\'o}pez}, M.~A., {Cepa}, J., {Bongiovanni}, A., {et~al.} 2010, \aap,
  521, L53

\bibitem[{{Larson}(1976)}]{Larson-76}
{Larson}, R.~B. 1976, \mnras, 176, 31

\bibitem[{{Law} {et~al.}(2015){Law}, {Yan}, {Bershady}, {Bundy}, {Cherinka},
  {Drory}, {MacDonald}, {S{\'a}nchez-Gallego}, \& {et al.}}]{Law-15}
{Law}, D.~R., {Yan}, R., {Bershady}, M.~A., {et~al.} 2015, \aj, 150, 19

\bibitem[{{Law} {et~al.}(2016){Law}, {Cherinka}, {Yan}, {Andrews}, {Bershady},
  {Bizyaev}, {Blanc}, {Blanton}, \& {et al.}}]{Law-16}
{Law}, D.~R., {Cherinka}, B., {Yan}, R., {et~al.} 2016, \aj, 152, 83

\bibitem[{{Li} {et~al.}(2005){Li}, {Wang}, {Zhou}, {Dong}, \& {Cheng}}]{Li-05}
{Li}, C., {Wang}, T.-G., {Zhou}, H.-Y., {Dong}, X.-B., \& {Cheng}, F.-Z. 2005,
  \aj, 129, 669

\bibitem[{{Li} {et~al.}(2015){Li}, {Wang}, {Lin}, {Bershady}, {Bundy},
  {Tremonti}, {Xiao}, {Yan}, \& {et al.}}]{Li-15}
{Li}, C., {Wang}, E., {Lin}, L., {et~al.} 2015, \apj, 804, 125

\bibitem[{{Lilly} \& {Carollo}(2016)}]{Lilly-Carollo-16}
{Lilly}, S.~J., \& {Carollo}, C.~M. 2016, \apj, 833, 1

\bibitem[{{Lilly} {et~al.}(2013){Lilly}, {Carollo}, {Pipino}, {Renzini}, \&
  {Peng}}]{Lilly-13}
{Lilly}, S.~J., {Carollo}, C.~M., {Pipino}, A., {Renzini}, A., \& {Peng}, Y.
  2013, \apj, 772, 119

\bibitem[{{Lin} {et~al.}(2017){Lin}, {Li}, {He}, {Xiao}, \& {Wang}}]{Lin-17}
{Lin}, L., {Li}, C., {He}, Y., {Xiao}, T., \& {Wang}, E. 2017, \apj, 838, 105

\bibitem[{{Liu} {et~al.}(2018){Liu}, {Wang}, {Lin}, {Gao}, {Liu}, {Berhane
  Teklu}, \& {Kong}}]{Liu-18}
{Liu}, Q., {Wang}, E., {Lin}, Z., {et~al.} 2018, \apj, 857, 17

\bibitem[{{Mahoro} {et~al.}(2017){Mahoro}, {Povi{\'c}}, \&
  {Nkundabakura}}]{Mahoro-Povic-Nkundabakura-17}
{Mahoro}, A., {Povi{\'c}}, M., \& {Nkundabakura}, P. 2017, \mnras, 471, 3226

\bibitem[{{Mannucci} {et~al.}(2010){Mannucci}, {Cresci}, {Maiolino}, {Marconi},
  \& {Gnerucci}}]{Mannucci-10}
{Mannucci}, F., {Cresci}, G., {Maiolino}, R., {Marconi}, A., \& {Gnerucci}, A.
  2010, \mnras, 408, 2115

\bibitem[{{Martig} {et~al.}(2009){Martig}, {Bournaud}, {Teyssier}, \&
  {Dekel}}]{Martig-09}
{Martig}, M., {Bournaud}, F., {Teyssier}, R., \& {Dekel}, A. 2009, \apj, 707,
  250

\bibitem[{{Matthee} \& {Schaye}(2019)}]{Matthee-Schaye-19}
{Matthee}, J., \& {Schaye}, J. 2019, \mnras, 484, 915

\bibitem[{{McNamara} \& {Nulsen}(2007)}]{McNamara-Nulsen-07}
{McNamara}, B.~R., \& {Nulsen}, P.~E.~J. 2007, \araa, 45, 117

\bibitem[{{Medling} {et~al.}(2018){Medling}, {Cortese}, {Croom}, {Green},
  {Groves}, {Hampton}, {Ho}, {Davies}, \& {et al.}}]{Medling-18}
{Medling}, A.~M., {Cortese}, L., {Croom}, S.~M., {et~al.} 2018, \mnras, 475,
  5194

\bibitem[{{Moore} {et~al.}(1996){Moore}, {Katz}, {Lake}, {Dressler}, \&
  {Oemler}}]{Moore-96}
{Moore}, B., {Katz}, N., {Lake}, G., {Dressler}, A., \& {Oemler}, A. 1996,
  \nat, 379, 613

\bibitem[{{Muratov} {et~al.}(2015){Muratov}, {Kere{\v s}},
  {Faucher-Gigu{\`e}re}, {Hopkins}, {Quataert}, \& {Murray}}]{Muratov-15}
{Muratov}, A.~L., {Kere{\v s}}, D., {Faucher-Gigu{\`e}re}, C.-A., {et~al.}
  2015, \mnras, 454, 2691

\bibitem[{{Nelson} {et~al.}(2016){Nelson}, {van Dokkum}, {F{\"o}rster
  Schreiber}, {Franx}, {Brammer}, {Momcheva}, {Wuyts}, {Whitaker}, \& {et
  al.}}]{Nelson-16}
{Nelson}, E.~J., {van Dokkum}, P.~G., {F{\"o}rster Schreiber}, N.~M., {et~al.}
  2016, \apj, 828, 27

\bibitem[{{Newman} {et~al.}(2012){Newman}, {Ellis}, {Bundy}, \&
  {Treu}}]{Newman-12}
{Newman}, A.~B., {Ellis}, R.~S., {Bundy}, K., \& {Treu}, T. 2012, \apj, 746,
  162

\bibitem[{{Noeske} {et~al.}(2007){Noeske}, {Weiner}, {Faber}, {Papovich},
  {Koo}, {Somerville}, {Bundy}, {Conselice}, \& {et al.}}]{Noeske-07}
{Noeske}, K.~G., {Weiner}, B.~J., {Faber}, S.~M., {et~al.} 2007, \apjl, 660,
  L43

\bibitem[{{Nulsen} {et~al.}(2005){Nulsen}, {McNamara}, {Wise}, \&
  {David}}]{Nulsen-05}
{Nulsen}, P.~E.~J., {McNamara}, B.~R., {Wise}, M.~W., \& {David}, L.~P. 2005,
  \apj, 628, 629

\bibitem[{{Omand} {et~al.}(2014){Omand}, {Balogh}, \&
  {Poggianti}}]{Omand-Balogh-Poggianti-14}
{Omand}, C.~M.~B., {Balogh}, M.~L., \& {Poggianti}, B.~M. 2014, \mnras, 440,
  843

\bibitem[{{Peng} {et~al.}(2012){Peng}, {Lilly}, {Renzini}, \&
  {Carollo}}]{Peng-12}
{Peng}, Y.-j., {Lilly}, S.~J., {Renzini}, A., \& {Carollo}, M. 2012, \apj, 757,
  4

\bibitem[{{Peng} {et~al.}(2010){Peng}, {Lilly}, {Kova{\v c}}, {Bolzonella},
  {Pozzetti}, {Renzini}, {Zamorani}, {Ilbert}, \& {et al.}}]{Peng-10}
{Peng}, Y.-j., {Lilly}, S.~J., {Kova{\v c}}, K., {et~al.} 2010, \apj, 721, 193

\bibitem[{{P{\'e}rez} {et~al.}(2013){P{\'e}rez}, {Cid Fernandes}, {Gonz{\'a}lez
  Delgado}, {Garc{\'{\i}}a-Benito}, {S{\'a}nchez}, {Husemann}, {Mast},
  {Rod{\'o}n}, \& {et al.}}]{Perez-13}
{P{\'e}rez}, E., {Cid Fernandes}, R., {Gonz{\'a}lez Delgado}, R.~M., {et~al.}
  2013, \apjl, 764, L1

\bibitem[{{Pezzulli} {et~al.}(2015){Pezzulli}, {Fraternali}, {Boissier}, \&
  {Mu{\~n}oz-Mateos}}]{Pezzulli-15}
{Pezzulli}, G., {Fraternali}, F., {Boissier}, S., \& {Mu{\~n}oz-Mateos}, J.~C.
  2015, \mnras, 451, 2324

\bibitem[{{Poggianti} {et~al.}(2017){Poggianti}, {Moretti}, {Gullieuszik},
  {Fritz}, {Jaff{\'e}}, {Bettoni}, {Fasano}, {Bellhouse}, \& {et
  al.}}]{Poggianti-17}
{Poggianti}, B.~M., {Moretti}, A., {Gullieuszik}, M., {et~al.} 2017, \apj, 844,
  48

\bibitem[{{Rodr{\'{\i}}guez-Puebla} {et~al.}(2016){Rodr{\'{\i}}guez-Puebla},
  {Primack}, {Behroozi}, \& {Faber}}]{Rodrguez-Puebla-16}
{Rodr{\'{\i}}guez-Puebla}, A., {Primack}, J.~R., {Behroozi}, P., \& {Faber},
  S.~M. 2016, \mnras, 455, 2592

\bibitem[{{Rowlands} {et~al.}(2018){Rowlands}, {Heckman}, {Wild}, {Zakamska},
  {Rodriguez-Gomez}, {Barrera-Ballesteros}, {Lotz}, {Thilker}, \& {et
  al.}}]{Rowlands-18}
{Rowlands}, K., {Heckman}, T., {Wild}, V., {et~al.} 2018, \mnras, 480, 2544

\bibitem[{{Salim} {et~al.}(2007){Salim}, {Rich}, {Charlot}, {Brinchmann},
  {Johnson}, {Schiminovich}, {Seibert}, {Mallery}, \& {et al.}}]{Salim-07}
{Salim}, S., {Rich}, R.~M., {Charlot}, S., {et~al.} 2007, \apjs, 173, 267

\bibitem[{{S{\'a}nchez} {et~al.}(2012){S{\'a}nchez}, {Kennicutt}, {Gil de Paz},
  {van de Ven}, {V{\'{\i}}lchez}, {Wisotzki}, {Walcher}, {Mast}, \& {et
  al.}}]{Sanchez-12}
{S{\'a}nchez}, S.~F., {Kennicutt}, R.~C., {Gil de Paz}, A., {et~al.} 2012,
  \aap, 538, A8

\bibitem[{{S{\'a}nchez} {et~al.}(2018){S{\'a}nchez}, {Avila-Reese},
  {Hernandez-Toledo}, {Cortes-Su{\'a}rez}, {Rodr{\'{\i}}guez-Puebla},
  {Ibarra-Medel}, {Cano-D{\'{\i}}az}, {Barrera-Ballesteros}, \& {et
  al.}}]{Sanchez-18}
{S{\'a}nchez}, S.~F., {Avila-Reese}, V., {Hernandez-Toledo}, H., {et~al.} 2018,
  \rmxaa, 54, 217

\bibitem[{{Schmidt}(1959)}]{Schmidt-59}
{Schmidt}, M. 1959, \apj, 129, 243

\bibitem[{{Shen} {et~al.}(2003){Shen}, {Mo}, {White}, {Blanton}, {Kauffmann},
  {Voges}, {Brinkmann}, \& {Csabai}}]{Shen-03}
{Shen}, S., {Mo}, H.~J., {White}, S.~D.~M., {et~al.} 2003, \mnras, 343, 978

\bibitem[{{Shi} {et~al.}(2011){Shi}, {Helou}, {Yan}, {Armus}, {Wu}, {Papovich},
  \& {Stierwalt}}]{Shi-11}
{Shi}, Y., {Helou}, G., {Yan}, L., {et~al.} 2011, \apj, 733, 87

\bibitem[{{Shi} {et~al.}(2018){Shi}, {Yan}, {Armus}, {Gu}, {Helou}, {Qiu},
  {Gwyn}, {Stierwalt}, \& {et al.}}]{Shi-18}
{Shi}, Y., {Yan}, L., {Armus}, L., {et~al.} 2018, \apj, 853, 149

\bibitem[{{Shibuya} {et~al.}(2015){Shibuya}, {Ouchi}, \&
  {Harikane}}]{Shibuya-Ouchi-Harikane-15}
{Shibuya}, T., {Ouchi}, M., \& {Harikane}, Y. 2015, \apjs, 219, 15

\bibitem[{{Silk}(1997)}]{Silk-97}
{Silk}, J. 1997, \apj, 481, 703

\bibitem[{{Silk} \& {Nusser}(2010)}]{Silk-Nusser-10}
{Silk}, J., \& {Nusser}, A. 2010, \apj, 725, 556

\bibitem[{{Smee} {et~al.}(2013){Smee}, {Gunn}, {Uomoto}, {Roe}, {Schlegel},
  {Rockosi}, {Carr}, {Leger}, \& {et al.}}]{Smee-13}
{Smee}, S.~A., {Gunn}, J.~E., {Uomoto}, A., {et~al.} 2013, \aj, 146, 32

\bibitem[{{Smethurst} {et~al.}(2015){Smethurst}, {Lintott}, {Simmons},
  {Schawinski}, {Marshall}, {Bamford}, {Fortson}, {Kaviraj}, \& {et
  al.}}]{Smethurst-15}
{Smethurst}, R.~J., {Lintott}, C.~J., {Simmons}, B.~D., {et~al.} 2015, \mnras,
  450, 435

\bibitem[{{Spindler} {et~al.}(2018){Spindler}, {Wake}, {Belfiore}, {Bershady},
  {Bundy}, {Drory}, {Masters}, {Thomas}, \& {et al.}}]{Spindler-18}
{Spindler}, A., {Wake}, D., {Belfiore}, F., {et~al.} 2018, \mnras, 476, 580

\bibitem[{{Tacchella} {et~al.}(2016{\natexlab{a}}){Tacchella}, {Dekel},
  {Carollo}, {Ceverino}, {DeGraf}, {Lapiner}, {Mandelker}, \&
  {Primack}}]{Tacchella-16a}
{Tacchella}, S., {Dekel}, A., {Carollo}, C.~M., {et~al.} 2016{\natexlab{a}},
  \mnras, 458, 242

\bibitem[{{Tacchella} {et~al.}(2016{\natexlab{b}}){Tacchella}, {Dekel},
  {Carollo}, {Ceverino}, {DeGraf}, {Lapiner}, {Mandelker}, \& {Primack
  Joel}}]{Tacchella-16b}
---. 2016{\natexlab{b}}, \mnras, 457, 2790

\bibitem[{{Tacchella} {et~al.}(2015){Tacchella}, {Carollo}, {Renzini},
  {Schreiber}, {Lang}, {Wuyts}, {Cresci}, {Dekel}, \& {et al.}}]{Tacchella-15}
{Tacchella}, S., {Carollo}, C.~M., {Renzini}, A., {et~al.} 2015, Science, 348,
  314

\bibitem[{{Talbot} \& {Arnett}(1975)}]{Talbot-Arnett-75}
{Talbot}, Jr., R.~J., \& {Arnett}, W.~D. 1975, \apj, 197, 551

\bibitem[{{Toft} {et~al.}(2007){Toft}, {van Dokkum}, {Franx}, {Labbe},
  {F{\"o}rster Schreiber}, {Wuyts}, {Webb}, {Rudnick}, \& {et al.}}]{Toft-07}
{Toft}, S., {van Dokkum}, P., {Franx}, M., {et~al.} 2007, \apj, 671, 285

\bibitem[{{Trujillo} {et~al.}(2006){Trujillo}, {F{\"o}rster Schreiber},
  {Rudnick}, {Barden}, {Franx}, {Rix}, {Caldwell}, {McIntosh}, \& {et
  al.}}]{Trujillo-06}
{Trujillo}, I., {F{\"o}rster Schreiber}, N.~M., {Rudnick}, G., {et~al.} 2006,
  \apj, 650, 18

\bibitem[{{Utomo} {et~al.}(2017){Utomo}, {Bolatto}, {Wong}, {Ostriker},
  {Blitz}, {Sanchez}, {Colombo}, {Leroy}, \& {et al.}}]{Utomo-17}
{Utomo}, D., {Bolatto}, A.~D., {Wong}, T., {et~al.} 2017, \apj, 849, 26

\bibitem[{{van den Bosch} {et~al.}(2008){van den Bosch}, {Aquino}, {Yang},
  {Mo}, {Pasquali}, {McIntosh}, {Weinmann}, \& {Kang}}]{vandenBosch-08}
{van den Bosch}, F.~C., {Aquino}, D., {Yang}, X., {et~al.} 2008, \mnras, 387,
  79

\bibitem[{{van der Wel} {et~al.}(2014){van der Wel}, {Franx}, {van Dokkum},
  {Skelton}, {Momcheva}, {Whitaker}, {Brammer}, {Bell}, \& {et
  al.}}]{vanderWel-14}
{van der Wel}, A., {Franx}, M., {van Dokkum}, P.~G., {et~al.} 2014, \apj, 788,
  28

\bibitem[{{Wake} {et~al.}(2012){Wake}, {van Dokkum}, \&
  {Franx}}]{Wake-vanDokkum-Franx-12}
{Wake}, D.~A., {van Dokkum}, P.~G., \& {Franx}, M. 2012, \apjl, 751, L44

\bibitem[{{Wang} {et~al.}(2018{\natexlab{a}}){Wang}, {Kong}, \&
  {Pan}}]{Wang-Kong-Pan-18}
{Wang}, E., {Kong}, X., \& {Pan}, Z. 2018{\natexlab{a}}, \apj, 865, 49

\bibitem[{{Wang} {et~al.}(2017){Wang}, {Kong}, {Wang}, {Wang}, {Lin}, {Gao}, \&
  {Liu}}]{Wang-17}
{Wang}, E., {Kong}, X., {Wang}, H., {et~al.} 2017, \apj, 844, 144

\bibitem[{{Wang} {et~al.}(2018{\natexlab{b}}){Wang}, {Li}, {Xiao}, {Lin},
  {Bershady}, {Law}, {Merrifield}, {Sanchez}, \& {et al.}}]{Wang-18a}
{Wang}, E., {Li}, C., {Xiao}, T., {et~al.} 2018{\natexlab{b}}, \apj, 856, 137

\bibitem[{{Wang} {et~al.}(2018{\natexlab{c}}){Wang}, {Wang}, {Mo}, {Lim}, {van
  den Bosch}, {Kong}, {Wang}, {Yang}, \& {et al.}}]{Wang-18b}
{Wang}, E., {Wang}, H., {Mo}, H., {et~al.} 2018{\natexlab{c}}, \apj, 860, 102

\bibitem[{{Wang} {et~al.}(2018{\natexlab{d}}){Wang}, {Mo}, {Chen}, {Yang},
  {Yang}, {Wang}, {van den Bosch}, {Jing}, \& {et al.}}]{Wang-18c}
{Wang}, H., {Mo}, H.~J., {Chen}, S., {et~al.} 2018{\natexlab{d}}, \apj, 852, 31

\bibitem[{{Wang} {et~al.}(2012){Wang}, {Kauffmann}, {Overzier}, {Tacconi},
  {Kong}, {Saintonge}, {Catinella}, {Schiminovich}, \& {et al.}}]{Wang-12}
{Wang}, J., {Kauffmann}, G., {Overzier}, R., {et~al.} 2012, \mnras, 423, 3486

\bibitem[{{Weinmann} {et~al.}(2006){Weinmann}, {van den Bosch}, {Yang}, \&
  {Mo}}]{Weinmann-06}
{Weinmann}, S.~M., {van den Bosch}, F.~C., {Yang}, X., \& {Mo}, H.~J. 2006,
  \mnras, 366, 2

\bibitem[{{Willett} {et~al.}(2015){Willett}, {Schawinski}, {Simmons},
  {Masters}, {Skibba}, {Kaviraj}, {Melvin}, {Wong}, \& {et al.}}]{Willett-15}
{Willett}, K.~W., {Schawinski}, K., {Simmons}, B.~D., {et~al.} 2015, \mnras,
  449, 820

\bibitem[{{Williams} {et~al.}(2010){Williams}, {Quadri}, {Franx}, {van Dokkum},
  {Toft}, {Kriek}, \& {Labb{\'e}}}]{Williams-10}
{Williams}, R.~J., {Quadri}, R.~F., {Franx}, M., {et~al.} 2010, \apj, 713, 738

\bibitem[{{Yan}(2018)}]{Yan-18}
{Yan}, R. 2018, \mnras, 481, 476

\bibitem[{{Yan} {et~al.}(2016){Yan}, {Tremonti}, {Bershady}, {Law}, {Schlegel},
  {Bundy}, {Drory}, {MacDonald}, \& {et al.}}]{Yan-16}
{Yan}, R., {Tremonti}, C., {Bershady}, M.~A., {et~al.} 2016, \aj, 151, 8

\bibitem[{{Yang} {et~al.}(2007){Yang}, {Mo}, {van den Bosch}, {Pasquali}, {Li},
  \& {Barden}}]{Yang-07}
{Yang}, X., {Mo}, H.~J., {van den Bosch}, F.~C., {et~al.} 2007, \apj, 671, 153

\bibitem[{{Zhang} {et~al.}(2017){Zhang}, {Yan}, {Bundy}, {Bershady}, {Haffner},
  {Walterbos}, {Maiolino}, {Tremonti}, \& {et al.}}]{Zhang-17}
{Zhang}, K., {Yan}, R., {Bundy}, K., {et~al.} 2017, \mnras, 466, 3217

\bibitem[{{Zolotov} {et~al.}(2015){Zolotov}, {Dekel}, {Mandelker}, {Tweed},
  {Inoue}, {DeGraf}, {Ceverino}, {Primack}, \& {et al.}}]{Zolotov-15}
{Zolotov}, A., {Dekel}, A., {Mandelker}, N., {et~al.} 2015, \mnras, 450, 2327

\end{thebibliography}

\clearpage

%%%%%%%%%%%%The End%%%%%%%%%%%%%%%%%%%%%%%%%%%%%%%%%%%%%%%%%
\label{lastpage}
\end{document}